\documentstyle[12pt,amsmath,amssymb,graphicx]{article}
\begin{document}

\begin{center}
{\large {\bf Formation of Jupiter and Conditions for Accretion of the 
Galilean Satellites}}
\end{center}

\begin{center}
{\bf P. R. Estrada, and I. Mosqueira}

{\bf SETI Institute} 
\end{center}

\vspace{0.05in}
\begin{center}
{\bf J. J. Lissauer, G. D'Angelo, and D. P. Cruikshank}

{\bf NASA Ames Research Center}
\end{center}

\noindent
\begin{center}
{\bf Abstract}
\end{center}

{\small We present an overview of the formation of Jupiter and its 
associated circumplanetary disk. Jupiter forms via a combination 
of planetesimal accretion and gravitational accumulation of gas from the 
surrounding solar nebula. The formation of the circumjovian gaseous
disk, or subnebula, straddles 
the transitional stage between runaway gas accretion and Jupiter's eventual 
isolation from the circumsolar disk. This isolation, which effectively signals 
the termination of Jupiter's accretion, takes place as Jupiter opens a 
deep gas gap in the solar nebula, or the solar nebula gas dissipates. 
The gap-opening stage is relevant to
subnebula formation because the radial extent of the circumjovian disk is 
determined by the specific angular momentum of gas that enters Jupiter's 
gravitational sphere of influence. Prior to opening a well-formed, deep gap
in the circumsolar disk, Jupiter accretes low specific angular momentum gas 
from its vicinity, resulting in the formation of a rotationally-supported compact 
disk whose size is comparable to the radial extent of the Galilean satellites. 
This process may allocate similar amounts of angular momentum to the planet and
the disk, leading to the formation of an {\it ab-initio} massive disk compared to
the mass of the satellites. 
As Jupiter approaches its final mass and the gas gap deepens, 
a more extended, less massive disk forms because the gas inflow, which must come
from increasingly farther away from the planet's semimajor axis,
has high specific angular momentum. Thus, 
the size of the circumplanetary gas disk upon inflow is dependent on whether or not 
a gap is present. 
We describe the conditions for accretion of the Galilean satellites,
including the timescales for their formation and mechanisms for their survival, 
all within the context of key constraints for satellite formation models. The
environment in which the regular satellites form is tied to the timescale for 
circumplanetary disk dispersal, which depends on the nature and persistence of
turbulence. In the case that subnebula turbulence decays as gas inflow wanes,
we present a novel mechanism for satellite survival involving gap opening by
the largest satellites. On the other hand, assuming that sustained turbulence
drives subnebula evolution on a short timescale compared to the satellite formation
timescale, we review a model that emphasizes collisional processes to explain
satellite observations.
We briefly discuss the mechanisms by which solids may be delivered to the 
circumplanetary disk. 
At the tail end of Jupiter's accretion, most of the mass in solids resides in 
planetesimals of size $> 1$ km; however, planetesimals in Jupiter's feeding zone 
undergo a period of intense collisional grinding, placing a significant amount 
of mass in fragments $< 1$ km.
Inelastic 
or gravitational collisions within Jupiter's gravitational sphere of 
influence allow for the mass contained in these planetesimal fragments to be 
delivered to the circumplanetary disk either through direct collisional/gravitational 
capture, or via ablation through the circumjovian gas disk. We expect
that planetesimal delivery 
mechanisms likely provide the bulk of material for satellite accretion.}

\newpage

\begin{center}
{\bf 1. INTRODUCTION}
\end{center}
 
The prograde, low inclination orbits and close spacing of the regular 
satellites of the four giant planets indicate that they formed within
a circumplanetary disk
around their parent planet. The fact that all of the giant planets of our
Solar System harbor families of regular satellites suggests that satellite 
formation may be a natural, if not inevitable, consequence of giant planet formation. 
Thus to understand the origins of the Jovian satellites, 
we must understand the formation of Jupiter and its circumplanetary disk.

Observations provide some basic constraints on Jupiter's origin. Jupiter must 
have formed prior to the dissipation of the solar nebula because H and He (the
planet's primary constituents) do not condense at any location in the 
protoplanetary disk; therefore these light elements were accumulated in gaseous 
form. Protoplanetary disks are observed to have lost essentially all of their 
gases in $< 10^7$ years ({\it Meyer et al.}, 2007), e.g., by photoevaporation
({\it Shu et al.}, 1993; {\it Dullemond et al.}, 2007), implying that Jupiter 
must have completed its formation by then. A fundamental 
question of giant planet formation is whether Jupiter formed  
as a result of a gravitational instability that occurred 
in the solar nebula gas, or through accretion of a solid core that eventually 
became large enough to accumulate a gaseous envelope directly from the nebula. 
The evidence supports the latter formation scenario in which the early stage 
of planetary growth consists of the accumulation of planetesimals, in a manner analogous 
to the consensus treatment of terrestrial planets ({\it Lissauer and Stevenson}, 
2007; cf. {\it Durisen et al.}, 2007). The heavy element core grows sufficiently 
large ($\sim 5-15$ M$_\oplus$, where M$_\oplus = 5.976\times 10^{27}$ g is an 
Earth mass) that it is able to accrete gas from the 
surrounding nebula ({\it Hubickyj et al}., 2005). The planet's final mass is determined 
by either the opening of a gas gap in the protoplanetary disk,
or the eventual dissipation of the 
solar nebula. Jupiter's atmosphere is enhanced in condensible material by a factor 
of $\sim 3-4$ times solar; Jupiter as a whole is enhanced in heavy elements by a
factor of $\sim 3-8$ compared to the Sun.
These observations suggest that the Jovian subnebula
could also have been enhanced in solids with respect to
solar composition mixtures, though the way in which the disk was enhanced may differ
from that of the planet (Sec. 4.2).

The enhancement of solids in Jupiter's gaseous envelope, the composition
of its core, and the source material that went into making the
Jovian satellite system, likely reflect the composition of planetesimals 
that formed within its vicinity as well as farther from the Sun in the circumsolar 
disk. 
The composition of outer solar system bodies such as comets contain
grains with a wide range of volatility 
({\it Brownlee et al.}, 2006; {\it Zolensky et al.}, 2006). 
Dust samples from Comet 81P/Wild 2 brought to Earth by the Stardust space 
mission show a great diversity of high- and moderate-temperature minerals, 
in addition to organic material (e.g., {\it Brownlee et al}., 2006; {\it 
Sandford et al}., 2006). 
The implication from these observations is that
there was a significant amount of radial mixing within the early solar 
nebula leading to the transport of high temperature minerals from the 
innermost regions of the solar nebula to beyond the orbit of Neptune.
These high temperature minerals could have been transported 
through ballistic transport away from the midplane ({\it Shu et al.}, 2001), or
turbulent transport in the midplane (e.g., {\it Scott and Krot}, 2005; 
{\it Cuzzi and Weidenschilling}, 2006; {\it Natta et al.}, 2007). 
Based on the ``X-wind'' model, 
Shu and colleagues predicted that there would be transport of CAIs (calcium- aluminum-rich
inclusions) from near the Sun to the outer edge of the Solar System where Comet
Wild 2 formed. Once transported, these minerals
were incorporated into the first generation of planetesimals 
that went into making Jupiter and the outer planets.  
The presence of 
super-solar abundances of more volatile species such as argon (which requires 
very low temperatures to condense) in Jupiter's atmosphere likely implies that, in 
addition to outward transport of high temperature material, there was also 
inward transport of {\it highly} volatile material within solid bodies 
(e.g., {\it Cuzzi and Zahnle}, 2004; 
see Sec. 2.4). Similar migration might also have characterized circumplanetary nebulae 
at some stage in their evolution.

The formation of giant planet satellite systems has been treated as a
poorly understood extension of giant planet formation
(see, however, early work by {\it Coradini et al}., 
1981; 1982). Thanks in great part to the success of the Voyager and Galileo 
missions, the Jovian satellite system has been studied extensively over the last
three decades; the emphasis has been on the Galilean satellites 
Io, Europa, Ganymede, and Callisto. 
The cosmochemical and dynamical properties of the Jovian satellites may provide
important clues about the late and/or post-accretional stages of Jupiter's 
formation ({\it Pollack and Reynolds}, 1974; 
{\it Mosqueira and Estrada}, 2003a,b; {\it Estrada and Mosqueira}, 2006).

The abundances and the chemical forms of constituents observed in the Jovian 
satellite system are a diagnostic of their source material (i.e., either the
source material condensed in the solar nebula or it was re-processed/condensed in 
the subnebula). However, interpretation of their cosmochemical properties is
complicated because objects the size of the Galilean 
satellites could have been 
altered through extensive resurfacing due to differentiation, energetic impacts, 
and other forms of post-accretional processing. Callisto and Ganymede 
have mean densities of $1.83$ and $1.94$ g cm$^{-3}$,
respectively; each may consist of $\sim 50\%$ rock, and 
$\sim 50\%$ water-ice by mass, with Ganyemede being slightly more rock rich 
({\it Sohl et al.}, 2002). In contrast, the solids
component of solar composition material may be closer to
$\sim 60\%$ water-ice, $\sim 40\%$ rock by mass, although the ice/rock ratio in the
solar nebula is uncertain since it depends on the carbon speciation
({\it Wong et al.}, 2008).
This variation from solar composition may indicate 
that Callisto
and Ganymede lost some of their total water inventory during their accretion. However, 
the total amount of 
water in Europa ($3.01$ g cm$^{-3}$) is only $\sim 10\%$ by mass ({\it Sohl et al.},
2002). If the initial composition was like that of its outer 
neighbors, its current state requires additional mechanisms to explain its 
water loss. On the other hand, Io has a mean density of $3.53$ g cm$^{-3}$ 
consistent with rock.
Thus, a key feature of the Galilean satellite system is the strong monotonic
variation in ice fraction with distance from Jupiter. 

Yet, despite this remarkable radial trend in water-ice fraction, the largest of
the tiny moons orbiting interior to Io, Amalthea, has
a density so low that even models for its composition that include 
high choices for its porosity require that water-ice be a major constituent 
({\it Anderson et al.}, 2005). 
The lack of mid-sized moons, which could provide additional clues to 
the relative abundances of rocky and icy material, complicates our attempts to 
understand the cosmochemical history of the Jovian system. Satellites 
in the Jovian system are found 
to contain evidence of hydrocarbons through their spectral signatures in the 
low-albedo surface materials (e.g., {\it Hibbitts et al}., 2000, 2003), 
and possibly CN-bearing materials ({\it McCord 
et al}., 1997). Whether these materials are intrinsic or extrinsic
(e.g., a coating from impactors) in origin is not known. Contrast this with 
the Saturnian system, where the average density of the regular, icy satellites 
($\sim 1.3$ g cm$^{-3}$) may allow us to infer a non-solar composition or the 
presence of a significant amount of low-density hydrocarbons (e.g., 
{\it Cruikshank et al}., 2007a,b). A caveat is that the 
Saturnian mid-sized satellites may have been subject to disruptive collisions 
(see Sec. 4.5.1). 

Is there accessible information on Ganymede and Callisto that can distinguish
between material processed in the Jovian subnebula and material processed in
the solar nebula? Both bodies show surface deposits of water ice ({\it Calvin
et al.}, 1995), and CO$_2$ is found in small quantities on both these Galilean
satellites ({\it Hibbitts et al.}, 2001). While the surface water ice is
probably native material, the CO$_2$ is largely associated with the non-ice
regions, and the CO$_2$ spectral band is shifted in wavelength from its nominal
position measured in the lab, indicating that it is probably complexed with
other molecules ({\it Chaban et al.}, 2006). The CO$_2$ is likely to have
been synthesized locally by the irradiation of water and a source of carbon,
perhaps from micrometeorites. If the chemistry in the subnebula occurred at
pressures in the range $10^{-3}$ to $10^{-4}$ bar as the gas cooled, CO gas
and water would have condensed early on (e.g., {\it Fegley}, 1993). The 
condensation of water may trap CO ({\it Stevenson and Lunine}, 1988) if the
pressure is high, thus altering the C/O ratio in the remaining gas and affecting
formation and growth of mineral grains.
Unfortunately, molecules such as CH$_4$, NH$_3$,
and native CO$_2$ and CO, which would give important clues to the conditions 
in which the Galilean satellites formed would help distinguish between gas-poor 
and gas-rich models, for example, are not detected on the surfaces of any
of these bodies (though we note that CO and CH$_4$ are too volatile in general to 
be stable on the surfaces of these bodies).
 
Although one can think of Jupiter and its satellites as a 
``miniature Solar System'', there are several significant differences between 
accretion in a circumplanetary disk and the solar nebula. In the circumplanetary
disk, thermodynamical properties,
length scales, and dynamical times are quite different from their circumsolar 
analog. Most of these differences can be attributed to the fact that
giant planet satellite systems are compact.  
For instance, the giant planets dominate the angular momentum budget of the 
Solar System; however, the analogy does not extend to the Jovian system 
where the majority of the system angular momentum resides in 
Jupiter's spin. Moreover, dynamical times, which can influence a number of 
aspects of the accretion process, are orders of magnitude faster in 
circumplanetary disks than in the circumsolar disk (accretion of planets in
the habitable zones of
M dwarf stars are an intermediate case, {\it Lissauer}, 2007). As 
an example, Europa's orbital period is $\sim 10^{-3}$ times that 
of Jupiter. 
 
Over the last decade, there have been a number of global models of the Jovian 
satellite system designed to fit the observational constraints provided 
by the Voyager and Galileo missions ({\it Mosqueira and Estrada}, 2000, 2003a,b; 
{\it Estrada}, 2002; {\it Canup and Ward}, 2002, and others; {\it Mousis and 
Gautier}, 2004; {\it Alibert et al.}, 2005a; {\it Estrada and Mosqueira}, 2006).
Key constraints for models of the Galilean 
satellite system are as follows:

\begin{enumerate}
\item The mass and angular momentum contained within the Galilean satellites serve
as firm constraints on the system. Aside from explaining the overall values, 
one must also address why there is so much mass and angular momentum in Callisto, 
yet no regular satellites outside its orbit. 

\item The increase in ice fraction with distance from the planet for the
Galilean satellites (a.k.a., compositional gradient) is often attributed to a 
temperature gradient imposed by the proto-Jupiter. 
However, it is debatable whether 
the compositions of Io and Europa, specifically, are primordial or altered 
by mechanisms (e.g., tidal heating, hypervelocity impacts) that would 
preferentially strip away volatiles. Moreover, Amalthea's low density 
strongly suggests that water-ice is a major constituent of this inner, small satellite.
 
\item The moment of inertia of Callisto suggests a partially differentiated 
interior provided hydrostatic equilibrium is assumed ({\it Anderson et al.}, 
1998; 2001). This would imply that Callisto formed 
slowly ($\gtrsim 10^5$ years, {\it Stevenson et al.}, 1986). However, 
it may be possible 
that a nonhydrostatic component in Callisto's core could be large enough to 
mask complete differentiation ({\it McKinnon}, 1997).

\item The nearly constant mass ratio of the largest satellites of Jupiter
and Saturn (and possibly Uranus) to their parent planet,
$\mu \sim 10^{-4}$, suggests that there may be a truncation 
mechanism that affects satellite mass ({\it Mosqueira and Estrada}, 
2003b). 

\end{enumerate}

This chapter is organized as follows. In Section 2, we present a detailed 
summary of the currently favored model for the formation of Jupiter. 
In Section 3, we discuss the 
formation of the circumjovian disk. 
In Section 4, we address the possible environments 
that arise for satellite formation during the late stages of Jupiter's accretion. 
We touch on the processes involved in the accretion of Europa and the other
Galilean satellites, the methods of solids mass delivery, timescales for 
formation, and discuss the key constraints in more detail. In Section 5, we 
present a summary. 
 
\vspace{0.25in}
\begin{center}
{\bf 2. FORMATION OF JUPITER}
\end{center}
 
Jupiter's accretion is thought to occur in a protoplanetary disk with
roughly the same elemental composition as the Sun; that is, primarily H and He
with $\sim 1-2\%$ of heavier elements. A lower bound on the mass of this 
solar disk of $\sim 0.01-0.02$ M$_\odot$ (where M$_\odot = 1.989\times 10^{33}$ g 
is the mass of the Sun) has been derived from taking 
the condensed elemental fractions of the planets of the Solar System and 
reconstituting them with enough volatiles to make the composition solar. 
This is historically referred to as the ``minimum mass solar nebula'' (MMSN, 
{\it Weidenschilling}, 1977b; {\it Hayashi}, 1981). The radial distribution of 
solids is found by smearing the augmented mass of the planet over a region 
halfway to each neighboring planet. This approach yields a trend in surface 
density (by design, both in solids and gas) with semimajor axis, $r$, of 
$\Sigma \propto r^{-3/2}$ between Venus and Neptune ({\it Weidenschilling}, 
1977b). 
The radial temperature of the photosphere of the solar nebula in these earlier
models has been commonly taken to be that derived from assuming that the disk is 
optically thin (to 
its own emission), and dust grains are in radiative balance with the solar 
luminosity ($T\propto r^{-1/2}$, {\it Pollack et al}., 1977; {\it Hayashi}, 
1981). This temperature dependence is one that might be expected from a scenario in 
which most of the dust has accumulated into larger bodies. 

In this picture, it is implicitely assumed that the planets 
form near their present locations. This means that planetary formation is 
characterized by ``local growth'' -- that is, the planets accreted from the 
reservoir of gas and solids that condensed within their vicinity. 
In order to explain the abundances above solar of heavy volatile elements
in Jupiter's atmosphere,
the distribution of solids at the location of Jupiter is often assumed to be
enhanced by a factor of several above the MMSN. 
However, the Jovian atmosphere contains enhancements above solar in 
highly volatile noble gases such as argon (see 
Sec. 2.4), so that even with a modification in the surface density, a local 
growth model likely does not do a good enough job of explaining the composition 
of Jupiter's atmosphere because it would require local temperatures much lower 
than what is predicted by an optically thin nebula model. 

Efforts to 
match the spectral energy distributions of circumstellar disks around T Tauri 
stars have spawned a number of alternative disk models that lead to different thermal disk structures (e.g., {\it Kenyon and Hartmann}, 1987; {\it Chiang and Goldreich}, 1997;
{\it Garaud and Lin}, 2007) that are considerably lower than what was assumed
previously. Even with these lower temperature models, though, the implication is
that Jupiter would need to have accreted material from much farther out in the
circumsolar disk. If planetesimals migrated over such large distances, 
it suggests that the $r^{-3/2}$ trend in surface density derived for the MMSN model 
is not a particularly useful constraint. These uncertainties in the most fundamental of 
disk properties underlie some of the reasons why the formation of giant planets 
remains one of the more scrutinized problems in planetary cosmogony today 
(see, e.g., recent reviews by {\it Wuchterl et al.}, 2000; 
{\it Hubbard et al.}, 2002; {\it Lissauer and Stevenson}, 2007).

Of the two formation models for giant planets that have received the most 
attention, the preponderance of evidence supports the formation of Jupiter 
via {\it core nucleated accretion}, which relies on a combination of planetesimal 
accretion and gravitational accumulation of gas. The alternative, the 
so-called {\it gas instability model}, would have Jupiter forming directly from 
the contraction of gaseous clump produced through a gravitational instability 
in the protoplanetary disk. In the gas instability model, the formation of Jupiter 
is somewhat akin
to star formation. Although numerical calculations have produced $\sim 1$ M$_J$
(where M$_J = 1.898\times 10^{30}$ g is a Jupiter mass)
clumps given sufficiently unstable disks (e.g., {\it Boss}, 2000; {\it Mayer et 
al}., 2002), these clumps do not form unless the disk is highly atypical (very massive, 
and/or very hot; {\it Rafikov}, 2005). Furthermore, unless there are 
processes that keep the disk unstable, weak gravitational instabilities 
lead to stabilization of the protoplanetary disk via the excitation of spiral 
density waves. These waves carry away angular momentum that spread the disk, lowering 
its surface density. Given the lack of observational support, along with 
theoretical arguments against the formation of Jupiter via fragmentation 
({\it Bodenheimer et al}., 2000a; {\it Cai et al.}, 2006), the gas instability 
model for Jupiter will not be considered in this chapter.
For more in-depth discussions of the gas 
instability model, see {\it Wuchterl et al.} (2000), {\it Lissauer and 
Stevenson} (2007), and {\it Durisen et al}. (2007).

In the core nucleated accretion model ({\it Pollack et al.}, 1996; 
{\it Bodenheimer et al.}, 2000b; {\it Hubickyj et al.}, 2005; {\it Alibert
et al.}, 2005b; {\it Lissauer et al.}, 2009), Jupiter's formation and evolution 
is thought to occur in the 
following sequence: (1) Dust particles in the solar nebula form planetesimals 
that accrete, resulting in a solid core surrounded by a low mass 
gaseous envelope. Initially, runaway accretion of solids occurs, and the 
accretion rate of gas is very slow. As the solid material in the planet's 
feeding zone is depleted, the rate of solids accretion 
tapers off. The ``feeding zone'' is defined by the separation distance between a 
massive planet and a massless body on circular orbits such that they never experience 
a close encounter ({\it Lissauer}, 1995).
The gas accretion rate steadily increases and eventually exceeds 
the accretion rate of solids. (2) Proto-Jupiter continues to grow as the gas 
accretes at a relatively constant rate. The mass of the solid core also 
increases, but at a slower rate. Eventually, the core and envelope masses 
become equal. (3) At this point, the rate of gas accretion increases in 
runaway fashion, and proto-Jupiter grows at a rapidly accelerating rate. 
The first three parts of the evolutionary sequence are referred to as  
the $\it{nebular}$ stage, because the outer boundary of the protoplanetary 
envelope is in contact with the solar nebula, and the density and temperature 
at this interface are those of the nebula. (4) The gas accretion rate reaches 
a limiting value defined by the rate at which the nebula can transport gas to 
the vicinity of the planet ({\it Lissauer et al.}, 2009). After this point, 
the equilibrium region of 
proto-Jupiter contracts, and gas accretes hydrodynamically into this 
equilibrium region.  This part of the evolution is considered to be the 
$\it{transition}$ stage. (5) Accretion is stopped by either the opening of a 
gap in the gas disk as a consequence of the tidal effect of Jupiter, 
accumulation of all nearby gas, or by dissipation of the nebula. Once accretion 
stops, the planet enters the $\it{isolation}$ stage. Jupiter then contracts 
and cools to the present state at constant mass. In the following subsections, 
we present a more detailed summary of this formation sequence.

\vspace{0.1in}
\noindent
{\bf 2.1. From Dust to Planetesimals}
\vspace{0.1in}

Sufficiently far away from the Sun where temperatures are low enough to 
allow for the condensation of solid material, 
grain growth begins with sticking of sub-micron sized dust, 
composed of surviving interstellar grains and condensates which formed within 
the protoplanetary disk. These small grains are dynamically coupled to the 
nebula gas, and collide at low (size-dependent) relative velocities that can 
be caused by a variety of mechanisms ({\it V\"{o}lk et al}., 1980; {\it 
Weidenschilling}, 1984; {\it Nakagawa et al}., 1986; {\it Weidenschilling and 
Cuzzi}, 1993; {\it Ossenkopf}, 1993; {\it Cuzzi and Hogan}, 2003; {\it Ormel 
and Cuzzi}, 2007). Low impact velocities (and thus low impact energies) tend 
to allow small particles to grow more efficiently because collisions can be 
completely inelastic (e.g., {\it Wurm and Blum}, 1998). However, as grains 
collide to form larger and larger agglomerates, the level of coupling between 
growing particles and the gas decreases, and particle velocities relative to 
both other particles and the gas increases. As a result, larger particles begin 
to experience higher impact energies during interparticle collisions which can 
lead to fragmentation or erosion, rather than growth.

When the level of coupling between particles and the gas decreases, dust grains,
which can be initially suspended at distances well above and below the solar 
nebula midplane, begin to gravitationally settle toward the midplane. As 
particles settle, they tend to grow. The time it takes for a particle to 
settle is a function of its size, and may be hundreds, thousands, or more orbital 
periods depending on ambient nebula conditions. Very small particles (e.g., 
sub-micron grains) have such long settling times relative to the lifetime of
the gas disk that they are considered to be fully ``entrained'' in the gas. 
On the other end, some particles may become large enough that their settling time
is comparable to their orbital period. 
Although technically these particles ``couple'' with the gas once per
orbit, they are essentially viewed as the transition size from a regime
in which particles collectively move with the gas, 
to a regime in which particles
would prefer to move at the local Kepler orbital speed. 
These transitional
particles are referred to as the ``decoupling size'', and as they settle to the 
nebular midplane they continue to grow by sweeping up dust and rubble 
({\it Cuzzi et al.}, 1993; {\it Weidenschilling}, 1997).
 
The fact that particles tend to drift relative to the surrounding gas indicates that 
the gas component itself does not rotate at the local Kepler orbital velocity,
$v_K$. This is because the nebula gas is not supported against the Sun's gravity 
by rotation alone. In a rotating frame 
there are three forces at work on a gas parcel in the nebula disk: a 
gravitational force radially directed toward the Sun; a centrifugal force 
directed radially outward from the nebula's rotation axis; and (because the 
gas is not pressureless) an outward directed pressure gradient force that works 
to counter the effective gravity. The condition for  
equilibrium 
then requires that the nebula gas orbit at a velocity slightly less than $v_K$. 
Solid objects, which might otherwise orbit at the local Keplerian orbital speed, 
are too dense to be 
supported by the pressure gradient, and are subject to a drag force that 
can systematically remove their angular momentum leading to orbital decay ({\it Weidenschilling}, 1977a). 

Thus, a more formal definition of the decoupling size is the size at which
the time needed for the gas drag force to dissipate a particles momentum relative
to the gas (known 
as the ``stopping time'') is similar to its orbital period. At Jupiter, this size 
is $R\sim 1$ m for the canonical MMSN model (e.g., {\it Cuzzi et al.}, 1993). 
Since decoupling-size particles encounter the strongest drag force, they tend 
to have the most rapid inward orbital migration. For example, such a particle 
(assuming 
no growth) would eventually spiral in to the Sun in $\sim 10^4-10^5$ years. This 
time is short compared to planetary accretion timescales 
($\gtrsim 10^6-10^7$ years); but, as objects grow larger, the drag force on 
them decreases and the stopping times can become quite long. Kilometer-sized 
planetesimals are mostly unaffected by the gas due to their greater 
mass-to-surface-area ratio. Thus, getting from decoupling-size objects to 
the safety of relatively immobile ($\gtrsim 1$ km) planetesimals is a key issue
in planet formation.

Whether growth beyond the decoupling size happens
depends on the turbulent viscosity $\nu$ in the disk. 
In general, the dissipation of viscous energy leads to the transport of mass 
inwards, facilitating further accretion onto the central object. Angular 
momentum is transported
outwards, causing the disk to spread (Sec. 4.1). The efficiency of the mechanism of 
angular momentum transport is most commonly characterized by a turbulent parameter 
$\alpha \propto \nu$ ({\it Shakura and Sunyaev}, 1973). 
The level of turbulence is important because collisional 
velocities for decoupling-size objects can reach tens of meters per second in 
turbulence, which has more of a tendency to fragment them than to promote growth. 
This suggests that, while the incremental growth of sufficiently small grains and 
dust may take place irrespective of the level of nebula turbulence, successful 
growth past the decoupling size may require very low levels of turbulence (e.g., 
{\it Youdin and Shu}, 2002; {\it Cuzzi and Weidenschilling}, 2004) which helps to 
overcome the two main obstacles for this critical size --  collisional 
disruption and rapid orbital migration. Several mechanisms have 
been proposed in the literature to explain planetesimal growth under non-turbulent 
(laminar) conditions ({\it Safronov}, 1960; {{\it Goldreich and Ward}, 1973; {\it Weidenschilling}, 1997; {\it Sekiya}, 1998; {\it Youdin and Chiang}, 2004). Yet, 
in spite of the difficulties associated with potentially destructive collisions, 
mechanisms that attempt to explain various stages of planetesimal formation 
under turbulent conditions have also recently been advanced 
(see {\it Cuzzi and Weidenschilling}, 2006; 
{\it Cuzzi et al.}, 2007; {\it Johansen et al.}, 2007). 

One can argue on the basis of these works that growth under 
non-turbulent conditions provides a relatively straightforward pathway to planetesimal 
formation, while growth in the turbulent regime, if feasible, is a more complicated process 
with many stages. 
Indeed, the transition from agglomerates to 
planetesimals continues to provide the major stumbling block in planetary 
origins ({\it Weidenschilling}, 1997; 2002, 2004; {\it Weidenschilling and 
Cuzzi}, 1993; {\it Stepinski and Valageas}, 1997; {\it Dullemond and Dominik}, 
2005; {\it Cuzzi and Weidenschilling}, 2006; {\it Dominik et al}., 2007).

\vspace{0.1in}
\noindent
{\bf 2.2. Core Accretion}
\vspace{0.1in}

The initial stage of Jupiter's formation entails the accretion of its solid 
core from the available reservoir of heliocentric planetesimals. 
Once planetesimals grow large enough, gravitational interactions and physical
collisions between pairs of 
solid planetesimals provide the dominant perturbation of their basic Keplerian 
orbits. At this stage, effects that were more influential in the earlier stages 
of growth such as electromagnetic forces, collective gravitational effects, and 
in most circumstances gas drag, play minor roles. These planetesimals continue to 
agglomerate via pairwise mergers, with the rate of solid body accretion by a 
planetesimal or planetary embryo (basically a very large planetesimal) being 
determined by the size and mass of the planetesimal/planetary embryo, the surface 
density of planetesimals, and the distribution of planetesimal velocities
relative to the accreting body. 

The planetesimal velocity distribution is probably 
the most important factor that controls the growth rate of planetary embryos into 
the core of a giant planet. As larger objects accrete, gravitational scatterings 
and elastic collisions can convert the ordered relative motions of orbiting 
planetesimals (i.e., Keplerian shear) into random motions, and can ``stir up'' 
the planetesimal random velocities up to the escape speed from the largest 
planetesimals in the swarm ({\it Safronov}, 1969). The effects of this gravitational
stirring, however, tend to be balanced by collisional damping, because inelastic 
collisions (and, for smaller objects, gas drag) can damp eccentricities and 
inclinations. 
 
If one assumes that planetesimal pairwise collisions lead to perfect accretion, 
i.e., that all physical collisions are completely inelastic (fragmentation, 
erosion, and bouncing do not occur), this stage of growth can be initially quite 
rapid. 
With this assumption, the planetesimal accretion rate, 
$\dot{M}_Z$, is:

\begin{equation}
                     \dot{M}_Z = \pi  R^2  \sigma_Z  \Omega  F_g ,
\end{equation}

\noindent
where $R$ is the radius of the accreting body, $\sigma_Z$ is the surface density 
of solid planetesimals in the solar nebula, $\Omega$ is the orbital frequency 
at the location of the growing body, and $F_g$ is the gravitational enhancement 
factor. The 
gravitational enhancement factor $F_g$ arises from the ratio of the distance of 
close approach to the asymptotic unperturbed impact parameter,
and in the 2-body approximation (ignoring the tidal effects of the Sun's 
gravity) it is given by:

\begin{equation}
                      F_g = 1 + \left(\frac{v_e}{v}\right)^2 .
\end{equation}

\noindent
Here, $v_e$ is the escape velocity from the surface of the planetary embryo, and 
$v$ is the velocity dispersion of the planetesimals being accreted. 
The evolution 
of the planetesimal size distribution is determined by $F_g$, which favors 
collisions between bodies having larger masses and smaller relative velocities. 
Moreover, they can
accrete almost every planetesimal they collide with (i.e., the perfect accretion
approximation works best for the largest bodies).

As the planetesimal size distribution evolves, planetesimals (and planetary
embryos) may pass through different growth regimes. These growth regimes are 
sometimes characterized as either {\it orderly} or {\it runaway}. When the 
relative velocities between planetesimals is comparable to or larger than the 
escape velocity of the largest body, 
$v \gtrsim v_e$, the growth rate $\dot{M}_Z$ is approximately 
proportional to $R^2$. This implies that the growth in radius is roughly constant 
(as can be easily derived from Eq. 1). Thus the evolutionary path of the 
planetesimals exhibits an orderly growth across the entire size distribution 
so that planetesimals containing most of the mass double their masses at least 
as rapidly as the largest particle. When the relative velocity is small, 
$v \ll v_e$, the gravitational enhancement factor $F_g \propto R^2$, and so
the growth rate $\dot{M}_Z$ is proportional to $R^4$. By virtue of its
large, gravitationally enhanced cross-section, the most massive
embryo doubles its mass faster than the smaller bodies do, and detaches itself 
from the mass distribution ({\it Levin}, 1978; {\it Greenberg et al.}, 1978;
{\em Wetherill and Stewart}, 1989; {\em Ohtsuki et 
al.}, 2002).  

Eventually the runaway body can grow so large (its $F_g$ can exceed $\sim 1000$)
that it transitions from dispersion-dominated growth to shear-dominated growth
({\em Lissauer}, 1987). This means that for these extremely low random 
velocities ($v\ll v_e$), the rate at which planetesimals encounter the growing 
planetary embryo is determined by the Keplerian shear in the planetesimal disk, 
and not by the random motions of the planetesimals. At this stage, larger embryos 
take longer to double in mass than do smaller ones, although embryos of all masses 
continue their runaway growth relative to surrounding planetesimals. This phase of 
rapid accretion of planetary embryos is known as {\it oligarchic} growth 
({\em Kokubo and Ida}, 1998).

Rapid runaway or oligarchic accretion requires low random velocities, and thus 
small radial excursions of planetesimals. The planetary embryo's feeding zone 
is therefore limited to the annulus of planetesimals which it can gravitationally 
perturb into embryo-crossing orbits. Rapid growth stops when a planetary embryo 
has accreted most of the planetesimals within its feeding zone. Thus, 
runaway/oligarchic growth is self-limiting in nature, which implies that massive 
planetary embryos form at regular intervals in semimajor axis. The agglomeration 
of these embryos into a small number of widely spaced bodies necessarily requires 
a stage characterized by large orbital eccentricities. The large velocities 
implied by these large eccentricities imply small collision cross-sections 
(Eq. 2) and hence long accretion times. Growth via binary (pairwise) collisions 
proceeds until the spacing of planetary orbits become dynamically isolated from 
one another, i.e., spacing sufficient for the configuration to be stable to gravitational
interactions among the planets for the lifetime of the system
({\em Safronov}, 1969; {\em Wetherill}, 1990; {\em Lissauer}, 1993, 1995; 
{\em Agnor et al.}, 1999; {\em Laskar}, 2000; {\em Chambers}, 2001). 

For shear-dominated accretion, the mass at which such a planetary embryo 
becomes isolated from the surrounding circumsolar disk via runaway accretion
is given by ({\it Lissauer}, 1993):

\begin{equation}
                      M_{iso} = \frac{(8\pi\sqrt3r^2\sigma_Z)^{3/2}}{(3M_{\odot})^{1/2}}
\approx 1.6\times 10^{25}\left(\sigma^{3/2}_Zr^3_{\rm{AU}}\right)\,\,\rm{g},
\end{equation}

\noindent 
where $r_{\rm{AU}}$ is the distance from the Sun in astronomical units ($\rm{AU} =
1.496\times 10^{13}$ cm). For the MMSN in 
which only local growth is considered, the mass at which runaway accretion would
have ceased in Jupiter's accretion zone is $\sim 1$ M$_\oplus$ ({\it Lissauer}, 1987),
which is likely too small to explain Jupiter's formation (see below). Rapid
accretion can persist beyond the isolation mass if additional solids can diffuse 
into its feeding zone ({\it Kary et al.}, 1993; {\it Kary and Lissauer}, 1995). 
%(thus continued growth is non-local). 
There are three 
plausible mechanisms for such diffusion: scattering between planetesimals, 
perturbations by neighboring planetary embryos, and migration of smaller 
planetesimals due to gas drag (Sec. 2.1). Other mechanisms that can lead
to the migration of embryos from regions that are depleted in planetesimals into
regions that are not depleted include gravitational torques resulting from the
excitation of spiral density waves in the gaseous component of the disk (see Sec.
4.3.3), and dynamical friction (if significant energy is transferred from the
planetary embryo to the protoplanetary disk; e.g., see {\em Stewart and Wetherill}, 
1988).

In the inner part of protoplanetary disks, Kepler shear is too great to allow the 
accretion of solid planets larger than a few M$_\oplus$ on any timescale 
unless the surface densities are considerably above that of the MMSN or a large 
amount of radial migration occurs. However, the fact that the relatively small 
terrestrial planets orbit deep within the Sun's potential well suggests that 
they likely were unable to eject substantial amounts of material from the inner
Solar System. Thus, the total amount of mass present in the terrestrial region 
during the runaway accretion epoch was probably not much more than the current 
mass of the terrestrial planets, implying that a high-velocity growth phase 
subsequent to runaway accretion was necessary in order to explain their present 
configuration ({\it Lissauer}, 1995). 

This high-velocity final growth stage takes O(10$^8$) years in the 
terrestrial planet zone ({\em Safronov}, 1969; {\em Wetherill}, 1980; 
{\em Agnor et al.}, 1999; {\em Chambers}, 2001), but would require 
O($10^9-10^{10}$) years in the giant planet zone ({\em Safronov}, 1969) if 
one assumes local growth in a MMSN disk. These timescales reflect a 
slowing down of the accretion rate during the late stages of planetary growth due 
to a drop in the planetesimals surface density ({\it Wetherill}, 1980, 1986).
The growth timescales quoted above are far longer than any modern estimates of the lifetimes 
of gas within protoplanetary disks ($\lesssim 10^7$ years, {\it Meyer et al.}, 2007),
implying that 
Jupiter's core must grow large enough during the rapid runaway/oligarchic growth. 
The epoch of runaway/rapid oligarchic growth lasts 
only millions of years or less near the location of Jupiter ($r = 5.2$ AU), but 
can produce $\sim$ 10 M$_\oplus$ cores if the circumsolar disk is enhanced in 
solids by only a few times the MMSN ({\em Lissauer}, 1987). The limits on the 
initial surface density of the disk are less restrictive in the giant planet 
region, because excess solid material can be ejected to the Oort cloud, or out 
of the Solar System altogether.  

The masses at which planets become isolated from the disk thereby terminating the 
runaway/rapid oligarchic growth epoch are likely to be comparably large at greater 
distances from the Sun. However, at these large distances, 
random velocities of planetesimals must remain quite small for accretion rates to 
be sufficiently rapid for planetary embryos to approach $M_{iso}$ within the 
lifetimes of gaseous disks ({\it Pollack et al.}, 1996). Indeed, if planetesimal 
velocities become too large, material is more likely to be ejected to interstellar
space than accreted by the planetary embryos.

\begin{figure}[t!]
 \resizebox{\linewidth}{!}{%
 \includegraphics{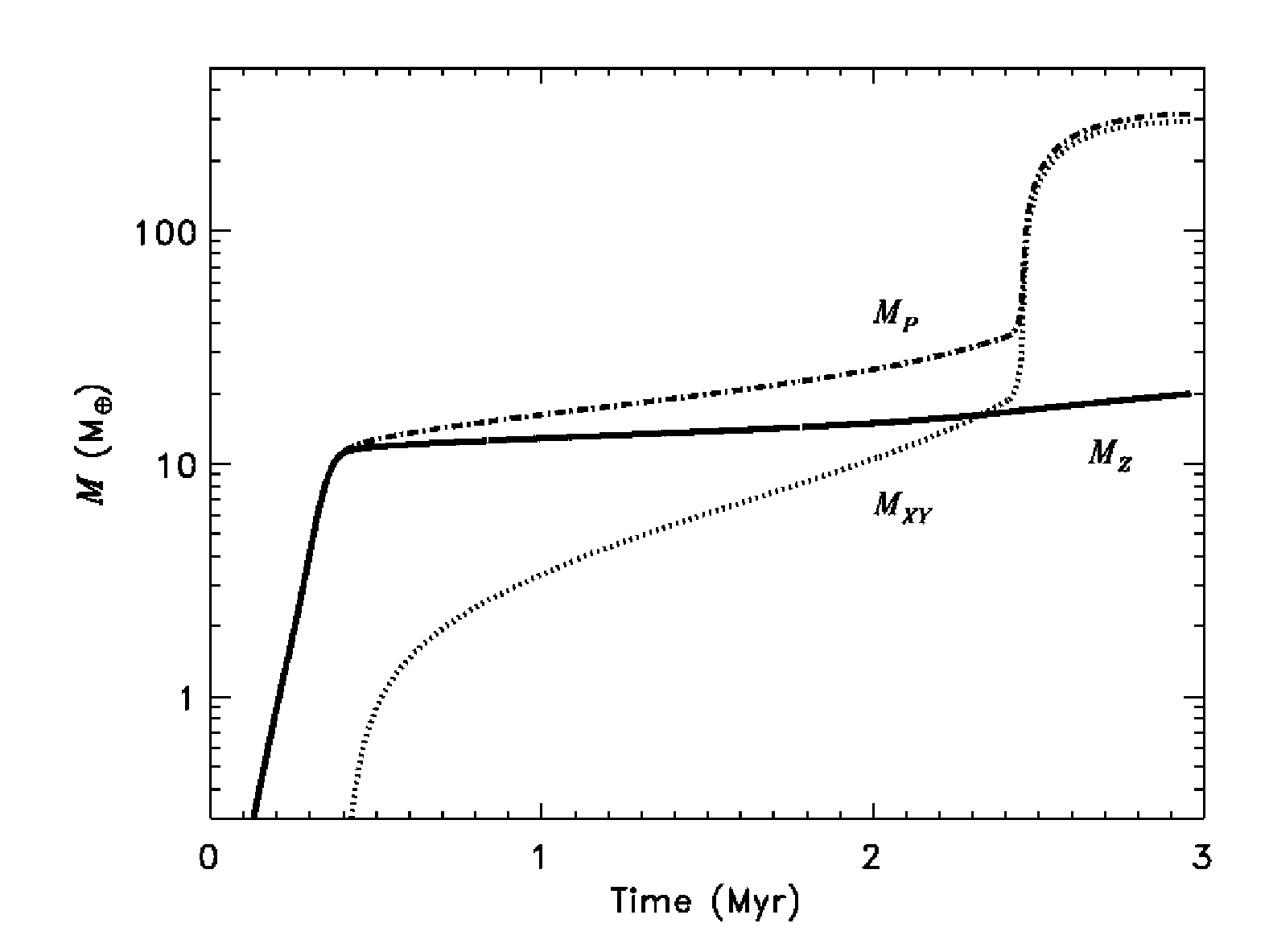}%
 \includegraphics{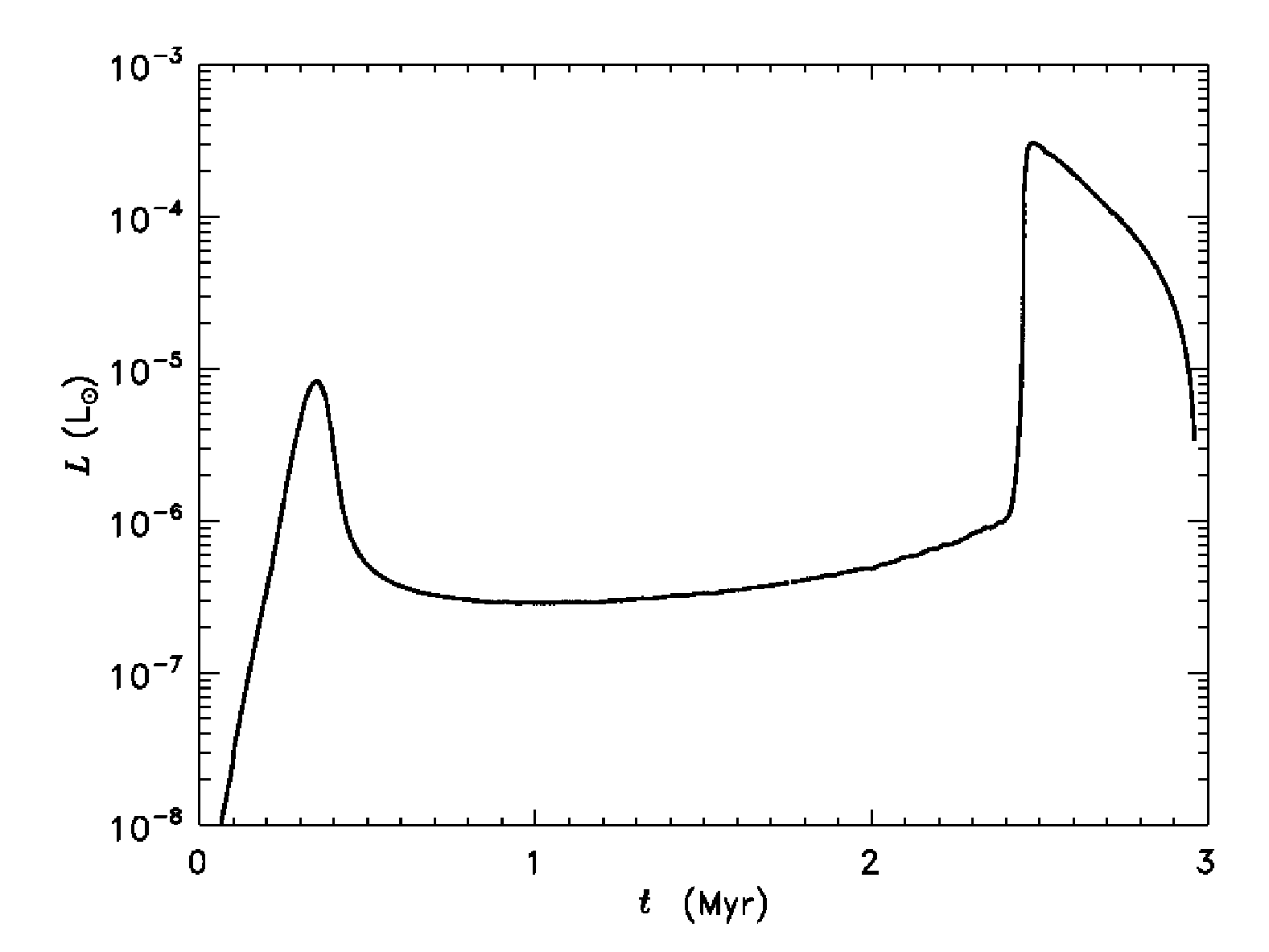}}
\caption{{\small Evolution of a proto-Jupiter within a protoplanetary disk with
surface density of solids
$\sigma_{Z} = 10$ g cm$^{-2}$ and grain opacity in the protoplanet's envelope
assumed to be at $2\%$ of the interstellar value.  
Details of the calculation are presented in {\em Lissauer et al.} (2009).
\underline{Left panel:} The mass of solids in the planet (solid curve), gas in
the planet (dotted curve), and the total mass of the planet (dot-dashed curve)
are shown as functions of time. Note the slow, gradually increasing, buildup of
gas, leading to a rapid growth spurt, and finally a slow tail off in accretion.
(Courtesy O. Hubickyj)
\underline{Right panel:} The planet's luminosity  is shown as a function of
time. The rapid contraction of the planet just before $t = 2.5$ Myr coincides
with the highest luminosity and the epoch of most rapid gas accretion. From
{\it Lissauer et al}., (2009).}}
\label{fig:planacc}
\end{figure}

\vspace{0.1in}
\noindent
{\bf 2.3. Gas Accretion}
\vspace{0.1in}

\vspace{0.1in}
{\it 2.3.1. Tenuous Extended Envelope Phase}. As the core grows, its 
gravitational potential well gets deeper, allowing its gravity to 
pull gas from the surrounding nebula towards it. At this stage, the core may 
begin to accumulate 
a gaseous envelope. A planet of order 1 M$_\oplus$ (the value for a specific
planet depends upon the mean molecular weight and opacity of the atmosphere;
{\it Stevenson}, 1983) is able to capture 
an atmosphere because the escape speed from its physical surface is large compared 
to the thermal velocity (or sound speed) $c_s$ of the surrounding gaseous 
protoplanetary disk. Initially, all gas with gravitational binding energy to the
planet larger than the thermal energy is retained as part of the planet
({\it Cameron et al.}, 1982; {\it Bodenheimer and Pollack}, 1986).
The radial extent of this region is $\sim GM_p/c_s^2$, where $G$ is the gravitational
constant. 
If the protoplanet is small, this bound region can be 
significantly smaller than the protoplanet's gravitational domain,
whose size is typically a significant fraction of the protoplanet's Hill sphere, 
$R_H$, which is given by:

\begin{equation}
                     R_H = \left(\frac{M_p}{3M_\odot}\right)^{1/3}r .
\end{equation}

\noindent
Here, $M_p$ is the mass of the protoplanet, and $r$ is the distance between the
protoplanet and the Sun. The Hill radius denotes the distance from a planet's
center along the planet-Sun line at which the planet's gravity equals the tidal 
force of the solar gravity relative to the planet's center.
As the protoplanet increases in mass, the region in which gas can be bound 
increases and may become a significant fraction of $R_H$.

An embryo begins to accrete gas slowly, so its gaseous envelope is initially 
optically thin and isothermal with the surrounding protoplanetary disk. 
As the envelope gains mass it becomes optically thick to outgoing thermal radiation, and 
its lower reaches can get much warmer and denser than the gas in the surrounding 
protoplanetary disk. As the protoplanet's gravity continues to pull in gas from 
the surrounding disk towards it, thermal pressure from the existing envelope
limits further accretion. This is because, for much of its 
gas accretion stage, the key factor limiting the protoplanet's accumulation of 
gas is its ability to radiate away the gravitational energy provided by the 
continued accretion of planetesimals and the contraction of the envelope; this 
energy loss is necessary for the envelope to further contract and allow more gas 
to enter the protoplanet's gravitational domain.  

As the energy released by the accretion of planetesimals and gas is radiated 
away at the protoplanet's photosphere, the photosphere cools and a subsequent 
pressure drop causes the envelope to contract.
This is referred to as a Kelvin-Helmholtz contraction. Compression 
heats the envelope and regulates the rate of contraction which, in turn, controls 
how rapidly additional gas can enter the planet's gravitational domain and be 
accreted.  

This suggests that the rate and manner in which a giant planet accretes solids 
can substantially affect its ability to attract gas. Initially accreted solids 
form the planet's core (Sec. 2.2), around which gas is able to accumulate. 
Calculated gas accretion rates are very strongly increasing functions of the 
total mass of the protoplanet, implying that rapid growth of the core is a key 
factor in enabling a protoplanet to accumulate substantial quantities of gas. 
Continued accretion of solids acts to reduce the protoplanet's growth time by 
increasing the depth of its gravitational potential well, but also counters 
growth by providing additional thermal energy to the envelope from solids that
sink to the core. Another hurdle to rapid growth that planetesimal accretion 
provides is the increased atmospheric opacity from dust grains that are 
released (ablated) in the upper parts of the envelope. If the opacity is 
sufficiently high, much of the growing planet's envelope transports energy via 
convection.  However, the distended very low density outer region of the envelope 
has thermal gradients that are too small for convection, but are large enough 
that they can act as an efficient thermal blanket if it is sufficiently dusty 
to be moderately opaque to outgoing radiation. This has the effect of slowing 
contraction and frustrating further accretion of gas, and lengthening the 
timescale for planet accretion.  

Figure~\ref{fig:planacc} shows the evolution of the mass and luminosity
from a recent model of Jupiter's formation ({\it Lissauer et al.}, 2009). 
During the 
runaway planetesimal accretion epoch (when the core is predominantly being 
formed), the protoplanet's mass increases rapidly. Although
at this point the gaseous atmosphere is quite tenuous, the 
internal temperature and thermal pressure of the envelope increases, which 
prevents substantial amounts of nebular gas from falling onto the protoplanet. 
When the rate of planetesimal accretion decreases (roughly around 
$M_p \gtrsim 10$ M$_\oplus$), gas falls onto the protoplanet more rapidly as the 
additional component of thermal energy contributed by the accreting planetesimals 
decreases. At this stage the envelope mass is a sensitive function 
of the total mass, with the gaseous fraction increasing rapidly as the planet 
accretes ({\em Pollack et al.}, 1996).  When the envelope 
reaches a mass comparable to that of the core, the self-gravity of the gas becomes 
substantial, and the envelope contracts when more gas is added. Eventually, increases 
in the planet's mass and the radiation of energy allow the envelope to shrink rapidly.
Further accretion 
is then governed by the availability of gas rather than thermal considerations. 
At this point, the factor limiting the planet's growth rate is the flow of gas 
from the surrounding protoplanetary disk ({\it Lissauer et al.}, 2009).

The time required to reach this stage of rapid gas accretion is governed primarily 
by three factors: the mass of the solid core; the rate of energy input from continued 
accretion of solids; 
and the opacity of the envelope. 
These three factors appear to be key in determining whether giant 
planets are able to form within the lifetimes of protoplanetary disks 
($\lesssim 10^7$ years). For example, in a disk with initial $\sigma_Z=10$ g 
cm$^{-2}$ at 5.2 AU from a $1$ M$_\odot$ star, a planet whose atmosphere has 
$2\%$ interstellar opacity forms with a 
$16$ M$_{\oplus}$ core in 2.3 Myr; in the same disk, a planet whose atmosphere 
has full interstellar opacity ($\sim 1$ cm$^2$ g$^{-1}$) forms with a $17$ 
M$_{\oplus}$ core in 6.3 Myr; a planet whose atmosphere has $2\%$ interstellar 
opacity but stops accreting solids at $10$ M$_{\oplus}$ forms in 0.9 Myr, whereas 
if solids  accretion is halted at $3$ M$_{\oplus}$ accretion of a massive 
envelope requires 12 Myr ({\em Hubickyj et al.}, 2005).  These results suggest that 
if Jupiter's core mass is significantly less than $10$ M$_{\oplus}$, then it 
presents a problem for formation models mainly because disk dispersal times 
are observed to be shorter than the time it takes for a smaller core mass to 
accrete a massive enough envelope (unless the opacity of the envelope to outgoing
radiation is significantly less than $2\%$ of the interstellar medium).

Thus, the key to forming Jupiter prior to the dispersal of the nebula is the 
rapid formation of a massive core coupled with a combination of a decreased
solids accretion rate and/or the outer regions of the giant planet envelope
being transparent to outgoing radiation. However, since there is little in the 
way of observational constraints, our understanding continues to be handicapped by
uncertainties in quantities such as the opacity and solids accretion rate that are 
derived from planet formation models. 
The compositions of the atmospheres of the giant planets may provide some insight.
As the envelope becomes more massive, 
late-accreting planetesimals (but, early-arriving in the context of
satellite formation, see Sec. 4) sublimate before they can reach the core, thereby 
enhancing the heavy element content of the envelope considerably. In Sec. 2.4, we
discuss more on the composition of Jupiter's envelope.

\begin{figure}[t!]
 \begin{minipage}[c]{0.5\textwidth}
\resizebox{\linewidth}{!}{%
 \includegraphics{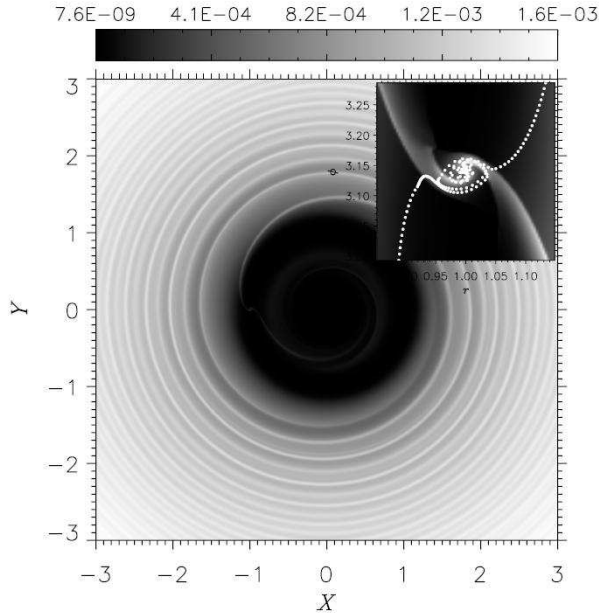}}
\label{fig:flow}
 \end{minipage}\hspace*{\fill}%
 \begin{minipage}[c]{0.5\textwidth}
\caption{{\small The surface mass density of a gaseous disk 
containing a Jupiter-mass planet on a circular orbit located 5.2 AU from a 
$1$ M$_\odot$ star.  The ratio of the scale height of the disk to the distance 
from the star is 1/20, and the dimensionless viscosity at the location of the 
planet is $\alpha = 4 \times 10^{-3}$. The distance scale is in units 
of the planet's orbital distance, and surface density of $10^{-4}$ 
corresponds to 33 g cm$^{-2}$.
The inset at the upper right shows a close-up of the disk region around the 
planet, plotted in 
cylindrical coordinates. The two series of white dots indicate actual 
trajectories (real
particle paths, not streamlines) of material that is captured in the 
gravitational well of the planet and eventually accreted by the planet. 
See {\em D'Angelo et al.}, (2005) for a description 
of the code used.  }
        }
\label{fig:flow}
 \end{minipage}
\end{figure}

\vspace{0.1in}
{\it 2.3.2. Hydrodynamic Phase}.
As demonstrated in Fig. \ref{fig:planacc}, a protoplanet accumulates gas 
at a gradually increasing rate until its gas component is comparable to its 
heavy element mass (i.e., the envelope and core are of comparable mass). At this 
point, the protoplanet has enough mass for its self-gravity to compress the 
envelope substantially. The rate of gas accretion then accelerates rapidly,
and a gas runaway occurs ({\em Pollack et al.}, 1996; {\em Hubickyj et al.}, 
2005). This accretion continues as long as there is gas in the vicinity of
the protoplanet's orbit. The ability of the protoplanet 
to accrete gas does not depend strongly on the outer boundary conditions 
(temperature and pressure) of the surrounding nebula, if there is adequate gas 
around to be accreted ({\it Mizuno}, 1980; {\it Stevenson}, 1982; {\it Pollack 
et al.}, 1996). Hydrodynamic limits allow quite rapid gas flow to the planet 
in an unperturbed disk. But in realistic scenarios, the protoplanet not 
only alters the disk by accreting material from it, but also by exerting 
gravitational torques on it (see Sec. 4). Both of these processes can lead to 
a formation of a gap in the circumsolar disk (e.g., {\it Lin and 
Papaloizou}, 1979) and isolation of the planet from the surrounding gas, thus 
prividing a means of limiting the final mass of the giant planet. 
 
Observationally, such gravitationally induced gaps have been 
observed around small moons within Saturn's rings ({\em Showalter}, 1991; 
{\em Porco et al.}, 2005). Numerically, gas gap formation has been studied 
extensively. For example, {\em D'Angelo et al.}, (2003b) used
a 3D adaptive mesh refinement code to follow the flow of gas onto 
accreting giant planets of various masses embedded within a gaseous protoplanetary 
disk. {\em Bate et al.}, (2003) have performed 3D simulations of this problem 
using the ZEUS hydrodynamics code. Using parameters appropriate for a moderately 
viscous MMSN protoplanetary disk at 5 AU
($\alpha \sim 4\times 10^{-3}$, see Sec. 3.3), both groups found that 
$< 10$ M$_\oplus$ planets don't perturb the protoplanetary disk enough to 
significantly affect the amount of gas that flows towards them. Gravitational 
torques on the disk by larger planets under these disk conditions drive away gas. 
In a moderately viscous disk, hydrodynamic limits on gas accretion reach 
to a $\rm{few}\times 10^{-2}$ M$_\oplus$ per year for planets in the 
$\sim 50 - 100$ M$_\oplus$ range, and then decline as the planet continues 
to grow.  An example of gas flow around/to a $1$ M$_J$ planet is 
shown in Figure~\ref{fig:flow}. In general, caution must be exercised
in the interpretation of these types of calculations when attempting to 
connect them with the formation of the giant planet itself. 
Thus, for example, these calculations do not include the thermal pressure 
on the nebula from the hot planet, which is found to be the major 
accretion-limiting factor for planets up to a few tens of M$_\oplus$ by the 
simulations discussed in Section 2.3.1 ({\it Hubickyj et al}, 2005; {\it
Lissauer and Stevenson}, 2007; {\it Lissauer et al.}, 2009). It should 
be noted that ability for a protoplanet to open a gap is dependent on the viscosity
of the disk. In nearly inviscid disks, for example, a $\sim 10$ M$_\oplus$ protoplanet
may be capable of opening a gap ({\it Rafikov}, 2002a,b; Sec. 4.3.3).

If the planet successfully cuts off its supply of gas by the opening of a gap
in the circumsolar disk, the planet effectively enters the isolation stage. Jupiter 
then contracts and cools to its present state at constant mass.

\begin{figure}[t!]
 \resizebox{\linewidth}{!}{%
 \includegraphics{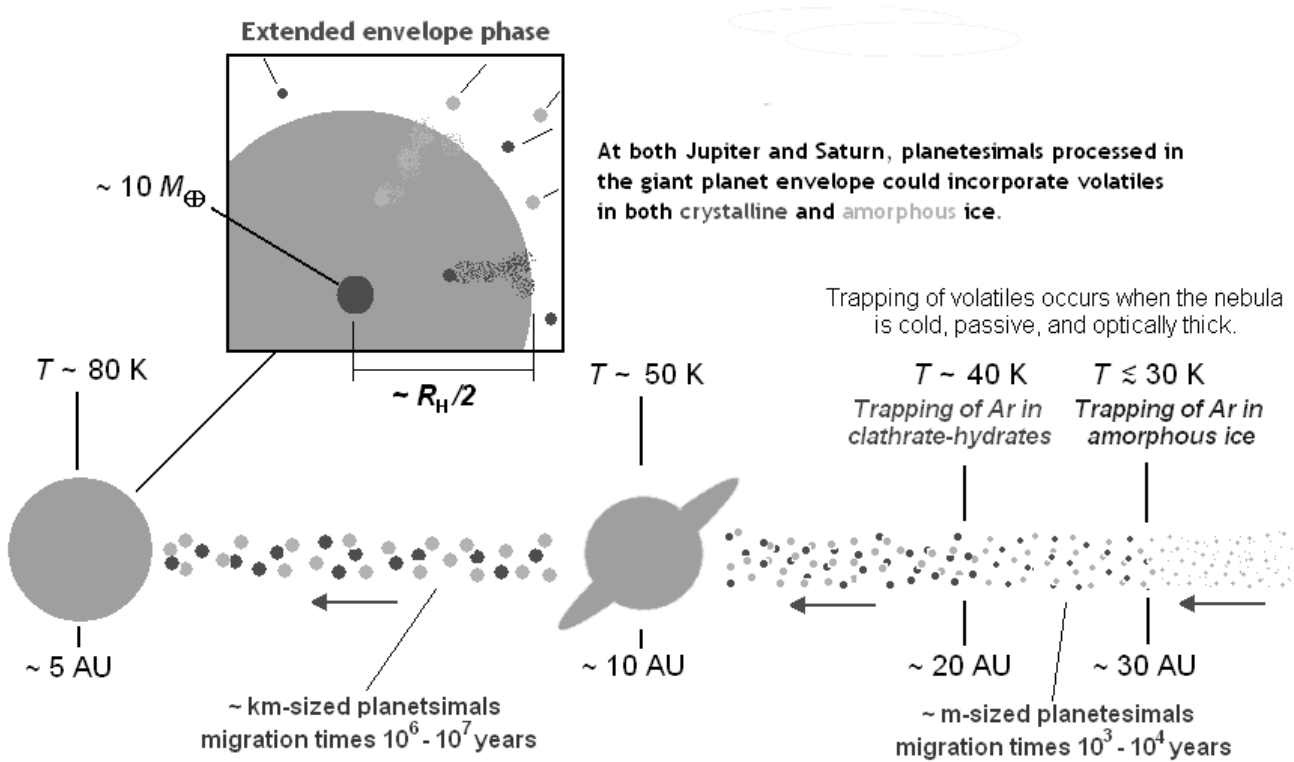}}
\caption{{\small Sufficiently far from the Sun ($\sim 30$ AU), amorphous ice
forming at low temperatures ($\lesssim 30$ K) can trap
volatiles ({\it Owen et al}., 1999). In warmer
regions of the nebula closer to the Sun, ice is
crystalline and volatiles may be trapped in clathrate
hydrates ({\it Lunine and Stevenson}, 1985;
{\it Gautier et al}., 2001a,b). Planetesimals that form in cold
regions and cross into warmer regions suffer a
transition from amorphous to crystalline ice at a rate
that depends on temperature ({\it Schmitt et al}, 1989;
{\it Mekler and Podolak}, 1994). The condition
that this transition takes longer than
the lifetime of the nebula defines a location outside of
which amorphous ice can mix in provided cold
planetesimals are delivered by gas drag migration in a
similar timescale (which applies to
planetesimals $< 1$ km for a MMSN). An ice grain may retain its
amorphous state for the age of the solar nebula ($\sim 10^7$ years)
provided that the temperature is $< 85$ K.
Thus, it may be possible
to deliver volatiles to the atmospheres of the forming
giant planets either in amorphous ice or crystalline
ice, depending on trapping efficiency in the two ice-phases,
the initial distribution of mass in the primordial nebula,
and the specifics of the growth, migration and thermal evolution
of planetesimals.
This point of view provides a natural fit for the existence of the outer
edge of the classical Kuiper Belt -- that is, primordial planetesimals located
outside $\sim 30$ AU may have migrated to the inner Solar System by gas drag,
delivering volatiles and enhancing the solid fraction in the planetary
region (this outer edge location refers to the time prior
to the outward migration of Neptune, which
could have pushed the Kuiper Belt out; {\it Levison and Morbidelli}, 2003).
Note that the temperature chosen at the location of Jupiter and Saturn is consistent with
passive disk models (e.g., {\it Chiang et al.}, 2001).}}
\label{fig:vign1}
\end{figure}

\vspace{0.1in}
\noindent
{\bf 2.4. The Composition of Jupiter's Envelope}
\vspace{0.1in}

{\it 2.4.1. Enhancement in Heavy Elements}.
The elemental abundances of gases in Jupiter's atmosphere that are quite volatile, 
but unlike H and He still condensible within the giant planet region of the solar 
nebula, are about $\sim 3-4$ times solar ({\it Atreya et al}., 1999; 
{\it Mahaffy et al}., 2000; {\it Young}, 2003). If the relative abundances of
all {\it condensible} elements in Jupiter's envelope are the same as in the Sun, then
such material must account for $\sim 18$ M$_\oplus$ ({\it Owen and Encrenaz}, 
2003). This suggests that solar ratio solids must have been abundant in the 
early Solar System. However, present day evidence for this material remains 
elusive, because no solid objects (e.g., comets, asteroids) have been found 
that have solar ratios of Ar, Kr, Xe, S and N relative to C, as does Jupiter
(although solar S/C was found for Comet Halley by Giotto, the detection of 
noble gas and N$_2$ abundances in cometary comae is quite challenging).

Explanation for this enhancement has been
attributed to one of two different mechanisms. The first idea relies on
the delivery of volatiles from the outer regions of the solar nebula that 
were trapped in amorphous ice and then incorporated into planetesimals.
Based on the laboratory work of {\it Bar-Nun et al}. (1988; also see 2007), 
this approach initially led to the expectation that Jupiter may not be enhanced 
in volatile elements like Ar which condenses at very low temperatures ($< 30$ K)
because planetesimals that formed near the snow line likely dominated the 
delivery of heavy elements to Jupiter's envelope. 
A second view for Jupiter's enhancement proposed by
{\it Gautier et al}. (2001a,b; also cf. {\it Hersant et al}., 2004), is that
volatiles were trapped in crystalline ices in the form
of clathrate-hydrates ({\it Lunine and Stevenson}, 1985) at different
temperatures in Jupiter's feeding zone which were then incorporated 
into the planetesimals that went into Jupiter. 
But even though argon can be trapped in clathrates at temperatures above its
condensation temperature,
clathration of Ar still requires very low
solar nebula temperatures $T\sim 36$ K, which is inconsistent with
temperatures at Jupiter's location even using cool passive disk models
({\it Chiang and Goldreich}, 1997; {\it Sasselov and Lecar}, 2000; {\it Chiang
et al.}, 2001). 
 
Given that Ar does not condense at temperatures higher than $\gtrsim 30$ K, one might 
expect that the ratio of Ar to H in Jupiter should be the same in Jupiter as 
the Sun, if indeed Jupiter's formation were characterized by local growth 
(see discussion at beginning of Sec. 2). However, there likely was considerable
migration of solids due to gas drag in the outer solar nebula 
(see Figure~\ref{fig:vign1}), so it cannot be assumed that material
(of solar proportions) remained at the location in the 
protoplanetary disk where it condensed. This argument is bolstered by 
high-resolution submillimeter continuum observations that indicate the 
average dust disk sizes around T Tauri stars are $\approx 200$ AU 
({\it Andrews and Williams}, 2007), with similar results being obtained via 
millimeter interferometry ({\it Kitamura et al}., 2002). 
This disk of solids eventually shrinks (even if the gas disk
spreads outwards) presumably due to coagulation with objects eventually growing large 
enough to decouple from the gas and migrate inwards (Sec. 2.1). 
Some planetesimal formation takes place at sufficient distances
that the circumsolar disk is very cold (see Fig. \ref{fig:vign1}),
perhaps cold enough
to allow for the trapping of volatiles within the interiors of planetesimals in 
either amorphous or crystalline
(in the form of clathrates) ice, depending on the trapping efficiency and kinetics. As
these planetesimals drift in, they likely encounter warmer regions.
Further growth to comet sizes
$\sim 1$ km occurs at some point ({\it Weidenschilling}, 1997;
{\it Kornet et al}., 2004), at which time they attain drag times comparable 
to the lifetime of the circumsolar gas disk ($\lesssim 10^7$ years).

The success of this solids migration mechanism depends on whether 
inwardly migrating planetesimals that
possibly included amorphous ice within their interiors at the 
outset ({\it Mekler and Podolak}, 1994), 
were altered by the higher temperatures of the inner nebula
and became mostly crystalline, losing the volatiles that would
have been trapped in them ({\it Bar-Nun et al}., 1985, 1987).
Amorphous ice undergoes a transformation to crystalline ice at
a rate that depends strongly on temperature ({\it Schmitt et al}., 1989; 
{\it Kouchi et al}., 1994; {\it Mekler and Podolak}, 1994). An ice grain may 
retain its amorphous state for the lifetime of the nebula ($\lesssim 10^7$ years)
provided the temperature is $ < 85$ K. That is, low temperatures favor 
the preservation of amorphous ice, while long time spans and even temporarily 
elevated temperatures drive the ice toward crystallization.  
 
Thus, while the temperature constraint for incorporation of Ar in grains is 
quite stringent, the temperature constraint for planetesimals to {\it preserve} 
volatiles they acquired in cold portions of the disk may be much less so. Indeed,
a temperature of $ < 85$ K is quite consistent with cool passive disk 
models at the location of Jupiter ({\it Chiang and Goldreich}, 1997; {\it Chiang et
al.}, 2001; see Fig. 3). However, if
such planetesimals incorporated
significant amounts of short-lived radionuclides such as $^{26}$Al, 
radioactive decay would provide heating that would
further complicate the ability for planetesimals to preserve their amorphous state 
({\it Prialnik and Podolak}, 1995). Nevertheless,
in the outer disk one would naturally expect longer accretion times,
which would result in weaker radioactive heating and lower temperatures.
It remains to be shown whether one can expect amorphous ice to be
preserved within migrating planetesimals
over the lifetime of the solar nebula, making it possible to deliver argon and
other volatiles to Jupiter.
 
\vspace{0.1in}
{\it 2.4.2. The Snow Line and Planetesimal Delivery}.
A likely consequence of this picture is that a shrinking (dust) disk
would lead to a higher solids fraction in the planetary region than given by 
a MMSN. As we noted in Sec. 2.2, this is consistent with the requirement that 
the circumsolar disk be enhanced in solids by at least a factor of few in order for
$\sim 10$ M$_\oplus$ cores to be formed during the runaway/oligarchic growth 
phase. Subsequently, some of the ``excess'' material, variously estimated between
$50-100$ M$_\oplus$ ({\it Stern}, 2003; {\it Goldreich et al}., 2004), may
wind up in the Oort cloud. Since Earth-sized objects may migrate (not via gas drag,
but by gravitational interaction with the gaseous disk, see Sec. 4.3.3)
in a time shorter than or comparable to the nebula dissipation time,
some of this solid material may also be lost to the Sun.

On the other hand, meter-sized objects have relatively short gas drag migration 
times ($\sim 10^4-10^5$ years at Jupiter), so it is possible that some fraction of the
solid content of the disk drifted in until it encountered the snow line.  
Inwardly drifting planetesimals 
might then sublimate at this (water-ice) evaporation front (e.g., {\it Stevenson and
Lunine}, 1988;
{\it Ciesla and Cuzzi}, 2006). 
Within the context of
the model proposed by these workers, the snow line might receive
most of the volatile enhancement, even for heavy elements more volatile
than water. If the solar nebula were turbulent, then diffusion due to turbulence 
might spread the effects of this evaporation front over a larger region. 
If the snow line is somewhere between $3-5$ AU 
({\it Morfill and V\"{o}lk}, 1984; {\it Stevenson and Lunine}, 1988; 
{\it Cuzzi and Zahnle}, 2004), then one might expect that the 
volatile heavy element enrichment in Jupiter's envelope is due to the high-volatile 
content of the nebula at its location. We should emphasize that this scenario is 
complicated by the likely temporal variation in the location of the snow line. The
snow line may have been much closer to the Sun 
as would be predicted by more recent thermal structure models (e.g.,
{\it Sasselov and Lecar}, 2000; {\it Chiang et al.}, 2001). 

Regardless of the snow line's
temporal variation, the condensation front scenario would imply that Jupiter's
enrichment should be more pronounced than Saturn's. Yet this conclusion 
appears to be in conflict with observations of Saturn's elemental ratio with respect 
to solar of C/H ($\sim 2-6$, {\it Courtin et al.}, 1984; {\it Buriez and de 
Bergh}, 1981) and N/H ($\sim 2-4$, {\it Marten et al}., 1980; {\it de Pater and 
Massie}, 1985). The enhancement at the water-ice evaporation front 
would be dependent on how much of this material is delivered and how this material
is incorporated into the Jupiter region. If most of the highly volatile content of 
planetesimals (e.g., argon) vaporizes prior to either being mixed into the planet
or somehow ``embedded'' in the circumplanetary subdisk, the enhancement in Jupiter's
atmosphere may
still be explained. The delivery of volatile-ladden, pristine planetesimals may be a 
more efficient process, especially if the temperature at Jupiter's location is more
consistent with a passive disk (Fig. 3). Thus, it is presently unclear whether it 
is possible to
deliver and enhance heavy elements more volatile than water simultaneously at 
Jupiter (and Saturn) by enriching the nebula gas instead of direct planetesimal 
delivery. The latter would likely predict a {\it higher} heavy element enhancement for 
planets that accreted {\it less} gas from the nebula. 
 
\vspace{0.25in}
\begin{center}
{\bf 3. FORMATION OF THE CIRCUMJOVIAN DISK}
\end{center}
 
As Jupiter's core mass grows, it obtains a substantial atmosphere through
the collection of the surrounding solar nebula gas. Early on (still in the
nebular stage), this gas falls onto a distended envelope that extends out to a 
significant fraction of the Hill sphere. 
Once the protoplanet reaches a mass of $\sim 50-100$ M$_\oplus$, the
envelope contracts rapidly (transition stage). Eventually,
Jupiter becomes massive enough to truncate the gas disk by the opening of a deep
gas gap in the solar nebula and/or because all of the gas in its feeding zone is 
depleted. Once all the gas within reach of the planet is depleted,
accretion ends (isolation stage). We now address how
the circumplanetary gas disk fits into this multi-stage process.

\vspace{0.1in}
\noindent
{\bf 3.1. Introduction}
\vspace{0.1in}

A very basic characteristic of the circumjovian disk is its radial extent: how
far does it extend from the planet? The size of the 
subnebula upon {\it inflow} depends on the specific angular momentum of gas flowing into 
the giant planet's gravitational domain ({\it Lissauer}, 1995; {\it Mosqueira 
and Estrada}, 2003a). Before and during much of the runaway gas accretion phase of 
Jupiter's formation (which spans a period prior to and after envelope contraction), 
the gas that enters the protoplanet's Roche lobe 
(equivalently, its Hill sphere, see Eq. 4) has low specific angular momentum.
This is because the mass of the protoplanet for much of the this period is
not sufficient to open a significant gas gap in the solar nebula for
typical circumsolar disk model parameter choices (see Sec. 3.4, and below).
As a result, most of the gas being accreted during this period
originates within the vicinity of 
the protoplanet (i.e., at heliocentric distances $\lesssim R_H$ from the planet). 
Prior to envelope contraction, gas accretes onto the giant planet's 
distended atmosphere. 

The contraction of the distended envelope happens early during the 
runaway gas accretion phase. Current models have Jupiter contracting prior to 
reaching $\sim 1/3$ of its final mass.
Moreover,
the timescale for envelope collapse (down to a few planetary radii) is relatively 
quick compared to the runaway gas accretion epoch
({see {\it Lissauer et al.}, 2009). This indicates that the
subnebula (circumplanetary disk)
begins to form relatively early in the planet formation process when
the planet is not sufficiently large to truncate the gas disk. As a result,
the gas that continues to flow into the Roche lobe has mostly low specific
angular momentum. Since the proto-Jupiter must still accrete  
$\gtrsim 200$ M$_\oplus$ after envelope collapse, this suggests that 
the gas mass deposited over time in the relatively compact subnebula could 
be substantial. Furthermore, this still preliminary picture may be consistent 
with the view that the 
planet and disk may receive similar amounts of angular momentum 
(e.g., {\it Stevenson et al.}, 1986; Sec. 4).
 
As the protoplanet grows more massive, it clears the 
surrounding nebula gas through the actions of gas accretion and tidal 
interaction, leading to the formation of a gas gap in the solar nebula 
(Sec. 2.3.2). Sufficiently massive objects may actually truncate the disk,
which we take to mean 
that almost all of the available gas in the vicinity of the giant planet 
has been accreted or shoved aside, so that the density in the (deep) gap is
$\lesssim 10^{-2}$ of the unperturbed density at that location. 
In a moderately viscous disk, a 
Jupiter mass planet (1 M$_J$), truncates the disk in a few hundred orbital 
periods (or $\sim 10^3$ years at 5.2 AU). Larger masses are required
to open a deep gap in a viscous disk in order to overcome the tendency of
turbulent diffusion to refill the gap.  

This picture applies to an accretion scenario for Jupiter 
that assumes that no other planets influence the evolution of the nebula gas. 
In a scenario in which the giant planets of the Solar System 
start out much closer to each other and subsequently migrate outwards
({\it Tsiganis et al}., 2005), 
the two planets jointly open a gap (if they grow almost simultaneously). In such 
a scenario, Saturn's local
influence would dominate over that of turbulence in a moderately viscous disk
({\it Morbidelli and Crida}, 2007). 
  
The process of gap-opening in the circumsolar gas disk
has direct bearing on the satellite formation 
environment. As the gap becomes deeper, the continued gas 
inflow through this gap can significantly alter the properties of the 
subnebula. In particular, as gas in the protoplanet's vicinity is depleted,
the inflow begins to be dominated by gas with specific 
angular momentum which is much higher than what previously accreted onto 
the distended envelope and/or compact disk. This
is because the gas must now come from farther away (heliocentric distances 
$\gtrsim R_H$). 
Since satellite formation is expected to begin at this time (see Sec. 4),
the character of the gas inflow during the waning stage of Jupiter's accretion 
is a key issue in determining the environment in which the Galilean satellites 
form.
 
It is worth noting that Europa and the other Galilean
satellites occupy a compact region (roughly $\sim 4\%$ of the Hill radius). The
mass (and angular momentum) of outermost Callisto 
is comparable to that of Ganymede.
%($j_{Gan} \sim 3.9 \times 10^{-3}$). 
This similarity in mass of these two Galilean satellites is
puzzling since one might expect the outermost satellite to have 
significantly {\it less} mass. This is because it is {\it a priori} 
difficult to envision how the surface density of the satellite disk
could have been large enough to make a massive
satellite such as Callisto at its location, but
insufficient to form other, smaller satellites outside
its orbit ({\it Mosqueira and Estrada}, 2003a). Callisto's large mass most likely
indicates that the circumjovian disk
extended significantly beyond Callisto's orbit; thus, the lack of regular 
satellites outside Callisto requires explanation.

Until recently, numerical
models that simulate gas accretion onto a ``giant planet'' embedded in a
circumstellar disk could not resolve
scales smaller than $\sim 0.1$ $R_H$ (e.g., {\it Lubow et al}., 1999; {\it Bate
et al}., 2003), a region several times larger than is populated by the regular
satellites. It is only with the advent of higher
resolution 2D and 3D simulations (described in Sec. 3.3) that it became 
possible to resolve structure on the scale of the radial extent of
the Galilean satellites. These recent simulations indicate
that the circumplanetary disk formed by the gas inflow through the gap --
irrespective of any subsequent viscous
evolution and spreading -- likely extended as much as
$\sim 5$ times the size of the Galilean system ({\it
D'Angelo et al}., 2003b). Thus, the specific angular
momentum of gas inflow through a gap is significantly
larger than that of the satellites themselves. 

\vspace{0.1in}
\noindent
{\bf 3.2. Analytical Estimates of Disk Sizes}
\vspace{0.1in}

We can obtain an estimate of the characteristic disk size formed
by the accretion of low specific angular momentum gas before gap-opening, using
angular momentum conservation. Assuming that Jupiter 
travels on a circular orbit, that the solar nebula gas moves in 
Keplerian orbits, and that {\it prior} to gap-opening Jupiter
accretes gas parcels with semimajor axes originating from up to
$R_H$ of its location, then the specific angular momentum 
$\ell$ of the accreted gas is approximately given by 
({\it Lissauer}, 1995):

\begin{equation}
\ell \approx -\Omega \frac{\int^{R_H}_{0} \frac{3}{2}
x^3 d x}
{\int^{R_H}_{0} x d x} + \Omega R_H^2 \approx
\frac{1}{4} \Omega R_H^2.
\label{equ:spang}
\end{equation}

\noindent
The expression for the specific angular momentum estimate given
above has two contributions. The first term is the specific
angular momentum flux flowing into the planet
due to Keplerian shear computed 
in the frame rotating at the planet's angular velocity. The second 
contribution is
a correction to translate back to an inertial frame
(see {\it Lissauer}, 1995 and references therein).
Equation (\ref{equ:spang}) neglects the gravitational
effect of the planet, and assumes that the angular
momentum of the inflowing gas is delivered to the
Roche lobe of the giant planet.
using conservation of angular momentum, balancing
centrifugal and gravitational forces $\ell^2/r_c^3 \approx GM_J/r_c^2$ 
we obtain the centrifugal
radius ({\it Cassen and Pettibone}, 1976; {\it Stevenson et al}., 1986; {\it
Lissauer}, 1995; {\it Mosqueira and Estrada}, 2003a):

\begin{equation}
r_c \approx R_H/48.
\end{equation}

\noindent
For a fully grown Jupiter, the centrifugal radius is located at $r_c
\approx 15$ R$_J$ (where R$_J = 71492$ km is a Jupiter radius)
just outside the position of Ganymede
(for Saturn, $r_c$ lies just outside of Titan, see
Fig. 1 of {\it Mosqueira and Estrada}, 2003a). While this calculation would
seem to employ unrealistic assumptions, recent 3D simulations indicate
that it provides a meaningful estimate ({\it Machida et al.}, 2008.
The resulting disk size is
consistent with the radial extent of the Galilean satellites.

{\it After} gap-opening, accretion may continue 
through the planetary Lagrange points, as seen in some simulations (e.g., 
{\it Artymowicz and Lubow}, 1996; {\it Lubow et al}., 1999). Specifically, 
accretion occurs through the $L_1$ and $L_2$ points, which are located at 
roughly a distance $R_H$ from the planet, and along the line connecting 
protoplanet and Sun (see Fig. 4). At these Lagrange points, the gravitational fields
of both the protoplanet and the Sun combined with the centrifugal force are in
balance. As seen in the rotating frame, a massless body placed at this 
location with zero relative velocity would remain stationary. 
We can obtain an estimate
of the specific angular momentum of the gas as it
passes through the Lagrange points by assuming that
the inflow takes place at a low velocity in the
rotating frame and it is directed nearly towards the
planet. This may be done by keeping only the change of frame
contribution of Eq. (\ref{equ:spang}) or $\ell \sim
\Omega R_H^2$. Again, using conservation of angular momentum,
the estimated characteristic disk size formed by the
inflow is significantly larger than before, roughly
$\sim R_H/3 \sim 260$ R$_J$ (e.g., {\it Quillen and
Trilling}, 1998; {\it Mosqueira and Estrada}, 2003a).

These estimates indicate that gas flowing through a gap in the circumsolar
disk brings with it significantly higher specific angular momentum, which
produces a larger characteristic circumplanetary disk size,
than does incoming gas when no gap is present. 
This information combined with the 
observed mass distribution of the regular satellites of Jupiter 
(and Saturn) can be used to argue in favor of a
two-component circumplanetary disk: (1) a compact,
relatively massive disk that forms over a period of time post 
envelope collapse and prior to disk truncation, and (2) a more 
extended, less massive 
outer disk that forms from gas flowing through a gap and at a lower
inflow rate (e.g., {\it Bryden et al}, 1999; {\it
D'Angelo et al}., 2003b). An idealization of this two-component subnebula
is shown in Fig. \ref{fig:semm}
(see left panel and caption). However, the details of the formation of 
the giant planet from the envelope collapse phase, to gap-opening and 
isolation remains to be shown using hydrodynamical simulations. 
Numerical simulations of giant planet formation tend 
either to treat the growth of the protoplanet in isolation (e.g., {\it
Pollack et al}., 1996; {\it Hubickyj et al}., 2005; see
Sec. 2), or to treat giant planets ($\sim 1$ M$_J$) embedded in
circumstellar disks in the presence of a well-defined, deep gap 
(e.g., {\it Lubow et al}., 1999; Kley 1999; {\it Bate et al}., 2003; 
{\it D'Angelo et al}., 2003b). Mainly because of the computational demands, 
the latter simulations do not model changing planetary or nebula conditions. 
Not surprisingly then, the transition between a distended planetary envelope 
and a subnebula disk has received scant attention. 

A consequence of continued gas inflow through the gap is that 
the subnebula will continue to evolve due to the turbulent 
viscosity generated by gas accretion onto the circumplanetary disk.
However, even weak turbulence can pose a problem for 
satellitesimal formation (Sec. 2.1). The turbulent circumplanetary disk environment 
generated by
the inflow likely means that satellite formation does not begin until late
in the planetary formation sequence when the gas inflow (through the gap) wanes, 
at which point turbulence in the subnebula may decay. The formation of the 
satellites at the stage where the planet approaches its final mass
is further supported by the fact that 
even weak, ongoing inflow through the gap can generate a substantial amount of 
heating due to turbulent viscosity, which would generally result in a 
circumplanetary disk that is too 
hot for ice to condense and satellites to form and survive (e.g.,
{\it Coradini et al}., 1989; {\it Makalkin et al}.,
1999; {\it Klahr and Kley}, 2006). 
As a result, a very low gas inflow rate
-- orders of magnitude lower than the accretion rates through 
gas gaps in numerical simulations ($\sim 10^{-2}$ M$_\oplus$ yr$^{-1}$,
e.g., {\it Lubow et al}., 1999) -- 
is probably a requirement of any
satellite formation model. 
Even so, some applicable conclusions can be drawn 
from existing simulations.

\vspace{0.1in}
\noindent
{\bf 3.3. Numerical Results in 2D/3D in the Presence of a Gap}
\vspace{0.1in}

Two-dimensional hydrodynamics calculations of a
Jupiter-mass planet embedded in a circumstellar disk
show prograde circulation of material within the planet's Roche lobe that is
reminiscent of a circumplanetary disk ({\it Lubow et
al}., 1999, {\it Kley}, 1999). These simulations show
that gas can flow through the gap formed by the giant
planet, depending on the value for the nebula
turbulence parameter ($\alpha \gtrsim 10^{-4}$, {\it Bryden et al}., 1999). 
A prominent feature
exhibited in these Roche-lobe flows or streams is a two-arm spiral wave
structure (see left panels of Fig. \ref{fig:3d_model}, which is a 3D simulation). 
As gas flow enters the Roche lobe near the planetary Lagrange points, 
these streams encircle the planet and impact one another on the opposite side
(from which they entered). 
The resulting collision shocks the material, and deflects the flow towards the 
planet (e.g., {\it D'Angelo et al.}, 2002). In 2D, the spiral wave structure is 
weaker (not as tightly wound) for decreasing protoplanet mass, and disappears 
altogether for $\sim 1$ M$_\oplus$ protoplanetary masses. In 3D simulations, 
these spiral waves are also less marked than in 2D as a consequence of the 
flow no longer being restricted to a plane (see, e.g., {\it D'Angelo et al.}, 2003a;
{\it Klahr and Kley}, 2006).
Despite the differences of the accretion flow, gas accretion rates in 2D and 3D
are similar.

Detailed simulations of such systems pose a significant
challenge from a numerical point of view since they
demand that both the circumstellar disk and the
hydrodynamics deep inside the planet's Roche lobe
must be resolved. This requirement means that length scales must be
resolved over more than two orders of magnitude, from
the planet's orbital radius, $r_{p}$, down to a few
per cent of the Hill radius, $R_{\mathrm{H}}$.
{\it D'Angelo et al}. (2002, 2003b) carry out a quantitative
analysis of the properties of circumplanetary disks
around Jovian and sub-Jovian mass planets. By treating
the circumstellar disk as a locally isothermal and viscous
fluid, and using a grid refinement technique known as
``nested grids'' that allow them to resolve length scales around the
planet on the order of $0.01$ R$_{\mathrm{H}}$ ($\sim 7$ R$_J$ for 1 M$_J$), 
these authors are able to show that the dynamical properties
of the material orbiting within a few tenths of
$R_{\mathrm{H}}$ from the planet are indeed consistent
with a disk in Keplerian rotation.

\begin{figure}[t!]
 \begin{minipage}[b]{0.5\textwidth}
 \resizebox{\linewidth}{!}{\includegraphics{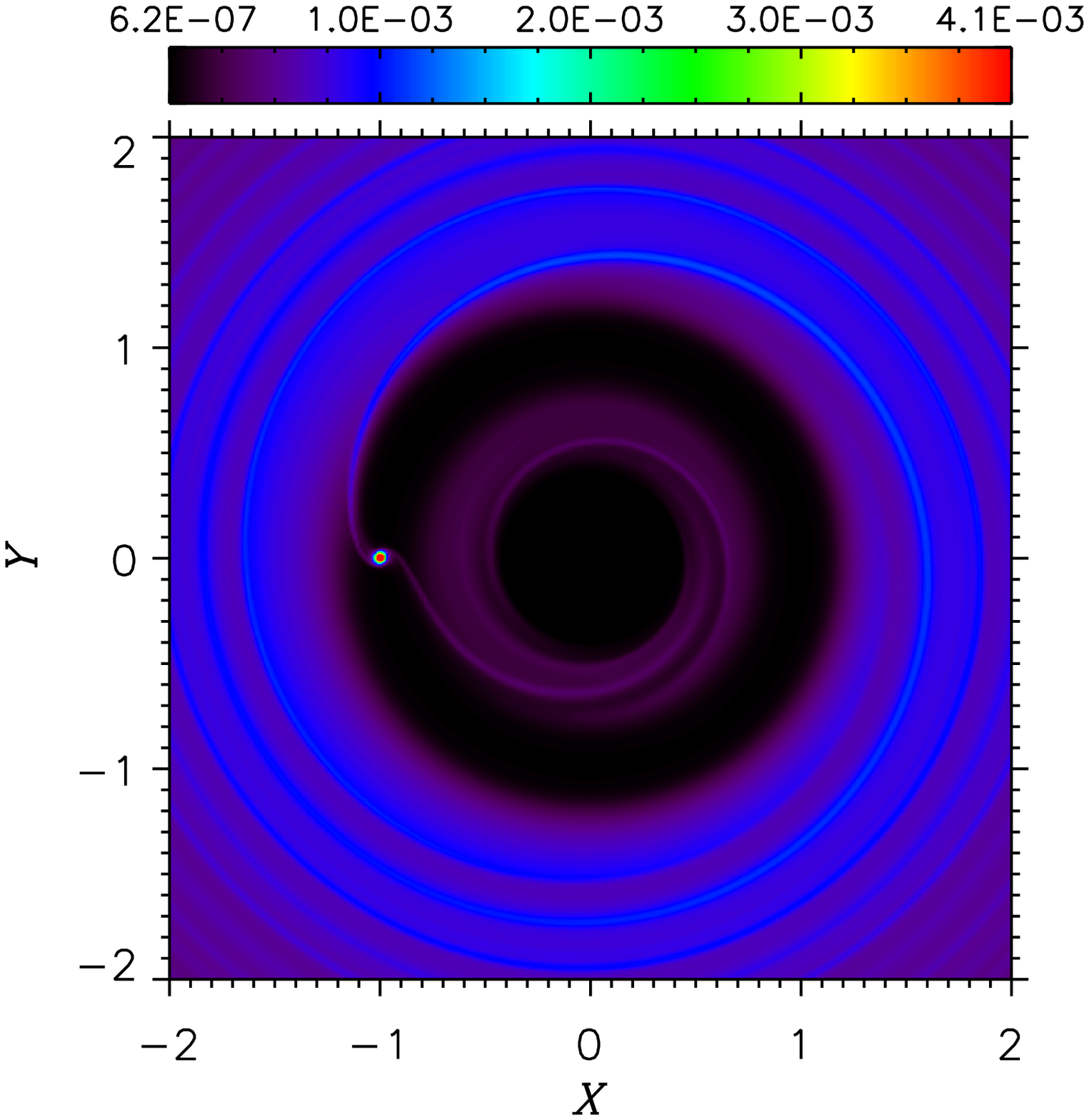}}
 \resizebox{\linewidth}{!}{\includegraphics{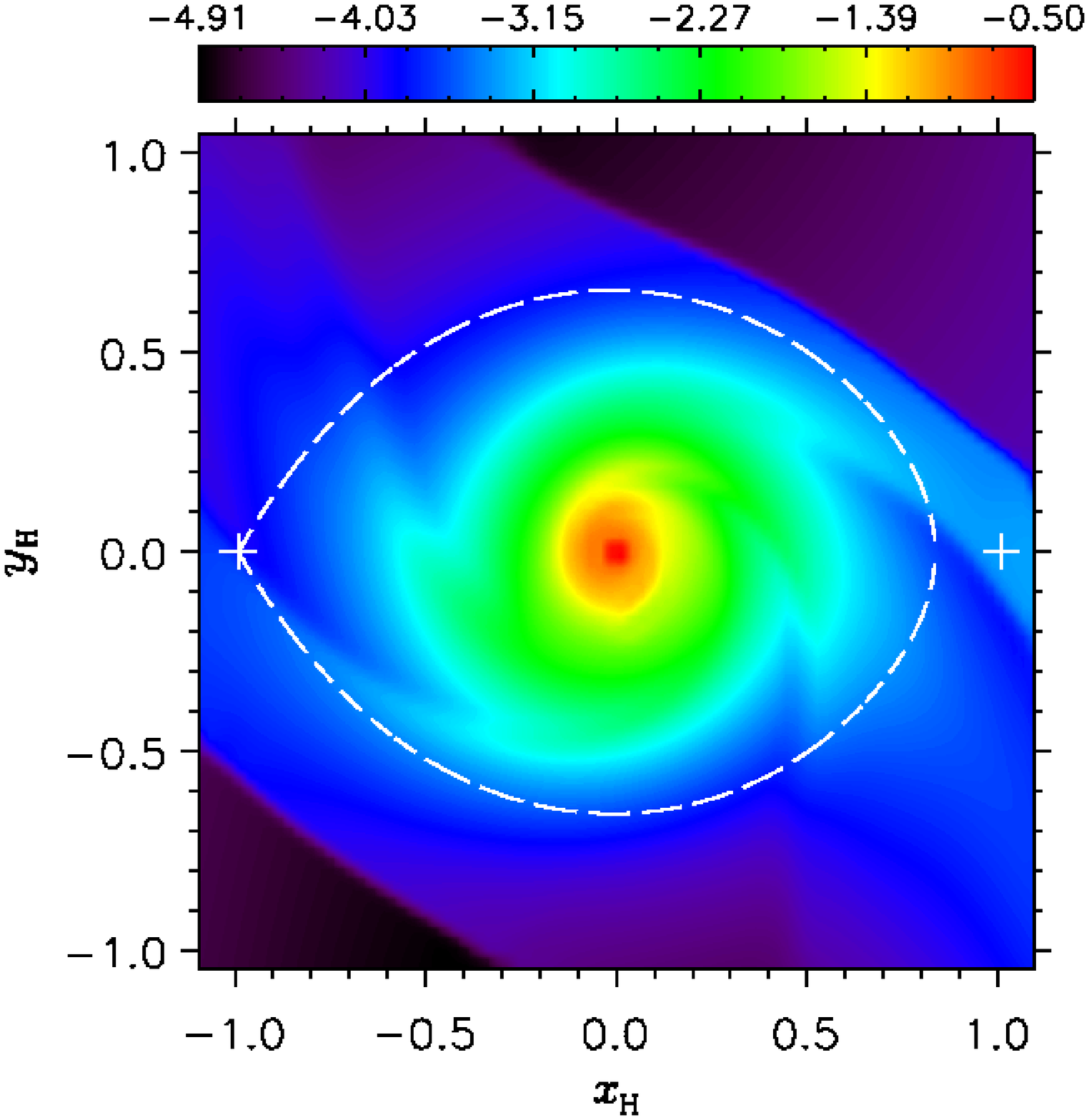}}
 \end{minipage}%
 \begin{minipage}[b]{0.5\textwidth}
 \resizebox{\linewidth}{!}{\includegraphics{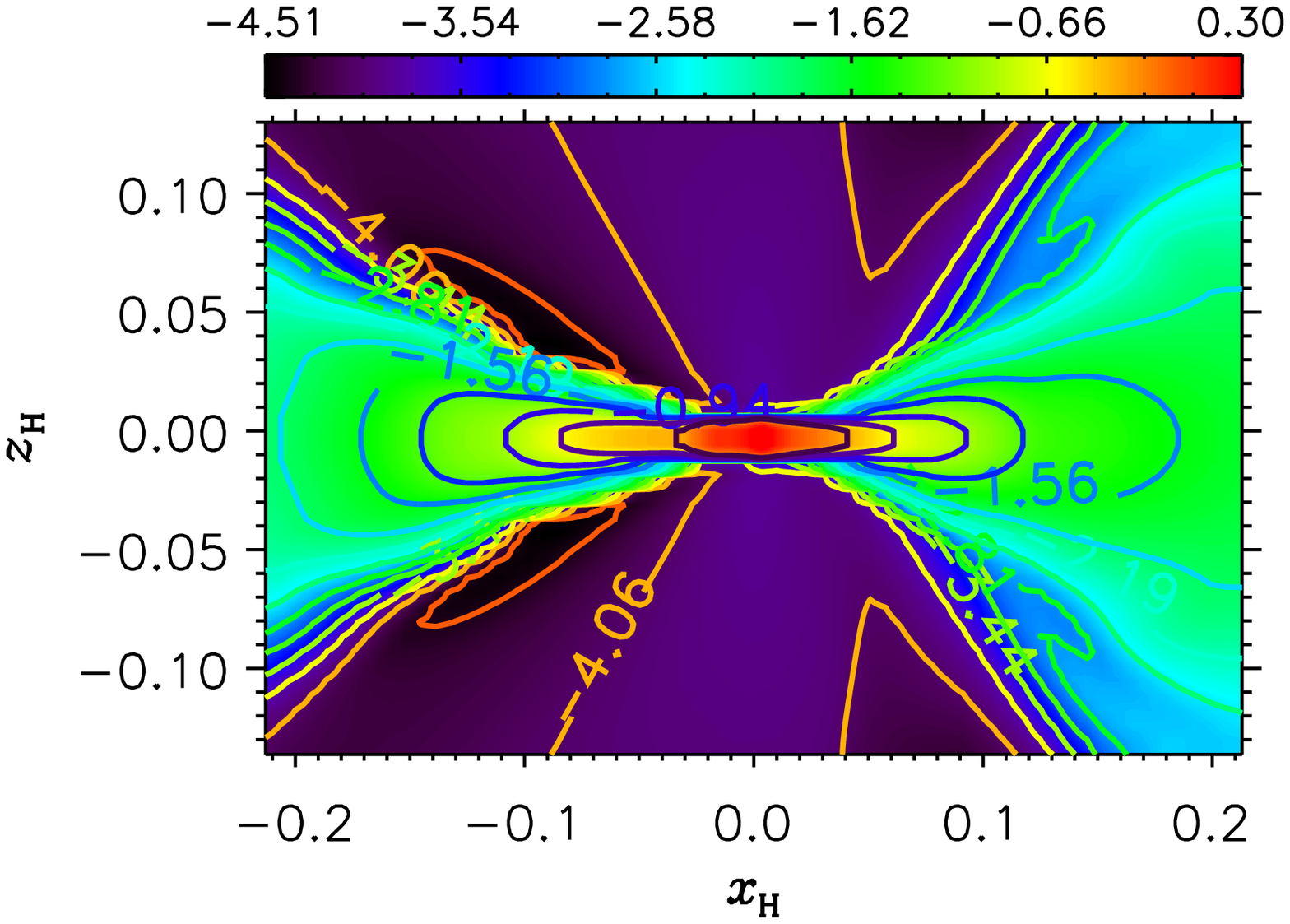}}
 \resizebox{\linewidth}{!}{\includegraphics{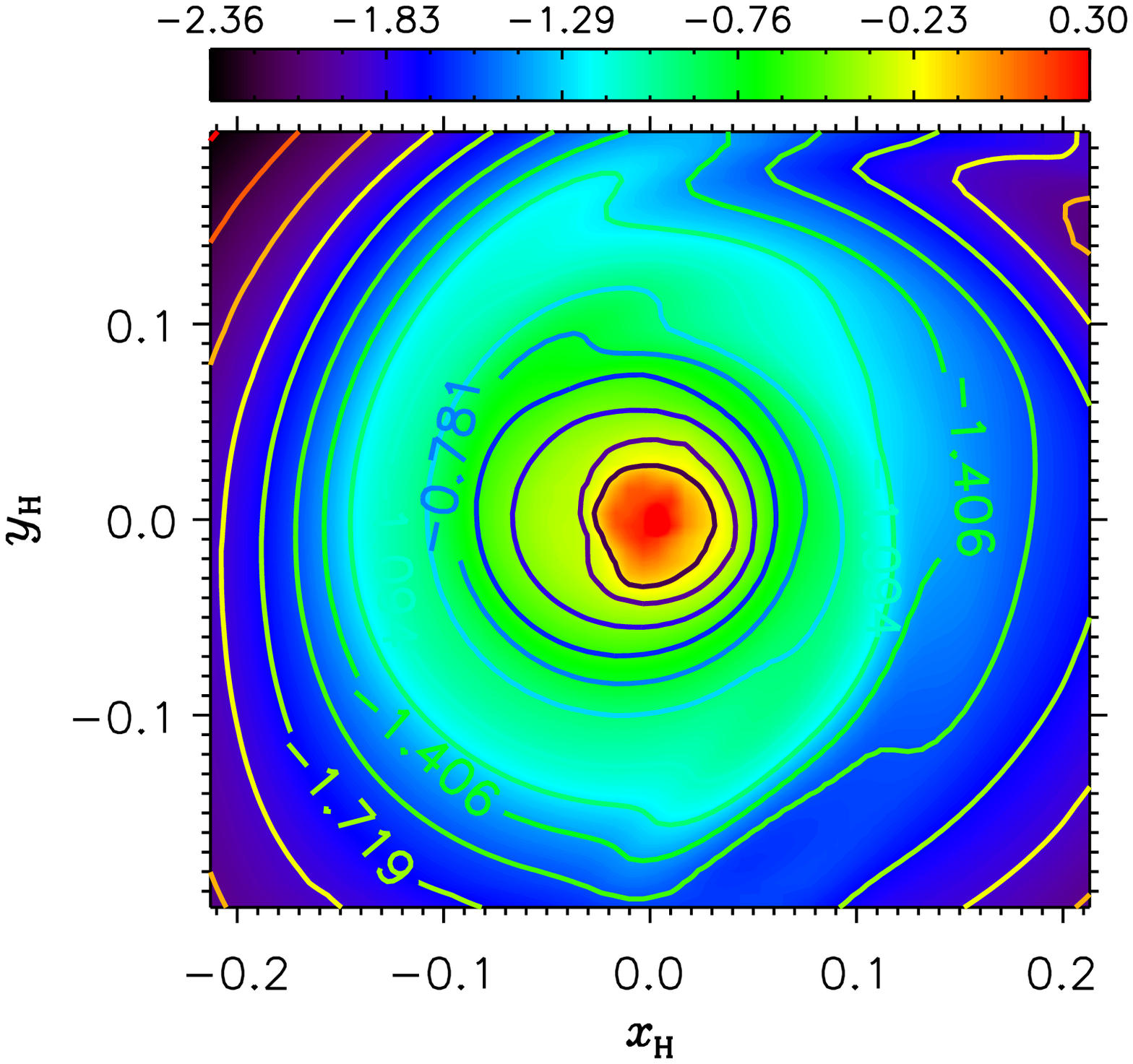}}
 \resizebox{\linewidth}{!}{\includegraphics{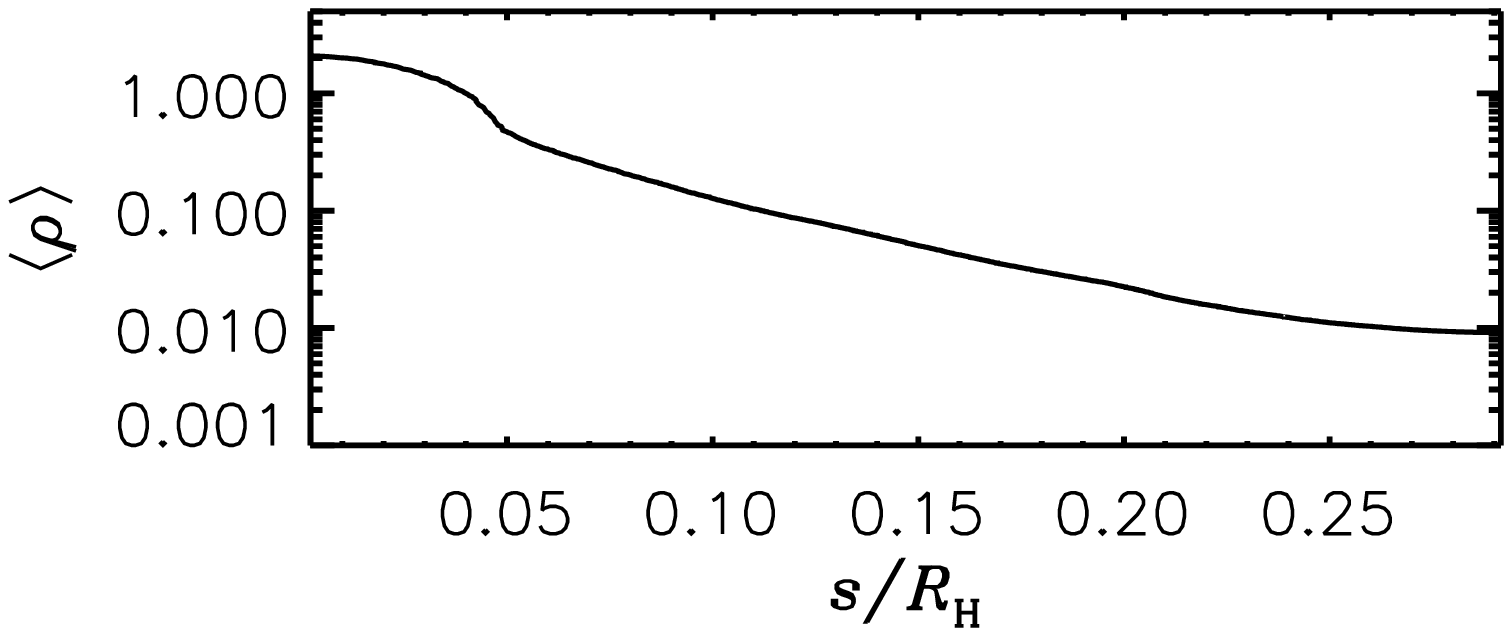}}
 \end{minipage}
\caption{{\small Formation of a circumplanetary disk around a Jupiter-mass
         planet in 3D. The top panel on the left shows the mass density
         distribution, $\rho$, in the circumstellar disk's 
         midplane. The bottom panel on the left as well as the top
         and center panels on the right show density distributions,
         in logarithmic scale, within the planet's Roche lobe
         (the tear-drop shaped region marked by the dashed line).
         Iso-density contours are also indicated in two panels on
         the right.
         The logarithm (base 10) of the azimuthally averaged density 
         in the disk's midplane
         is shown in the bottom panel on the right, where $s$ 
         represents the distance from the planet. 
         The units on the axes are either the planet's orbital
         radius, $r_{p}$ ($X$ and $Y$ coordinates), or the Hill 
         radius, $R_{\mathrm{H}}$ ($x_{\mathrm{H}}$, 
         $y_{\mathrm{H}}$, and $z_{\mathrm{H}}$ coordinates).
         The units of $\rho$ are such that $10^{-3}$ corresponds to
         $10^{-12}\;\mathrm{g}\,\mathrm{cm}^{-3}$.}
        }
\label{fig:3d_model}
\end{figure}

Figure~\ref{fig:3d_model} shows the mass density
distributions from a three-dimensional, local isothermal model (see
{\it D'Angelo et al}., 2003b for details). The temperature at
5.2 AU is assumed to be $T\simeq 110\;\mathrm{K}$ 
(if the mean molecular weight of the gas is about $2.2$) and the
kinematic viscosity, $\nu$, in this case is assumed to be constant in space
and time
(see discussion at the end of Sec. 3.2).
The aspect ratio of the circumstellar disk, which is given by the 
ratio of the disk's semi-thickness (generally denoted by the pressure 
or nebula scale height $H$) at the location of the planet to the planet's semi-major 
axis, is $H/r_p \sim 0.05$. For the disk parameters chosen,
$\nu$ is comparable to a turbulent viscosity with an $\alpha$-parameter 
of $4\times10^{-3}$ at $r_{p}$. The top panel on the left
illustrates the circumstellar disk and the density gap
produced by the planet that exerts gravitational
torques on the disk material. The other panels show
the mass density, in logarithmic scale, over lengths
$\sim \rm{R}_{\mathrm{H}}$ (left) and $\sim 0.1$
$\rm{R}_{\mathrm{H}}$ (right). The bottom panel on the right
displays the density in the disk's midplane,
azimuthally averaged around the planet, as a
function of the distance from the planet, $s$. The models shown in 
Fig. \ref{fig:3d_model} can be rescaled by the 
unperturbed value of the mass density $\rho$ (i.e., that of the circumstellar disk 
when the planet is not present) at $r_{p}$, which is a consequence of
the locally isothermal approach. Therefore,
the calculated density {\it structure} in the disk is
independent of the unperturbed value of $\rho$.
For the value of $\rho$ chosen in Fig. \ref{fig:3d_model}, 
the disk mass within $0.2$ R$_H$ is $\sim 10^{-4}$ M$_J$.

{\it D'Angelo et al}. (2003a) present thermo-dynamical
models of circumjovian disks in two dimensions. In
these calculations, the energy budget of the disk
accounts for advection and compressional work, viscous
dissipation and local radiative dissipation.
Characteristic temperatures
and densities in these models depend mainly upon
viscosity, opacity tables, and initial mass of the
circumstellar disk. In this case, the results are not scalable by the 
unperturbed mass density, because the opacity depends on 
the value of $\rho$ chosen.

\begin{figure}[t!]
 \resizebox{\linewidth}{!}{%
 \includegraphics{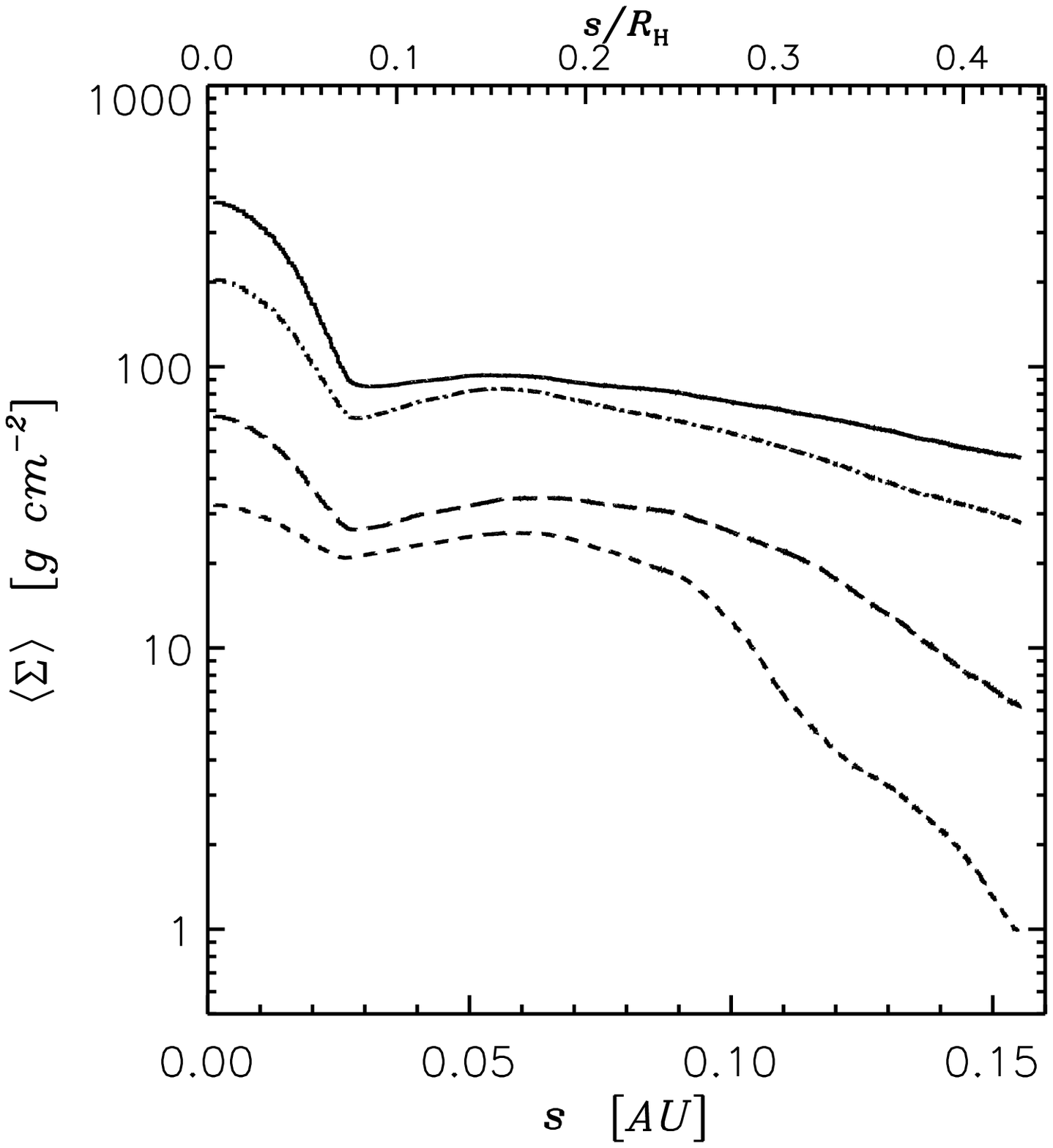}%
 \includegraphics{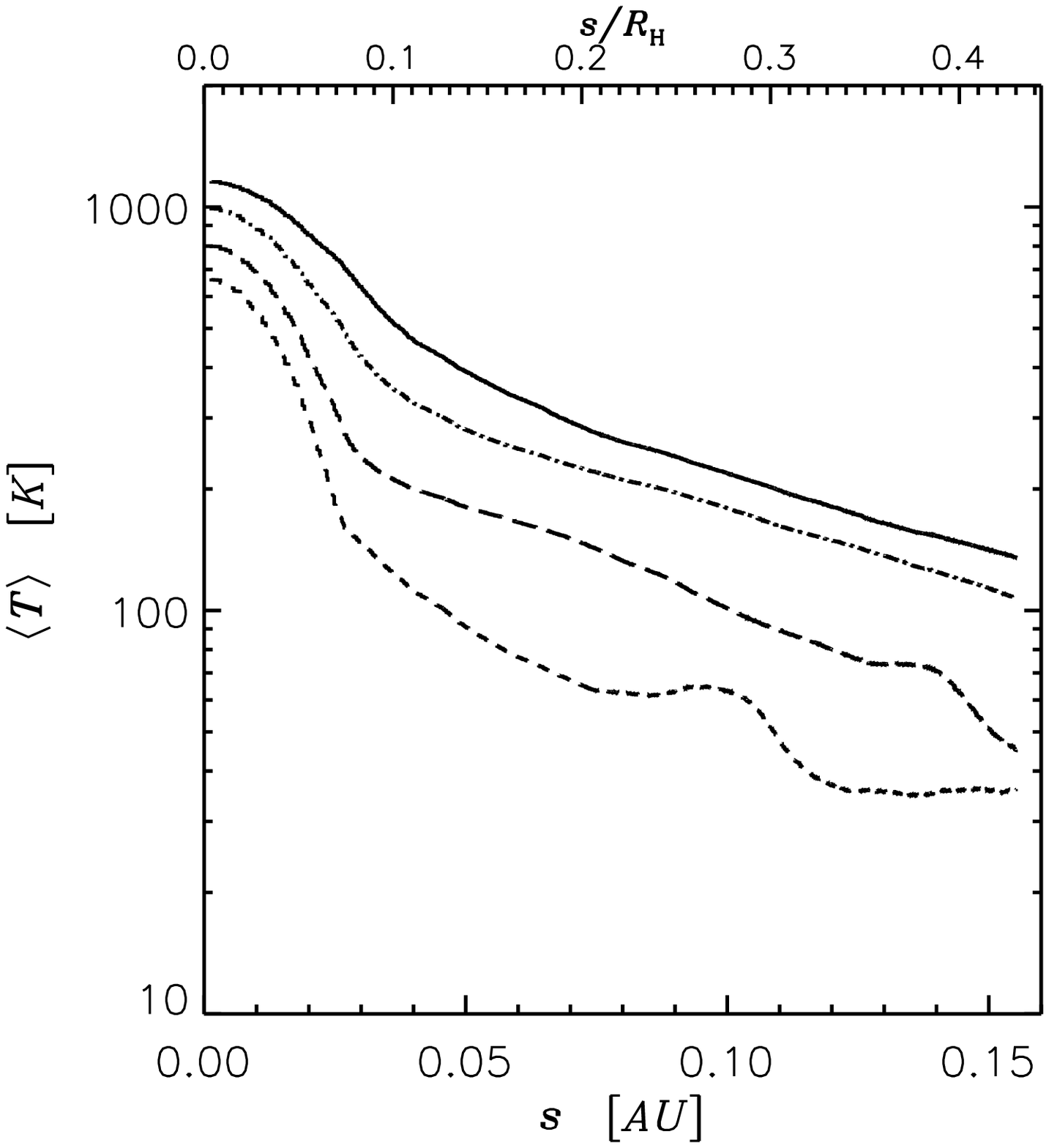}}
\caption{{\small Surface density (left) and temperature (right) of
         two-dimensional circumjovian disk models with viscous
         heating and radiative cooling (see text for further
         details). The quantities represent azimuthal averages
         around the planet. The models differ in the adopted
         viscosity prescription. The calculation which produces the
         highest density and temperature (solid curve) assumes a constant 
         $\nu = 10^{15}\;\mathrm{cm}^{2}\,\mathrm{s}^{-1}$.
         The other calculations assume an $\alpha$-type viscosity,
         $\nu = \alpha c_sH$ (the same value of $\alpha$ applies to 
	 nebula and subnebula), so $\nu$ is
         therefore is space- and time-dependent.
         For increasing density and temperature, models have 
         $\alpha=10^{-4}$ (short dash), $10^{-3}$ (long dash), and $10^{-2}$ (dot-dash), 
         respectively.}
        }
\label{fig:S_T}
\end{figure}

Figure~\ref{fig:S_T} displays surface density (left) and temperature 
profiles (right) obtained from calculations with different 
prescriptions and magnitude of the kinematic viscosity.
These models rely on the opacity tables of {\it Bell and Lin} (1994) and
assume that the initial unperturbed surface density, $\Sigma$, at 
$5.2\;\mathrm{AU}$ is roughly $100\;\mathrm{g}\,\mathrm{cm}^{-2}$.
The circumstellar disk contains about $4.8$ Jupiter masses within 
$13\;\mathrm{AU}$ of the star. 
The model with highest density and temperature (black curves) has a constant
(in space and time) kinematic viscosity 
$\nu=10^{15}\;\mathrm{cm}^{2}\,\mathrm{s}^{-1}$. The other models
assume an $\alpha$-viscosity $\nu=\alpha c_{s} H$, where the sound speed
$c_{s}$ and the pressure scale height $H$ are
a function of space and time while $\alpha$ is a constant. In
Fig.~\ref{fig:S_T}, for increasing density and temperature,
models have $\alpha=10^{-4}$ (short-dashed curves), $10^{-3}$ (long-dashed curves), and 
$10^{-2}$ (dot-dashed curves), respectively. The value of $\alpha$ applies to
both the circumsolar and circumplanetary disks.
In the cases shown in the Figure, 
the amount of mass within about $0.2$ R$_{H}$ of the planet is
in the range between $\sim 10^{-5}-10^{-4}$ M$_J$.

\begin{figure}[t!]
 \begin{minipage}[c]{0.5\textwidth}
 \resizebox{\linewidth}{!}{%
 \includegraphics{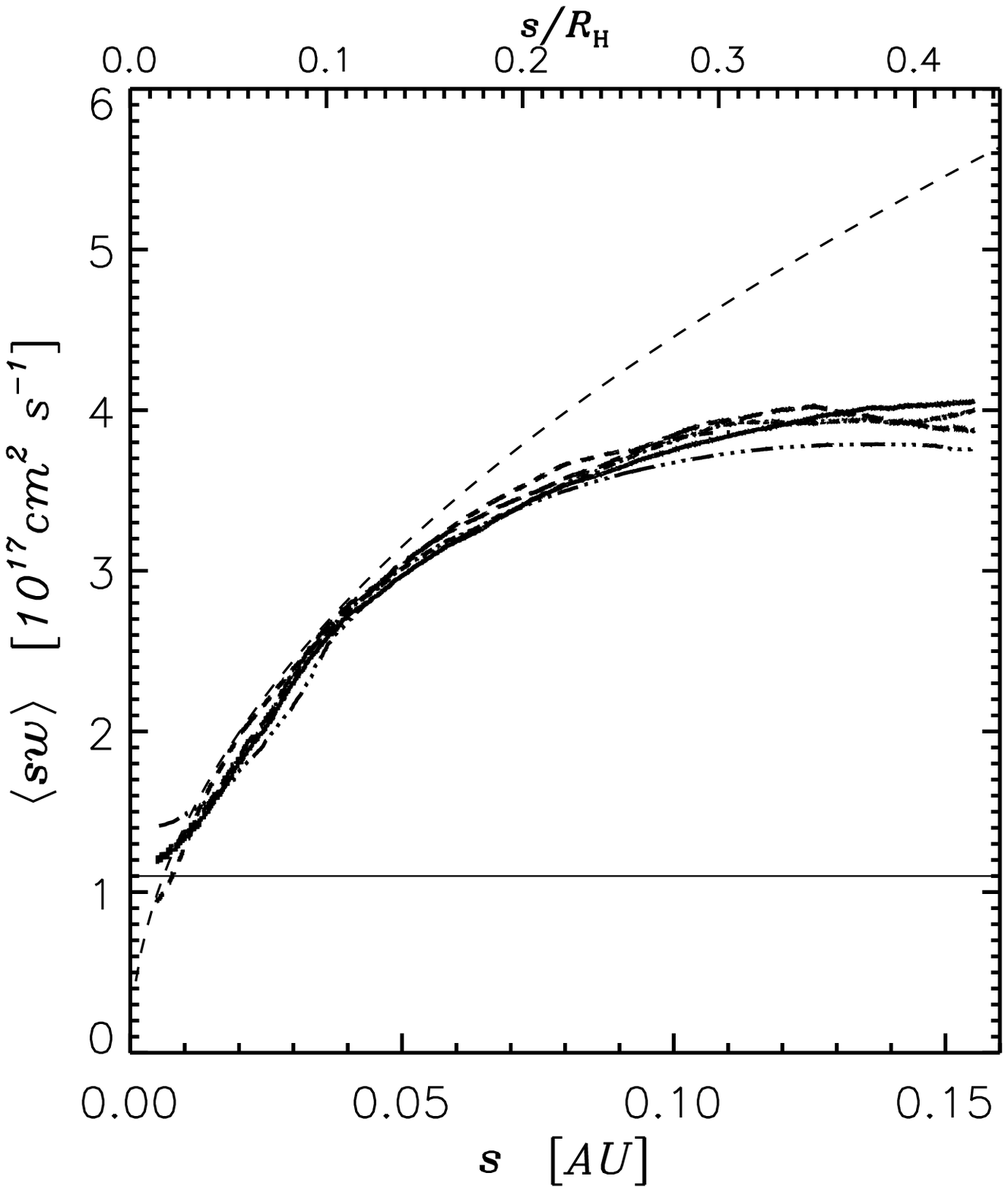}}
\label{fig:sw}
 \end{minipage}\hspace*{\fill}%
 \begin{minipage}[c]{0.5\textwidth}
\caption{{\small Specific angular momentum of circum-Jovian disk models,
         in two and three dimensions, azimuthally averaged around
         the planet. The quantity $w$ is the azimuthal velocity
         around the planet. Results from models
         in Figure~\ref{fig:3d_model} and Figure~\ref{fig:S_T} are
         displayed. The multiple dot-dash curve represents the 3D isothermal case. 
	 The less bold dashed line 
	 represents the Keplerian angular momentum.
         The horizontal line is the specific angular momentum
         of the Galilean satellites.}
        }
\label{fig:sw}
 \end{minipage}
\end{figure}

The specific angular momentum from three dimensional as well as two
dimensional models for a $1$ M$_J$ mass planet is plotted in 
Fig.~\ref{fig:sw}. Within $\sim 0.15$ R$_{H}$ of the planet, the rotation 
of the disk follows the rotation curve of a Keplerian disk (dashed curve). 
The curves correspond to the models in Fig. \ref{fig:S_T}, while the 
multiple dot-dashed curve represents the 3D case.
It is interesting to note in the latter
case that, even though the subnebula temperature is assumed to be
constant, the specific
angular momentum distribution is consistent with that of the 2D models 
which are determined by means of calculations that allow for
heating and cooling processes. Thus, it appears that for large 
planetary masses in which a deep gap is present in the circumstellar gas disk,
2D and 3D simulations give comparable results for the specific angular momentum. 
This is because the gas flowing across such a deep gap into the Roche lobe
is coming from much farther away than $R_H$ (see upper left panel of Fig. 
\ref{fig:3d_model}). Although the flow pattern in 2D and 3D differ, by the time
the gas reaches the planet the specific
angular momentum delivered with the inflow 
in both cases is qualitatively similar.

Finally, for comparison,
the specific angular momentum of the Galilean satellites is also
indicated (solid horizontal line). These simulations indicate that gaps
correspond to higher specific angular momentum, and a
larger characteristic disk size formed by the inflow, than
that of the regular satellites themselves.

\vspace{0.1in}
\noindent
{\bf 3.4. Connecting the Planet, its Disk, and the Satellites}
\vspace{0.1in}

As was pointed out in Sec. 3.2, planetary formation models tend to
focus on either the growth of the planet in isolation (Sec. 2), or a 
protoplanet of fixed mass embedded in a circumstellar disk in the presence of a 
well formed gap (Sec. 3.3). As a result, our understanding of the formation
of the circumjovian gas disk remains incomplete, as no simulations
have yet been done that model disk formation starting after the giant planet's 
envelope contraction to the time at which inflow from the circumsolar disk
ceases.

During the period over which circumjovian disk formation occurs, the characteristic 
disk size may 
be roughly estimated by a balance between gravitational and centrifugal forces 
(Sec. 3.2). 
Because 
the protoplanet likely accretes most of its gas mass after envelope collapse, a
significant fraction of this gas may end up in the circumjovian gas disk
leading to an initially massive subnebula. 
A legitimate concern that arises
from this scenario is why is it that Jupiter is not rotating near break-up velocity.
The origin of Jupiter's current spin angular 
momentum remains poorly understood. A massive circumplanetary disk
may be a requirement for despinning the planet (see e.g., {\it Korycansky et al.}, 1991;
{\it Takata and Stevenson}, 1996), although Jupiter's spin may require 
consideration of the role of magnetic fields. The question of Jupiter's spin thus represents
a key piece of the Jovian system puzzle that is in need of further investigation.
 
A compact massive disk will have important implications for satellite formation
models. The characterization of this disk component will require the 
marriage of isolated growth models and embedded planet simulations, a milestone 
that is just beginning to be explored. Indeed, recent simulations of gas 
accretion onto a low-mass protoplanet (i.e., a deep gap is not present) embedded 
in a circumstellar disk indicate that gas may not be bound to the planet outside 
of $\sim 0.25$ R$_H$ ({\it Lissauer et al.}, 2009).
Measurements of the specific angular momentum contained within the 
bound region for different 
pre-gap-opening protoplanet masses are consistent with Eq. \ref{equ:spang}
yielding $\ell \lesssim \Omega R_H^2/4$ (it is important to note that a parcel of 
gas that crosses within $0.25$ R$_H$ of the planet may originate from a radial
distance as far as R$_H$).
 
Recently, {\it Machida et al.} (2008) use 3D hydrodynamical simulations to model
the angular momentum accretion into the giant planet Hill sphere when a partially
depleted gap is present. These authors find for a range of planetary masses that a 
significant 
fraction of the total angular momentum may contribute to the formation of a compact 
circumplanetary disk. These results are inapplicable for masses comparable to $1$ M$_J$.
This is because the local simulation used by these workers is inappropriate to 
treat gap formation
because of the radial boundary ({\it Miyoshi et al.}, 1999), so that the depth and width of
the gap depends on the size of the simulation box used. However, as we pointed out
earlier, an important conclusion 
that may be drawn  from the results of {\it Machida et al}. is that a circumplanetary 
disk formed when a partial gap is present is compact due to the lower specific angular 
momentum of the inflow.

For large planetary masses in the presence of a 
deep gap, the circumplanetary disk size is qualitatively insensitive to the inflow 
rate, or whether the flow is treated in 2D or 3D. However, 3D simulations
are required early on in the accretion of the protoplanet when a deep gap is not present. 
We can understand this by noting that material in the nebula midplane whose semimajor 
axis is close to the planet ($\lesssim$ R$_H$) actually does not accrete onto the planet, 
but instead undergoes horseshoe orbits (e.g., {\it Lubow et al.}, 1999;
{\it Tanigawa and Watanabe}, 2002). Three-dimensional flows are then necessary to
allow for low angular momentum gas from radial distances $\lesssim R_H$ and {\it away} from
the midplane to be accreted directly onto the planet or into a compact disk 
({\it Bate et al}., 2003; {\it D'Angelo et al}., 2003a). 
Moreover, {\it D'Angelo et al}. (2003b) point out that because of the flow
circulation away from the disk midplane, less angular momentum is carried inside the 
Roche lobe by the midplane flow, so that in general 3D simulations are required in 
order to properly account for the angular momentum.  

What does this all mean for the formation of the Galilean satellites, which
lie in a quite compact region close to the planet?
When a well-formed gap is present, the characteristic circumjovian
disk size formed by the inflow is $> 0.1$ R$_H$, or $> 70$ R$_J$ (by 
comparison Ganymede is 
located at $\sim 15$ R$_J$). From Fig. \ref{fig:sw}, it can be seen that
the specific angular momentum of the inflow is about a
factor of $3-4$ larger than that of the Galilean
satellites ($ \sim 1.1 \times 10^{17}$ cm$^2$ s$^{-1}$), which means
that the gas will achieve centrifugal balance at a
radial location of $\sim 200$ R$_J$. While this disk
is compact compared to $R_H$, it is
extended in terms of the locations of the Galilean
satellites, and possibly linked to the location of the
irregulars (the innermost ones of which lie near $\sim 150$ R$_J$). Thus, these
results indicate that the inflow through a gap in the circumsolar disk results in an
extended circumplanetary disk of characteristic size
$\sim 70 - 200$ R$_J$, implying a mismatch between
the size of the circumplanetary disk formed by gas
inflow through the giant planet's gap and the compact region
where the regular satellites are found. 

This mismatch has important consequences for satellite
formation. Taken at face value, it may mean one of two
things: (i) the solid material coupled to the gas
coming through the gap did not provide the bulk of the
material that formed the regular satellites --
planetesimals that were not coupled to the gas
provided this source instead (see Sec. 4.2);
or (ii) the regular satellites migrated
distances considerably larger than their current distances from Jupiter. The
first option would effectively preclude gas inflow through
the gap as the source of solids for regular satellite formation.
The second option would make questionable any model
that explains the Galilean satellite compositional
gradient with subnebula ``snow line'' arguments; that
is, all satellites would presumably start out far from
the planet and outside the snowline, and receive their
full complement of ices. 

These options assume that after gap-opening the disk becomes cool enough, 
and thus the inflow is weak enough, that icy objects can form. Prior
to gap-opening, the inflow is likely to be fast, so that the circumplanetary
disk would be too hot (and turbulent) for the concurrent formation of ice-rich satellites
like Callisto and Ganymede. Therefore, it is not
possible to form a compact disk concurrently with the accretion of
ice-rich, close-in satellites, indicating that any satellite formation in a rotationally
supported circumplanetary disk likely doesn't 
begin until the gas inflow wanes and turbulence decays, or the subnebula gas
disk has dissipated.
 
\vspace{0.25in}
\begin{center}
{\bf 4. CONDITIONS FOR SATELLITE FORMATION}
\end{center}
 
In the core nucleated accretion model, Jupiter's formation is thought to occur in 
three stages (Sec. 2): {\it nebular}, {\it transition}, and 
{\it isolation}. 
The circumjovian gas disk forms during 
the transition stage of Jupiter's accretion and
should be viewed as a drawn out process that begins after the contraction of the envelope 
and ends when Jupiter is isolated from the solar nebula:
(1) During the {\it nebular} stage, most of 
the solids reside in large planetesimals, some of which may dissolve in the 
growing envelope during the latter part of this stage. Most of the high-$Z$ mass
delivery takes place before the ``cross-over'' time when the mass of the gaseous envelope
grows larger than the core.
This stage of growth may 
be followed by a {\it dilution} as the gas accretion rate accelerates.
(2) The envelope eventually becomes sufficiently massive to contract and
accrete gas from the circumsolar disk 
hydrodynamically ({\it transition} stage). 
Contraction down to a few planetary radii happens quickly relative to the
runaway gas accretion timescale. (3) After envelope collapse, the protoplanet's mass is still 
too small for it to clear a significant gap in
the surrounding solar nebula, so most of the gas flowing into the
Roche lobe accretes onto both the planet and a 
rotationally-supported compact disk. 
As the protoplanet becomes more massive, it depletes the gas 
in its vicinity of the solar nebula, truncating the disk. (4) The nebula 
gas continues to flow 
through the Lagrange points (see Fig \ref{fig:3d_model}), leading to the formation of an
extended disk. Meanwhile, 
planetesimals in Jupiter's feeding 
zone undergo an intense period of collisional grinding, leading to a fragmented 
population (Sec 4.2). Continued solids enhancement of the entire circumjovian 
disk occurs due to ablation of disk-crossing planetesimal fragments 
(see Fig. \ref{fig:vign2}).
The accretion of the satellites is expected to occur towards the tail end of 
Jupiter's formation, but how satellite formation proceeds largely 
depends on the level and persistence of turbulence both in the circumsolar and 
circumplanetary nebulae.

\vspace{0.1in}
\noindent
{\bf 4.1. Turbulence and its Implications for Satellite Accretion}
\vspace{0.1in}
 
For disks to accrete onto the central object, angular momentum must 
be transported outwards. For the case of disks around young stellar objects 
(YSOs), magneto-hydrodynamic (MHD) and self-gravitating mechanisms have been 
investigated in some detail (e.g., {\it Gammie and Johnson}, 2005). 
The self-gravitating mechanism eventually turns 
itself off as the gas disk dissipates, at which point MHD may gain relevance. 
Differentially rotating disks are subject to a local instability
referred to as a magneto-rotational instability, or MRI ({\it Balbus and Hawley}, 1991).
Significant portions of the disk 
(specifically the planet formation regions) may be insufficiently ionized
for MRI to be effective, creating a 
``dead zone'' ({\it Gammie}, 1996; {\it Turner et al}., 2007), or region of inactivity.

In the absence of an MHD mechanism, one would have to resort
to a purely hydrodynamic mechanism. However, it is now known that hydrodynamic Keplerian
disks are stable to linear perturbations (e.g., {\it Ryu and Goodman}, 1992; 
{\it Balbus et al.}, 1996).
Possible sources of turbulence, such as convection 
({\it Lin and Papaloizou}, 1980) and baroclinic effects ({\it Li et al}., 
2000; {\it Klahr and Bodenheimer}, 2003) may provide inadequate, decaying 
transport in 3D disks ({\it Barranco and Marcus}, 2005; {\it Shen et al}., 
2006), subside as the disk becomes optically thin, 
and may fail to apply to isothermal portions of the disk. 
A number of analytical studies have suggested
transient growth mechanisms for purely hydrodynamic turbulence
that would lead to the excitation of
non-linear behavior ({\it Chagelishvili et al}., 2003; {\it Umurhan and Regev}, 2004; 
{\it Afshordi et al}., 2005). However, 
numerical simulations ({\it Hawley et al}., 1999; 
{\it Shen et al}., 2006) and laboratory experiments ({\it Ji et al}., 2006) 
cast doubt on the ability of purely hydrodynamic turbulence to transport 
angular momentum efficiently in Keplerian disks, even for high Reynolds number 
({\it Lesur and Longaretti}, 2005). 
Although the evidence that Keplerian disks are 
laminar is not conclusive because the Reynolds numbers in disks are much 
larger than those accessible to computers or experiments, it is fair to say 
that what makes disks around YSOs accrete remains an open problem. 

While there is a consensus that both the nebula and the subnebula undergo
turbulent early phases, we presently lack a mechanism that can sustain
turbulence in a dense, mostly isothermal subnebula
({\it Mosqueira and Estrada}, 2003a,b). In order to 
sidestep this problem, one is then forced to postulate a low density gas disk 
({\it Estrada and Mosqueira}, 2006), 
and invoke MRI turbulence to sustain turbulence in such a disk. But here 
again the likely presence of dust complicates the situation, even in the low 
density case. Hence, a sufficiently general mechanism for sustaining 
turbulence in poorly ionized disks has yet to be found, suggesting that
alternative mechanisms of disk removal need to be explored. In 
particular, the role that the planets and satellites themselves play in driving 
disk evolution has only begun to be explored ({\it Goodman and Rafikov}, 2001; 
{\it Mosqueira and Estrada}, 2003b; {\it Sari and Goldreich}, 2004).

\vspace{0.1in}
\noindent
{\bf 4.2. Methods of Solids Delivery}
\vspace{0.1in}

There are several ways in which solids can be delivered to the circumjovian 
disk. Although all of these mechanisms likely
played a role in satellite formation, it should be emphasized that at the 
time of giant planet formation, most of the available mass of solids are
in planetesimals in sizes $\gtrsim 1$ km
({\it Wetherill and Stewart}, 1993; {\it Weidenschilling}, 1997;
{\it Kenyon and Luu}, 1999; {\it Charnoz and Morbidelli}, 2003). 
In the Jupiter-Saturn region the collisional timescale for kilometer-sized 
objects is similar to the ejection timescale ($\lesssim 10^5$ years,
{\it Goldreich et al}., 2004), so that a significant fraction of the mass 
of solids are fragmented into objects smaller than $\sim 1$ km
({\it Stern and Weissman}, 2001; {\it Charnoz and Morbidelli}, 2003). 
Sufficiently small planetesimals ($\sim 1$ m) are protected from further 
collisional erosion by gas-drag and by collisional eccentricity and 
inclination damping. 
Given that fragmentation likely plays a 
significant role in the continued evolution of the heliocentric planetesimal
population following the formation of Jupiter ({\it Stern and Weissman}, 2001;
{\it Charnoz and Morbidelli}, 2003), the $1$ m $-$ $1$ km size range
of planetesimals likely plays a key role in satellite formation.

\vspace{0.1in}
{\it I. Break-up and dissolution of planetesimals in the extended envelope}.
Since the giant planet envelope
probably filled a fair fraction of its Roche lobe during a significant fraction
of its gas accretion phase, its cross section would 
have been greatly enlarged ({\it Bodenheimer and Pollack}, 1986;
{\it Pollack et al}., 1996), so that early arriving planetesimals 
would break up and/or dissolve in the envelope, enhancing its
metallicity. In the earliest stages of growth when the envelope
mass is low, most planetesimals may reach the core intact. As the
gaseous envelope becomes more massive, planetesimals begin to deposit
significant amounts of their mass in the distended envelope 
(e.g., {\it Podolak et al}., 1988). Some dust and debris deposited during the
extended envelope and relatively rapid envelope collapse phases would have been 
left behind in any the subnebula.
 
{\it II. Ablation and gas drag capture of planetesimals through the circumjovian 
gas disk}. 
Depending on the gas surface density of the subnebula, disk crossers 
can either ablate, melt, vaporize or be captured as they pass 
through the disk.
Planetesimal fragments in the size range meter-to-kilometer may ablate and
be delivered to the subdisk.
The total mass budget depends on the
surface density of solids in the solar nebula, as well as the deposition
efficiency, but it may be possible to deliver more than the mass of
the Galilean satellites, a fraction of which could have been lost later due to the
inefficiencies of satellite accretion. 

{\it III. Collisional capture of planetesimals}.
Once the planetesimal population has been perturbed into planet-crossing 
orbits ({\it Gladman and Duncan}, 1990), both gravitational 
and inelastic collisions between planetesimals within Jupiter's Hill sphere occur.
If inelastic collisions occur between
planetesimals of similar size, the loss of kinetic energy through their
collision may lead to capture, and eventually to the formation of a circumplanetary 
accretion disk. Gravitational collisions between planetesimals also leads to
mass capture.
 
{\it IV. Dust coupled to the gas inflow through the gap}. Essentially all
numerical models of giant planet formation indicate that there is flow of gas 
through the gap, with the strength of the flow dependent on the
assumed value of the solar nebula turbulence (e.g., {\it Artymowicz and
Lubow}, 1996; {\it Lubow et al}., 1999; {\it Bryden et al}., 2000;
{\it Bate et al}., 2003; {\it D'Angelo et al}., 2003b). None of
these studies incorporated dust in their simulations;
however, there are numerous
arguments as to why dust inflow cannot be the 
dominant source of solid material (for more discussion, see 
{\it Mosqueira and Estrada}, 2003a; {\it Estrada and Mosqueira}, 2006).
First, little mass remains in dust at the time of planet formation
({\it Mizuno et al}., 1978; {\it Weidenschilling}, 1997; {\it Charnoz and
Morbidelli}, 2003). Second, as was discussed in Sec. 3.3, the
specific angular momentum of the
inflow through a gap forms an extended disk, which would lead to the formation of
satellites far from the planet where they are not observed. Third, the edges
of gaps opened by giant planets act as effective filters restricting the
size and amount of dust that can be delivered this way ({\it Paardekooper and
Mellema}, 2006; {\it Rice et al}., 2006). Entrained particle sizes would be
orders of magnitude smaller in radius than the local decoupling 
size ($\sim 1$ m at Jupiter). 
Fourth, the dust content of the inflowing gas may be
substantially decreased with respect to solar nebula gas. 
Embedded planet simulations in 3D show that the gas inflow comes from
lower density regions (e.g., see Fig. \ref{fig:3d_model}) above and below the 
midplane (e.g., {\it Bate et al}., 2003; {\it D'Angelo et al}., 2003b). The 
typical maximum flow velocity occurs at a pressure scale height. This
further restricts particle sizes entrained in the gas inflow. 
 
\vspace{0.1in}
\noindent
{\bf 4.3. Gas-rich Environment}
\vspace{0.1in}

\begin{figure}[t!]
 \resizebox{\linewidth}{!}{%
 \includegraphics{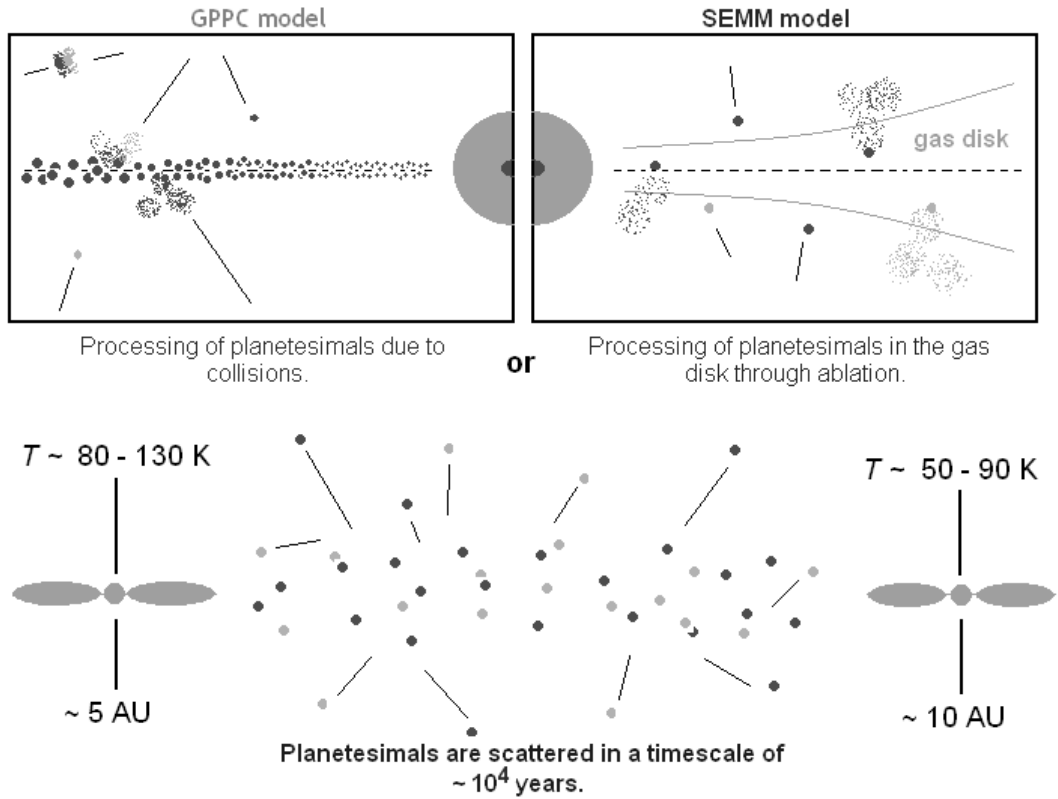}}
\caption{{\small Vignette of the dominant planetesimal delivery mechanisms
(Sec. 4.2) for the two satellite formation environments discussed. In the
left window is the gas-poor planetesimal capture (GPPC) model for the
circumjovian disk in which processing of planetesimals occurs through
planetesimal-planetesimal, and planetesimal-satellitesimal collisions. In
the right window is the solids-enhanced minimum mass (SEMM) model for the
circumjovian disk in which processing of heliocentric planetesimals occurs
through their ablation as they pass through the Jovian subnebula. The
different scenarios have implications for the compositional evolution of 
the Galilean satellites. In either case, it is expected that the planetesimal
population in the feeding zone of Jupiter (or between Jupiter and Saturn, 
if Saturn is present) will mostly be scattered away in $\sim 10^4$ years.
Some of this material ends up in the circumplanetary disk through 
collisional capture and/or through direct passage across the disk plane
prior to being scattered.}} 
\label{fig:vign2}
\end{figure}

In Section 3, we discussed the formation of a massive circumplanetary disk.
Traditionally, the approach has
been to calculate (akin to the MMSN) a minimum mass model for the 
circumjovian disk. In such a model, the total disk mass 
(gas + solids) is set 
by spreading the re-constituted mass (accounting for lost volatiles) of the 
Galilean satellites over the disk and adding enough gas 
to make the subnebula solar in composition (typically a factor of $\sim 100$, e.g., 
{\it Pollack et al}., 1994). 
The total mass of the circumplanetary disk that results from the approach described 
above is $\sim 10^{-2}$ M$_J$ (note that the disk-to-primary mass ratio for Jupiter is
similar to Sun). Interestingly, the total angular 
momentum of this gaseous disk is 
comparable to the spin angular momentum of Jupiter ({\it Stevenson et al}., 
1986; see Table 3, {\it Mosqueira and Estrada}, 2003a).  
Equipartition of angular momentum between planet and disk would result in
a massive subnebula.
 
\vspace{0.1in}
{\it 4.3.1. Decaying Turbulence Satellite Formation Scenario}.
As long as gas inflow through the gap continues, 
the circumplanetary disk should remain turbulent (driven by the inflow itself) and 
continue to viscously evolve.
The gas inflow from the circumsolar disk wanes in the
gap opening timescale of $< 10^5$ 
years in a weakly turbulent solar nebula ({\it Bryden et al}., 1999, 2000; 
{\it Mosqueira and Estrada}, 2006; {\it Morbidelli and Crida}, 2007). 
As the subnebula evolves, the gas surface density may decrease. 
Once the gas inflow ceases,
a different driving mechanism is needed to facilitate further disk evolution.

This circumplanetary disk environment in which the regular satellites
form has been dubbed the Solids Enhanced Minimum Mass (SEMM) disk
({\it Mosqueira and Estrada}, 2003a,b), because satellite formation occurs
once sufficient {\it gas} has been removed from an 
initially massive subnebula and turbulence in the circumplanetary disk
subsides. There are a number of processes that may raise the solids-to-gas
ratio and lead to a SEMM disk. In
particular, preferential removal of gas (e.g., {\it Takeuchi and Lin}, 
2002) during the inflow-driven subnebula evolution phase may lead to enhancement
of solids in the circumplanetary disk. Ablation of heliocentric planetesimals
crossing the disk may result in further enhancement of solids.
Such a disk may then
be enhanced in solids by a factor of $\sim 10$ above solar consistent with the 
solids enhancement observed in the Jovian atmosphere. This factor fits with
the solids enhancement in Jupiter's atmopshere, and is consistent 
with theoretical constraints based on the condition
for gap-opening and satellite formation and migration in such a disk (Sec. 4.3.3).
We stress that the
properties of a SEMM model are distinct from those of a minimum mass model
in terms of disk cooling, and satellite formation and migration timescales.

\begin{figure}[t!]
 \resizebox{\linewidth}{!}{%
 \includegraphics{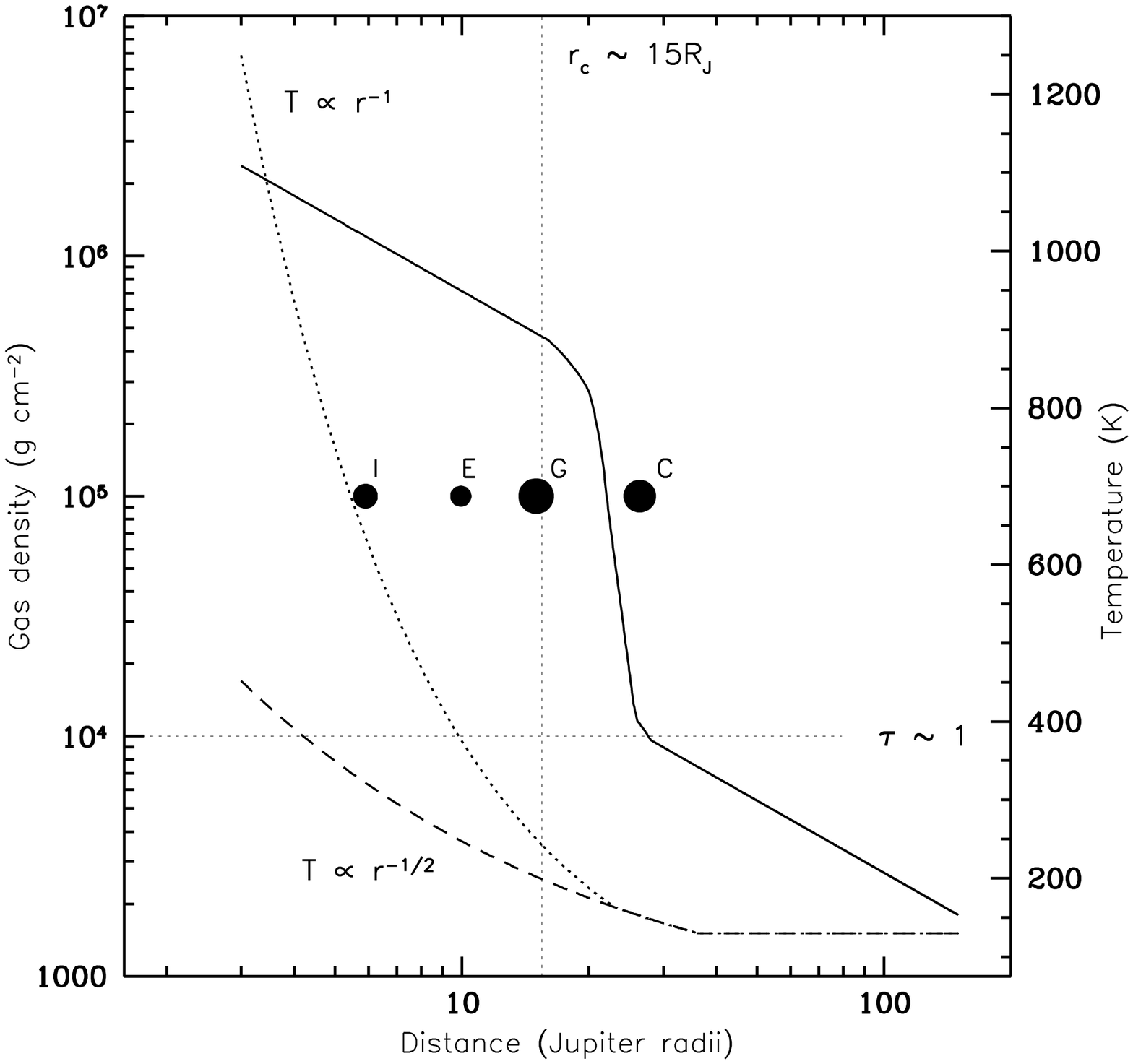}%
 \includegraphics{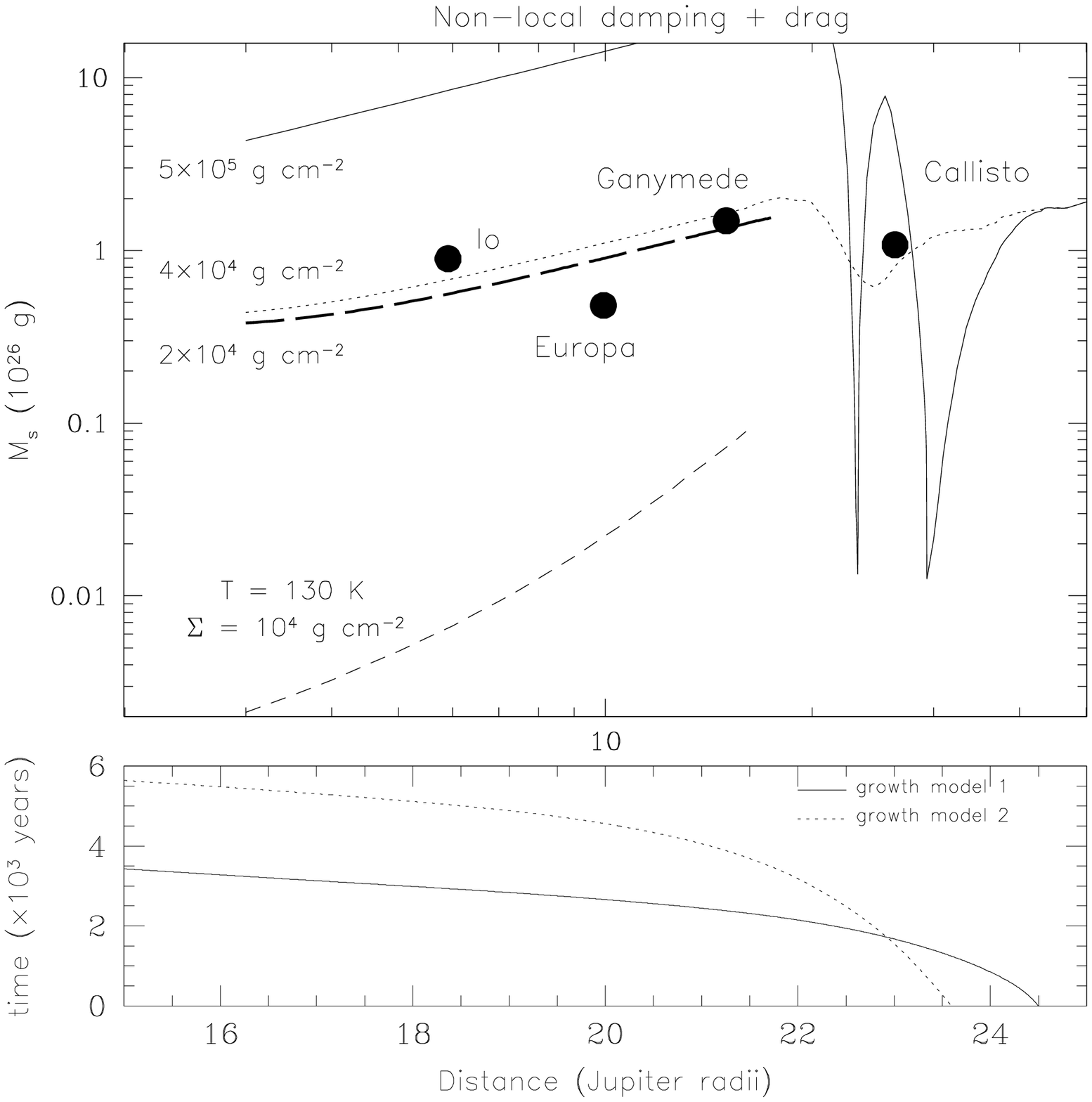}}
%\begin{center}
%$\begin{array}{cc}
%\psfig{figure=Estrada_fig8_1.ps,width=3.25in,height=3.25in} &
%\psfig{figure=Estrada_fig8_2.ps,width=3.25in,height=3.25in} \\
%\end{array}$
%\end{center}
\caption{{\small \underline{Left}: 
Idealization of the initial minimum mass $\Sigma$ and assumed photospheric $T$
profiles for the circumjovian subnebula. 
The re-constituted mass of Io, Europa, and 
Ganymede determine the mass of the optically thick inner disk, while the mass 
of Callisto is spread out over the optically thin outer disk. Ganymede
lies just inside the centrifugal radius $r_c$, while Callisto lies outside
a transition region that separates the inner and outer disks. 
The temperature is set to agree with the compositional constraints of the 
Galilean
satellites (e.g., {\it Lunine and Stevenson}, 1982;
{\it Mosqueira and Estrada}, 2003a,b), 
which implies a Jovian luminosity of $\sim 10^{-5}$ L$_\odot$,
for a planetary radius of $\sim 1.5-2$ R$_J$ consistent with planet formation 
models ({\it Hubickyj et al}., 2005). 
\underline{Upper right}: Critical
mass at which migration stalls as a function of Jupiter radii for various
$\Sigma$-profiles using both vertically thermally stratified (solid and
dotted curves), and vertically isothermal models (bold dashed curve).
Gas drag is included. The solid
curve corresponds to the minimum mass model, while the dotted and dashed
curves correspond to the SEMM model. The short-dashed curve is a constant
$\Sigma$ and $T$ model. \underline{Lower right}: Migration and growth models
for proto-Ganymede. The full sized Ganymede is evolved backward in time from
the location where it opens a gap to the point where it reaches embryo
size ($\sim 1000$ km) for a SEMM disk. Two models for growth are used.
Solid curve: linear growth model. Dotted curve: growth rate proportional to
the disk surface density. Growth is consistent with the limited migration
of Ganymede. See {\it Mosqueira and Estrada}, 2003b for detailed descriptions.}}
\label{fig:semm}
\end{figure}

\vspace{0.1in}
{\it 4.3.2. Satellite Growth in the Circumjovian Disk}.
The growth of satellitesimals and embryos in the circumplanetary disk
is controlled first by sweepup of dust and rubble 
(e.g., {\it Cuzzi et al}., 1993; {\it Weidenschilling}, 1997). As the
inflow from the circumsolar disk wanes and turbulence decays, the inner 
more massive region becomes weakly turbulent (while the outer extended disk becomes 
isothermal and quiescent).
Once this occurs,
dust coagulation and settling are assumed to proceed rapidly,
given that dynamical times are $\gtrsim 10^3$ times faster in the inner 
disk of Jupiter than in the local solar nebula. 

Once a significant fraction
of the solids in the disk have aggregated into objects large enough to
decouple from the gas (radii $R \gtrsim 10-100$ m for $\Sigma \gtrsim 10^4$ g) 
and settle to the disk 
midplane, they drift inward at different rates due to gas drag, leading to further
``drift-augmented'' accretion. In the weak turbulence regime, the ratio of the sweepup 
time (which assumes most of mass of the disk is initially in small particles 
entrained in the gas) to gas drag is

\begin{equation}
\frac{\tau_{sweep}}{\tau_{gas}} \approx \frac{4\rho_sR}{3\bar{\rho}_p
\Delta v}/\frac{4\rho_sRv_K}{3C_D\rho(\Delta v)^2} \sim 
C_D\eta \frac{\rho}{\bar{\rho}_p} < 1,
\label{equ:sweepgas}
\end{equation}

\noindent
in the inner (and outer) disk.
Here, $\bar{\rho}_p$ is the average solids density in the midplane,
$\rho_s$ is the satellitesimal density, $C_D$ the gas drag coefficient,
and $\eta = \Delta v/v_K$ (e.g., $\eta\sim 10^{-2}$ at Ganymede) measures
the difference between $v_K$ and the pressure supported gas
velocity. 
Equation (\ref{equ:sweepgas}) implies that it is
possible to form satellites/embryos of any size $< 1000$ km at any radial
location on a faster timescale then their inward migration due to gas drag.

Although perfect sticking is assumed in the sweepup model described above,
it is likely that some fragmentation and erosion occurred. In particular,
the relative velocities between decoupling particles 
and larger, relatively ``immobile'' satellitesimals are considerably greater
in the circumplanetary disk compared to the solar nebula due to much faster
dynamical times, which likely resulted in erosive impacts (turbulence exacerbates 
this problem, Sec. 2.1). In addition, other
factors may have contributed to less favorable growth (see {\it Mosqueira and Estrada},
2003a, for more discussion). However, in the SEMM model $\rho/\bar{\rho}_p \lesssim 10$ 
due to particle settling, so that Eq. (7) may remain satisfied even for  
inefficient growth. But,
it should be noted that a detailed simulation of the accretion of satellites 
from satellitesimals in circumplanetary nebulae remains to be done. 

Eventually, as the reservoir of dust and rubble is depleted, sweepup is
less efficient, and growth of 
the embryo begins to be controlled by gas drag drift-augmented accretion of 
satellitesimals and smaller embryos. 
Satellite embryos pose an effective barrier for inwardly drifting satellitesimals
due to high impact probabilities ({\it Kary et al}., 
1993; {\it Mosqueira and Estrada}, 2003a). 
Thus embryos choke off of the supply of material to the inner satellites. 
The protosatellite formation timescale is determined by the inward drift of 
the characteristic size of satellitesimals (or embryos) it accretes.
In this picture Ganymede (as well as Io and Europa) forms in $\sim 10^3-10^4$ years, 
whereas Callisto takes significantly longer 
($\sim 10^6$ years) because it derives solids from the extended low-density
outer disk. Nevertheless, the processes that lead to satellite formation are
essentially the same in the outer and inner disks.
It is important to keep in mind that in the SEMM model, the
delivery of material to the circumplanetary disk (either gas or solids) takes place 
in a $\sim 10^4$ year timescale
which is comparable to Ganymede's formation time, but shorter than that of Callisto.

We emphasize that while satellite embryos form quickly, the SEMM model
has full-sized satellites forming on a longer timescale. This is because, unlike
traditional minimum mass models, the SEMM model is not a local growth model; i.e.,
a full-sized satellite formation timescale is controlled by the timescale over
which the feeding zone of the embryo is replenished by other embryos or satellitesimals.
Thus, the mass of the satellites is spread out over the entire disk, and full-sized
satellites must accrete material from well outside their feeding zones.
In fact, most of the present day satellite disk is empty, presumably due to 
gas drag clearing of satellitesimals.

\vspace{0.1in}
{\it 4.3.3. Satellite Survival in a Gas-rich Disk}.
A gas-rich disk promotes accretion of satellites; however,
such a disk can also lead
to orbital decay or even loss of satellites on timescales much faster than
it would take the circumplanetary gas disk to dissipate. On the
one end, gas drag migration (Sec. 2.1) dominates 
for smaller (e.g., {\it Weidenschilling}, 1988),
and is the primary mechanism for drift-augmented accretion. 
On the other end of the size scale, the migration rate of larger objects 
is determined by the gravitational interaction with a gaseous disk at Lindblad 
resonances, or gas tidal torque (see {\it Goldreich and Tremaine}, 1979). 
Intermediate-sized
protosatellites ($\sim 1000$ km) must contend both with gas drag migration and
gas tidal torques that may lead to catastrophically fast (generally inward)
migration rates.
 
As a satellite grows in size, its migration
speed increases (the torque is proportional to the mass squared).
However, sufficiently large satellites (mass ratio of satellite to planet of
$\mu \sim 10^{-4}$) may stall and 
begin to open a gap
({\it Ward}, 1997; {\it Rafikov}, 2002b; {\it Mosqueira and Estrada}, 2003b). 
As a consequence of the satellite's tidal interaction with the disk (and concomitant
angular momentum transfer),  
the satellite can actually drive the evolution of the disk 
(e.g., {\it Sari and Goldreich}, 2004) by producing a local effective viscosity 
({\it Goodman and Rafikov}, 2001). 
Admittedly, the physics of disk-satellite interactions is complex. We simply note that 
other satellite formation models do not rely on gap opening for satellite survival
because the gas disk dissipates on a timescale comparable to the satellite formation
timescale ({\it Canup and Ward}, 2002; {\it Alibert et al}., 2005a; {\it Estrada and
Mosqueira}, 2006).

{\it Mosqueira and Estrada} (2003b) explored static models 
of the Jovian subnebula 
and determined the conditions under which the largest satellites may stall, 
open gaps, and survive under gas-rich conditions.
Figure \ref{fig:semm} (upper right panel) shows results from these
calculations.  First, these results indicate that a ``minimum mass disk'' 
(solid curve) is 
likely too massive to allow for the survival of any of the inner disk Galilean 
satellites, and that a significant decrease in the gas surface density is
required for satellite migration to stall (dotted and bold long-dashed curves, SEMM disk).
Thus, the gap-opening condition itself argues in favor of a subnebula solids 
enhancement of a factor of $\sim 10$ with respect to solar composition.
Second, the critical mass for a satellite to stall increases with 
distance from Jupiter. This allows a
satellite to drift in until it finds an equilibrium position, so long as it
is sufficiently massive to stall somewhere in the disk. In a regime of limited satellite
migration, the largest satellites would tend to be located near the centrifugal radius. 
Finally, due to the gradient in the disk 
temperature (taken to be controlled by Jupiter's luminosity, see Fig. 1), 
the slopes of the curves in Fig. \ref{fig:semm} are shallow, 
limiting the the range of masses that may stall  
to mass ratios of $\mu \sim 10^{-4}$. Additionally,
the density waves launched by objects of this size shock-dissipate in a length scale smaller
than their semimajor axis ({\it Goodman and Rafikov}, 2001), which allows the largest
satellites to drive the evolution of the disk, open a cavity and stall.

It must be stressed that there are a number of assumptions involved in the results 
described above. The initial gas surface density has been obtained by placing
Callisto in the outer disk, and Ganymede, Europa, and Io in the inner disk, separated
by an assumed transition between the outer and the inner disks. However, there
are presently no detailed simulations of the formation of such a disk. 
Furthermore, the temperature structure of the disk is heuristic. 
Nevertheless, we stress that the overall survival mechanism need not be dependent
on the specifics of the model. For instance, Callisto may stall because its migration
is halted by the change in the surface density due to the presence of Ganymede.
Alternatively, satellites may open gaps collectively. This latter option would be 
particularly relevant
if satellite-disk interactions can lead to the excitation and growth of satellite
eccentricities which would
result in spatially extended gap formation. Another process to consider is
photoevaporation, which may take place in a timescale comparable to that of the
formation and migration of Callisto (and Iapetus for Saturn).
The key point here
is that the largest satellites of each satellite system, and not disk dissipation due
to turbulence, may
be responsible for giant planet satellite survival.
 
Finally, it is useful to evaluate whether migration times
are consistent with the growth timescales of the Galilean satellites. This calculation
serves to provide an estimate of how far a satellite
may have migrated. By associating the current locations of the satellites
with their stalling location (Fig. \ref{fig:semm}, upper right panel), 
{\it Mosqueira and Estrada} (2003b) integrated the migration of the full-sized
Ganymede back in time to embryo size using different growth models (Fig. \ref{fig:semm},
lower right panel). Here again, migration and formation of full-sized satellites
is taken to occur in a subnebula enhanced in solids by a factor of $\sim 10$ with
respect to solar composition mixtures.
These results imply that SEMM disks are consistent with the limited
migration of at least Ganymede, so that Ganymede likely
formed close to the location of the centrifugal radius, $r_c$.

\vspace{0.1in}
\noindent
{\bf 4.4. Gas-poor Environment}
\vspace{0.1in}

In the gas-poor scenario
sustained turbulence (possibly hydrodynamic turbulence) or some other 
mechanism removes the 
gaseous circumplanetary disk quickly compared to the accretion
timescale of the satellites (which is tied to the timescale for clearing heliocentric
planetesimals from the giant planet's feeding zone). By construction, in this
model this timescale is $\sim 10^5-10^6$ years. The issue arises whether delivery
of planetesimals can last for this long. In Sec. 4.5.2, we discuss possible ways
to lengthen the planetesimal delivery timescale. Thus, satellite 
formation is taken to be somewhat akin to the formation of the terrestrial planets
as the gas surface density is taken to be low but is left unspecified.

The gas-poor environment does not face 
the survival issues associated with the presence of significant amounts of gas; 
yet, the remnant circumplanetary gas disk may still circularize orbits, 
or clear the disk of collisional debris. 
The way in which the solid material that makes up the
satellites is delivered to the 
circumplanetary disk must differ significantly from the gas-rich case. 
This scenario 
relies on the formation of an accretion disk resulting from the capture into
circumplanetary orbit of heliocentric planetesimals undergoing inelastic collisions
({\it Safronov et al}., 1986; {\it Estrada and Mosqueira}, 2006; {\it Sari
and Goldreich}, 2006), or gravitational scatterings({\it Goldreich et al}., 
2002; {\it Agnor and Hamilton}, 2006) within the Jovian Hill sphere (see Sec. 4.2).
Furthermore, here the angular momentum of the satellite system is 
largely determined by circumsolar planetesimal capture dynamics.
 
\vspace{0.1in}
{\it 4.4.1. The Circumplanetary Swarm}. 
The idea that the regular satellites could form out of a 
collisionally-captured, gravitationally-bound swarm of circumplanetary
satellitesimals has been suggested and explored in a number of classic 
publications ({\it Schmidt}, 1957; {\it Safronov and Ruskol}, 1977; 
{\it Ruskol}, 1981, 1982; {\it Safronov et al}., 1986). However, it has only
been recently that a GPPC (gas-poor planetesimal capture) model
has been advanced ({\it Estrada and Mosqueira}, 2006).  
Collisional capture mechanisms have been explored in terms of general
accretion ({\it Sari and Goldreich}, 2006), and applied to the
formation of Kuiper Belt binaries ({\it Schlichting and Sari}, 2007).

The GPPC formation scenario
is separated into two stages: an early stage in which the circumplanetary disk
is initially formed, and a late stage in which a quasi-steady state accretion disk 
is in place around the giant planet. The basic idea is that in the early stage,
inelastic and gravitational collisions 
within the Hill sphere of the giant planet 
lead to the creation of a protosatellite ``swarm'' of both retrograde and prograde 
satellitesimals extending 
out as far as circumplanetary orbits are stable, $\sim R_H/2$.
At present, 
the complicated process of circumplanetary swarm generation remains to be modeled.

{\it Estrada and Mosqueira} (2006) treat the late stage of satellite formation
in a gas-poor environment in which a circumplanetary accretion disk 
is assumed to be already present, focusing on inelastic collisions between 
incoming planetesimals with {\it larger} satellitesimals within the accretion disk
as the mass capture mechanism. Planetesimal-satellitesimal collisions
can lead to the capture of solids if the incoming planetesimal
encounters a mass comparable to, or larger than itself ({\it Safronov et al.},
1986). 

Since the circumplanetary disk has a significant surface area, the total 
mass of planetesimals that crosses the subdisk may be substantial.
A reasonable estimate of the amount of mass in residual 
planetesimals in Jupiter's feeding zone 
is $\sim 10$ M$_\oplus$ $\sim 10^{29}$ g. These planetesimals may cross the
circumplanetary disk a number of times before being scattered away by Jupiter, and thus 
may be subject to capture. The amount of mass captured depends
on the size distribution, and total mass of satellitesimals in the accretion disk,
as well as the timescale over which heliocentric planetesimals are fed into the
system.

Not all of the mass delivered to the circumplanetary disk is accreted by 
the regular satellites. Bound objects may be dislodged by passing
planetesimals, or during close encounters between the giant planets during the
excitation of the Kuiper Belt, as in the Nice
model ({\it Tsiganis et al}., 2005). 
Also, satellitesimals may be accreted onto the
planet or lost by an evection resonance ({\it Nesvorny et al.} 2003).
The timescale constraint imposed by Callisto's partially differentiated state 
would require that Callisto must form in $\gtrsim 10^5$ years, so that
the feeding timescale of planetesimals should be at least this long.
A condition for the GPPC model to satisfy the Callisto constraint is that the mass of
the extended disk of solids at any one time (excluding satellite embryos)
is typically a small fraction of a Galilean satellite ({\it Estrada and Mosqueira},
2006).  
 
Collisions in the circumplanetary disk can lead to fragmentation, accretion, 
or the removal of material from the outer to the inner portions of the disk, 
where the satellites form. This collisional removal of
material (to the inner disk) is assumed to be replenished by 
the collisional capture of heliocentric planetesimal fragments.
Removal of material from the outer regions to the
inner regions can occur because the 
net specific angular momentum of the
satellitesimal swarm (which consists of both retrograde and prograde satellitesimals) 
results in a much more compact prograde disk.

\begin{figure}[t!]
 \resizebox{\linewidth}{!}{%
 \includegraphics{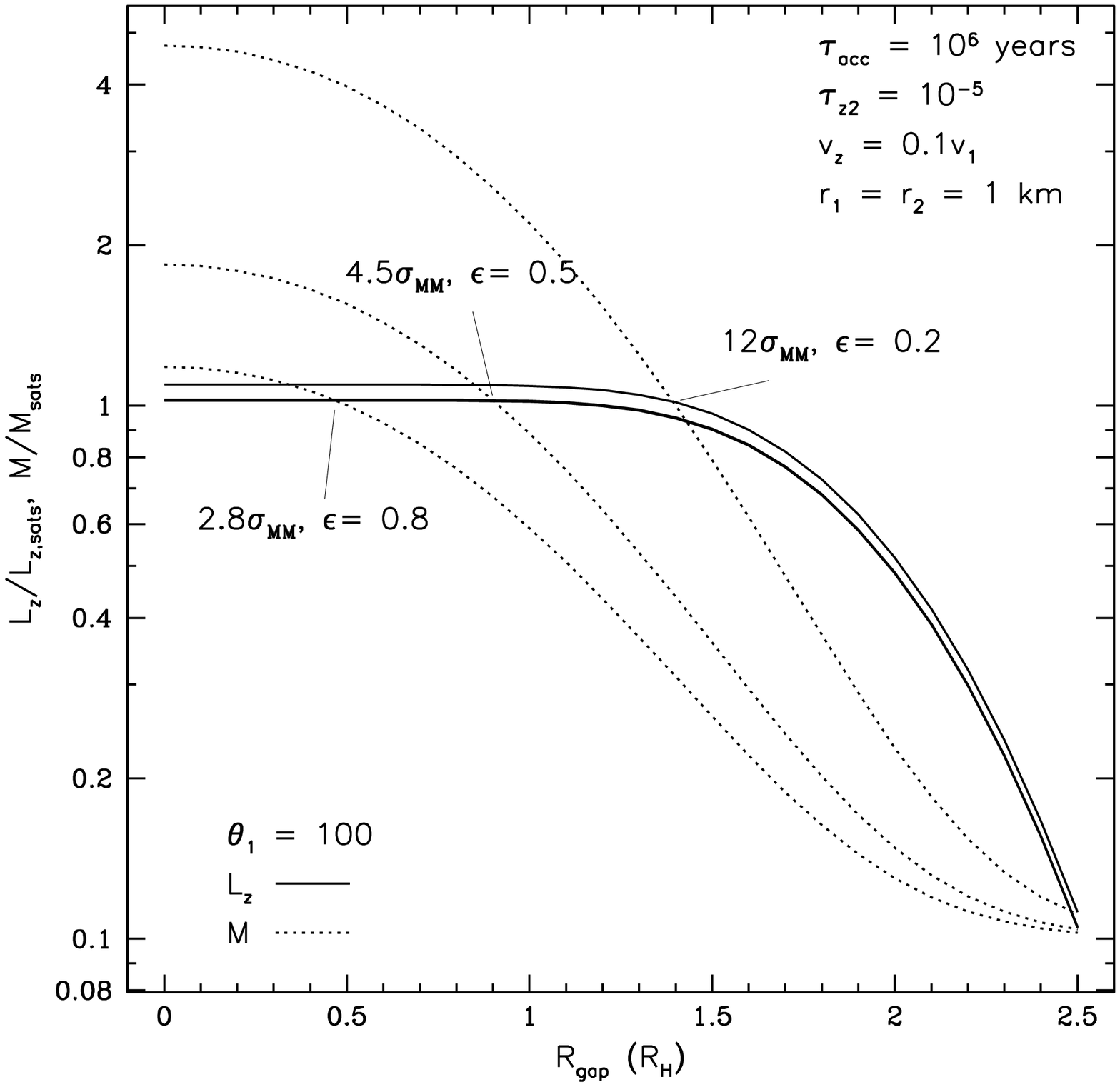}%
 \includegraphics{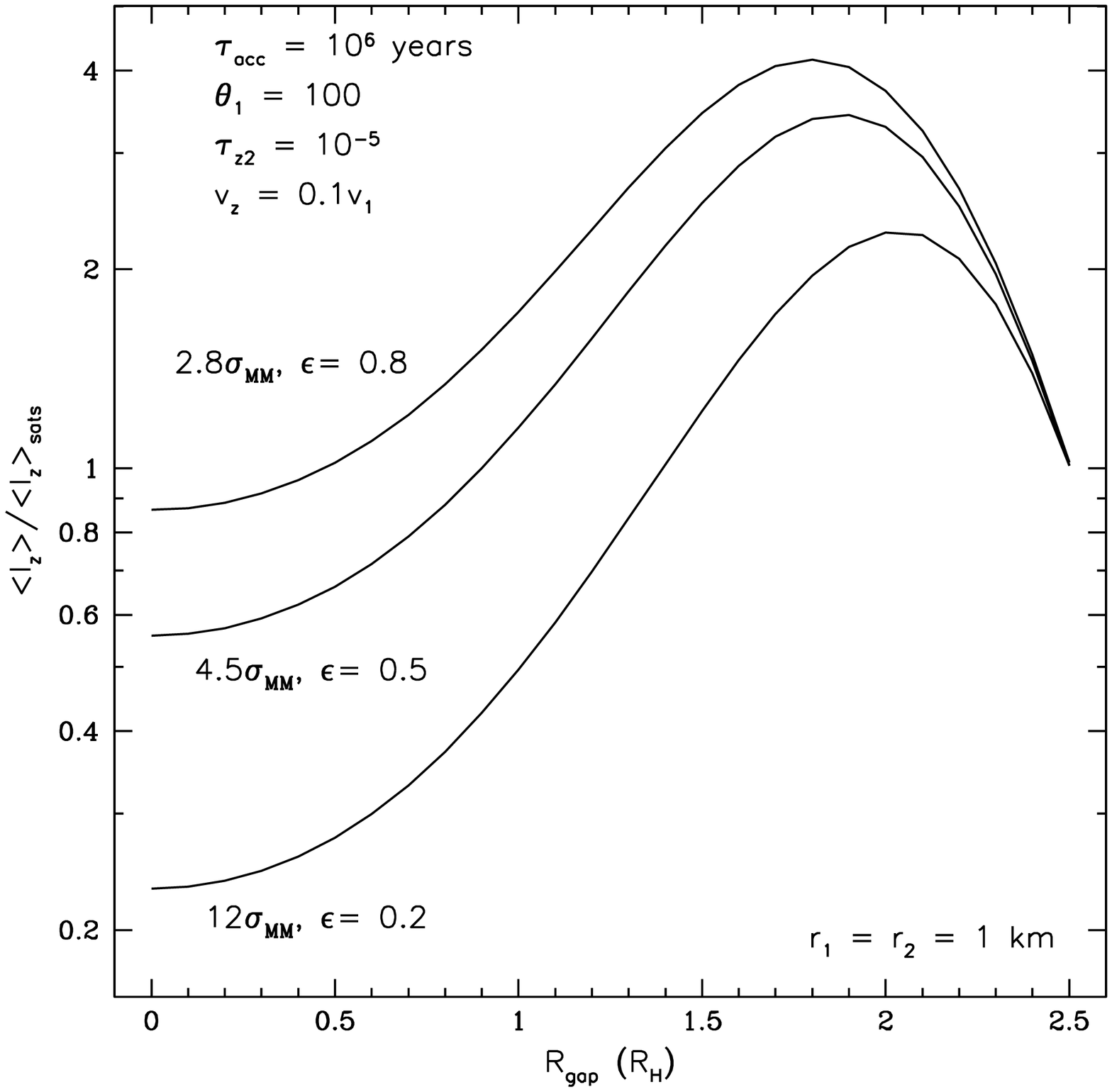}}
%\begin{center}
%$\begin{array}{cc}
%\psfig{figure=Estrada_fig9_1.ps,width=3.25in,height=3.25in} &
%\psfig{figure=Estrada_fig9_2.ps,width=3.25in,height=3.25in} \\
%\end{array}$
%\end{center}
\caption{\small \underline{Left}: Mass (dotted lines) and angular momentum 
(solid lines) 
normalized
to the values characteristic of the Galilean satellites ($M_{sats}\sim 4\times
10^{26}$ g, $L_{z,sats} \sim 4\times 10^{43}$ g cm$^2$ s$^{-1}$) delivered to
the circumplanetary disk in $\tau_{acc}=10^6$ years as a function of gap
size for several values of the solids surface density of the solar nebula. 
Surface density is expressed in terms of the MMSN value, $\sigma_{MM} = 3.3$ g cm$^{-2}$
(cf. Fig. 1). A cold planetesimal population 
($\theta_1\sim 100$ where $\theta_1 = 0.5(v_e/v_1)^2$ is
the Safronov parameter; cf. Eq. 2) is assumed. 
Solutions for the given parameters are indicated by a pointer to where
corresponding lines of $M$ and $L_z$ intersect. \underline{Right}: 
Corresponding solutions
for the specific angular momentum. In these plots, the ``optical depth''
$\tau_{z2}$ is the collision probability between the largest satellitesimals in
the circumplanetary disk, and the planetesimal scale height is $\sim v_z/\Omega$.
See {\it Estrada and Mosqueira}, 2006 for details.}
\label{fig:gppc}
\end{figure}
 
\vspace{0.1in}
{\it 4.4.2. Angular Momentum Delivery}. 
Initially, the specific angular momentum $\ell_z$ is small, but as the planet 
begins to clear its
feeding zone, sufficient planetesimals are fed from the outermost regions 
of the feeding zone where inhomogeneities in the circumsolar planetesimal disk
significantly increases $\ell_z$ of the circumplanetary swarm
({\it Lissauer and Kary}, 1991; {\it Dones and Tremaine}, 1993; 
{\it Estrada and Mosqueira}, 2006), whether they may or may not be captured (see
below). {\it Estrada and Mosqueira} (2006) posited that these collisional 
processes may deliver enough $\ell_z$
to account for the total mass and angular momentum contained in the 
Galilean satellites. 
Solutions in Fig. \ref{fig:gppc} were found for a range of gap 
sizes in the heliocentric planetesimal population, $R_{gap} \sim 0.5-1.5$ R$_H$. 
In this case, a gap refers to a depletion of solids in the circumsolar disk,
analogous to a gas gap.
As can be seen from Fig. \ref{fig:gppc} (right panel), 
the specific angular momentum contribution increases as the solids gap grows larger.
The total amount of mass and angular momentum that can be delivered is limited
by the size of the planet's feeding zone, roughly $R_{gap} \sim 2.5$ R$_H$. Thus, the
angular momentum is seen to reach a maximum before beginning to decrease
sharply as the gap size chokes off the mass inflow. The mass inflow will drop
to nearly zero {\it unless} there are mechanisms that can replenish the solids in
the planet's feeding zone. 
 
\vspace{0.1in}
\noindent
{\bf 4.5. Satisfying the Constraints}
\vspace{0.1in}
 
{\it 4.5.1. The Compositional Gradient}. Io is rocky, Europa is $\sim 90$\% rock,
$\sim 10$\% water-ice, whereas Ganymede and Callisto may be only $\sim 50$\% 
rock, and $\sim 50$\% water-ice by mass ({\it Sohl et al.}, 2002).
There are three main explanations for 
this observation. The first ascribes the high-silicate fraction of Europa relative
to Ganymede and Callisto to the subnebula temperature gradient due to Jupiter's
luminosity at the time of satellite formation
(e.g., {\it Pollack et al}., 1976; {\it Lunine and Stevenson}, 1982). In this view,
Ganymede's temperature is typically set at $T\sim 250$K to allow for the condensation
of ice at its location. Closer in, a persistently hot subnebula
would prevent the condensation of volatiles close to Jupiter 
({\it Pollack and Reynolds}, 1974; {\it Lunine and Stevenson}, 1982). In 
the optically thick case, this may imply a planetary luminosity 
$\sim 10^{-5}$ L$_\odot$ (where $L_\odot = 3.827\times 10^{33}$ ergs s$^{-1}$ is the
solar luminosity) for a planetary radius of $1.5-2$ R$_J$. 
However, this picture runs into trouble because the inner 
small satellite
Amalthea (located at $\sim 2.5$ R$_J$, Io is at $\sim 6$ R$_J$) has such a low mean 
density ($0.86$ g cm$^{-3}$) that models for its
composition require that water-ice be a major constituent, even for 
improbably high values of porosity ({\it Anderson et al.}, 2005). 

An alternative explanation
argues that the compositional gradient may be due to the increase of impact
velocities and impactor flux of Roche-lobe interlopers deep in the planetary
potential well, leading to preferential volatile depletion in the cases of Io and
Europa ({\it Shoemaker}, 1984). In this view, all of the Galilean satellites would
start out ice-rich, but some would lose more volatiles than others. 
One might expect a stochastic compositional component deep in the
planetary potential well due to high speed $\gtrsim 10$ km s$^{-1}$ impacts with
large (perhaps $\sim 10-100$ km) Roche-lobe interlopers. Such hyperbolic collisions
might conceivably remove volatiles from the mantle of a differentiated satellite and
place them on neighboring satellites (possibly analagous to the impact that may have
stripped Mercury's mantle; {\it Benz et al}., 1988). This might then explain the 
volatile depletion in Europa relative to Ganymede and Calliso, but needs to be
quantitatively evaluated. A third possibility is that Io, and possibly Europa, can 
lose their volatiles due to the Laplace resonance alone. 
A similar argument may apply to 
Enceladus and other mid-sized saturnian satellites. Amalthea then may
be a remnant of such a collisional process. 
The gas-poor satellite accretion environment
(Sec. 4.4) fits with this scenario, although a stochastic component
may also apply to the gas-rich case.

\vspace{0.1in}
{\it 4.5.2. The Callisto Constraint}. If Callisto is partially differentiated, it
argues that Callisto's formation took place over a timescale $\gtrsim 10^5$ years
({\it Stevenson et al.}, 1986; {\it Mosqueira et al}., 2001; 
{\it Mosqueira and Estrada}, 2003a,b; {\it Canup and Ward}, 2002;  {\it Alibert et al}., 2005a; {\it Estrada and Mosqueira}, 2006).
A long formation timescale may be required because the energy of accretion must be radiated 
away ({\it Safronov}, 1969; {\it Stevenson et al}., 1986) to avoid melting the interior.
Proper treatment of this problem should include the effects of an atmosphere
({\it Kuramoto and Matsui}, 1994), which allows for hydrodynamic and
collisional blow-off, or convective
transport to the subnebula ({\it Lunine and Stevenson}. 1982). Yet, 
non-hydrostatic mass anomalies at the 
boundary of the rocky core could still imply a differentiated state
({\it McKinnon}, 1997; {\it Stevenson et al.}, 2003). 

Assuming hydrostatic 
equilibrium, the simplest interpretation for Callisto's moment of
inertia is a satellite structure consisting of a $300$ km rock-free ice shell 
over a homogeneous rock and ice interior ({\it Anderson et al}., 2001; 
{\it Schubert et al},. 2004).  
The magnetic induction results may imply that an 
ocean occupies the lowermost part of Callisto's icy shell (\it Zimmer et al.}, 2000).
If Callisto's internal structure is primordial,
and it was accreted homogeneously, impacts during the 
late-stages of its growth can be used to raise the temperature of the surface 
regions of the satellite to the melting point and to supply the latent heat 
of melting (and vaporization), possibly leading to partial differentiation of just
the surface layers. 

On the other hand,  
the energy liberated by the sinking rocky component can eventually lead
to a runaway process ({\it Friedson and Stevenson}, 1983). Moreover,
both radiogenic heating and the presence of ammonia in the interior can result 
in a more stringent constraint on Callisto's thermal history. The 
relevance of these factors is uncertain: although ammonia may help to sustain 
an ocean in Callisto 
(inferred to be present from Galileo magnetometer data;
{\it Zimmer et al}., 2000), salts concentrated in
the liquid layer and/or a satellite surface regolith
can also help in this regard ({\it Spohn and Schubert},
2003); and the addition of $^{26}$Al, as specified by
CAIs, depends on the
assumption of spatial and temporal homogeneity (see
{\it Wadhwa et al}., 2007 and references therein;
{\it Castillo-Rogez et al}., 2007). 
We stress that the likely presence of an ocean means that melting did take
place, presumably during satellite accretion. 
Furthermore, the possible presence of ammonia need not result in full
differentiation ({\it Ellsworth and Schubert}, 1981).

A number of explanations for the Callisto-Ganymede
dichotomy have been offered that rely on fine-tuning uncertain parameters 
({\it Schubert et al}., 1981; {\it Lunine and Stevenson}, 1982; {\it Friedson 
and Stevenson}, 1983; {\it Stevenson et al}., 1986).
{\it Showman and Malhotra} (1997) proposed an explanation based on the Laplace
resonance; yet, it is unclear that Ganymede suffered sufficient
tidal heating to explain the differences between the two satellites.

The differences between Ganymede and Callisto can instead be a
consequence of the disk formation and evolution  
scenario described in Sec. 4.3 in which Callisto's formation time is 
long compared to that of Ganymede. In this SEMM model,
Callisto's accretion timescale is tied
to the disk clearing timescale ($\sim 10^6$ years; {\it Mosqueira
and Estrada}, 2003a). That is, in this view Callisto
derives its full mass from satellitesimals in the
extended outer disk that are brought into its feeding zone
by gas drag migration. On the other hand,
because Callisto poses an effective barrier for
inwardly migrating objects, Ganymede must
derive most of its mass from a more compact, denser
region inside of Callisto's orbit, resulting in a much
shorter accretion timescale ($\lesssim 10^4$ years). This
means that while Callisto may have enough time to radiate
away its energy of accretion Ganymede does not.
This explanation for the Ganymede-Callisto dichotomy does not require
special pleading for Callisto, but rather relies on its outer 
location to explain its accretional and thermal history. Note that a
similar formation timescale would also apply to Iapetus in the Saturnian
system for the same reason, i.e., its outermost location would lead to
long disk clearing times, which is directly tied to the satellite accretion time.

We stress that in the SEMM model, a natural outcome of a gas surface density
distribution with a long tail out to the location of the irregular satellites
is the formation of some satellites in dense inner portions, and others in
extended regions of the disk. In this context, the large separation between Titan
and Iapetus provides strong indirect evidence for a two-component subnebula.
Therefore, it is not surprising that the same can be said for both
Jupiter's and Saturn's regular satellites; i.e., Ganymede and Titan formed
in more compact, higher density regions of the disk than did Callisto and
Iapetus. Finally, we point out that it is implausible to argue that the region 
between Callisto and the irregular satellites was empty. The solids that must have r
esided there had to be cleared and most ended up accreting onto Callisto, which 
accounts for its longer formation timescale.

An additional concern is that impacts with large embryos
could result in Callisto's differentiation ({\it McKinnon},
2006). In the SEMM model, typical radii of inwardly 
migrating embryos that form in the outer disk are $\sim 200-500$ km. 
However, these
embryo sizes should {\it not} be confused with the typical sizes of
the objects that accrete onto Callisto. Characterizing typical
impactor sizes requires
treatment of the interaction between a late-stage
protosatellite and a swarm of satellitesimals,
including the effects of collisional fragmentation.

Work by {\it Weidenschilling and Davis}, (1985) shows that a
combination of gas drag and perturbations due to mean
motion resonances with a planet (or satellite) has
important consequences for the evolution of
planetesimals (or satellitesimals/embryos). Inwardly migrating embryos
can be readily captured into resonance before reaching Callisto (note that
Hyperion is in such a resonance with Titan).
{\it Malhotra} (1993) points out, though, that resonance trapping 
is vulnerable to mutual planetesimal (or satellitesimal) interactions, so that 
as the embryos approach a satellite collisions
among the embryos are expected to be destructive (see {\it Agnor and
Asphaug}, 2004 for an analogous argument for
planets), grinding them down and knocking them out of resonance.
In this view, Hyperion (radius $\sim 150$ km) might
represent a collisional remnant or survivor of such a collisional cascade.
It is reasonable to expect 
that Hyperion-like or smaller objects might be typical impactors in
the late-stage formation of Callisto (and Titan). Furthermore,
one might expect such impactors to
be porous, of low density ($\sim 0.5$ g cm$^{-3}$), and likely
undifferentiated, all of which tend to favor shallower
energy deposition during satellite accretion. The question then becomes 
whether Callisto's $\sim 10^6$ year accretion from Hyperion-like 
(or smaller), porous impactors striking the satellite
at its escape speed would result in a partially differentiated state. 
This problem remains to be tackled in detail.

Alternatively, models that make Ganymede and Callisto in the
same timescale rely on fine-tuning unknown parameters to explain the
differences between these two satellites.
This also applies to the gas-poor GPPC model, which forms 
both Callisto and Ganymede by the delivery of
small collisional fragments and debris to the circumplanetary disk over a long 
timescale.
Yet, the planetesimal clearing timescale in the Jupiter-Saturn
region is $\sim 10^4$ years ({\it Charnoz and Morbidelli}, 2003). 
Note that in the Nice model ({\it Tsiganis et al}., 2005), the timescale for
planetesimal delivery could be even shorter since Jupiter and Saturn are initially
closer together. Therefore,
a challenge in the GPPC model is to deliver enough mass at later
times to lead to the formation of a partially differentiated Callisto.
Possible ways to lengthen the planetesimal delivery timescale include stirring
of the circumsolar disk by Uranus and Neptune, and planetesimal replenishment 
by gas-drag inward migration of meter-sized bodies. 

In conclusion, it is important to point out that neither of the disk models
discussed in Sections 4.3 and 4.4 hinges on Callisto's internal state. If
Callisto turns out to be differentiated, both models remain viable even
if Callisto forms in a shorter timescale.
 
\vspace{0.25in}
\begin{center}
{\bf 5. SUMMARY}
\end{center}

We have attempted to provide a broad picture of the origin
of the Jovian system from the accumulation of the
first generation of planetesimals to the birth of the magnificently complex 
system we see today. We have summarized our current understanding
of the various stages of Jupiter's formation with the support of the most current 
numerical models of its accretion. There are several issues that still need to
be addressed in models that treat Jupiter's growth 
(see, {\it Lissauer and Stevenson}, 2007); most notably, more realistic opacity 
models for the giant planet envelope need to be implemented
(e.g., {\it Movshovitz and Podolak}, 2007). 
This issue is 
tied to requirement that Jupiter form faster than the nebula dispersal time.
Yet, the level of sophistication in these growth models continues to improve.

We have identified where the
circumplanetary disk, a by-product of Jupiter's later stages of accretion, fits
into Jupiter's formation history. There are two components to this disk:
a more radially compact disk that forms as a result of the 
accretion of low specific angular momentum gas during a period after envelope
contraction and prior to opening a deep gap in the solar nebula;
and, a more radially extended disk which 
arises as a result of continued gas inflow through a deep gap as Jupiter
approaches its final mass.
The resulting 
disk may be initially massive, as one 
might expect from a rough
equipartition of angular momentum between the planet and disk
(e.g., {\it Stevenson et al}., 1986), but is expected to evolve at least until
the gas inflow wanes. The subnebula may thus play
a key role in determining the final spin angular momentum of Jupiter. 

Presently, planetary formation models tend to focus on either the growth of the
planet in the spherically symmetric approximation, or a protoplanet embedded in a circumstellar disk
in the presence of a well-formed gap. As a result, our understanding of the formation
of the circumplanetary disk remains incomplete. Furthermore, no
simulations yet exist that model disk formation from envelope contraction 
to the isolation of the giant planet, when accretion ends. Thus, caution must be 
exercised in interpreting current numerical results. One important result from recent numerical simulations, however, is that the size of the 
circumplanetary disk formed is dependent on the specific angular momentum of the gas 
flowing into the planet's Roche lobe, which in turn depends on whether a deep gap 
is present or not.
Gap-opening may take place towards the tail end of the runaway gas accretion phase
as Jupiter approaches its final mass. If the architecture of the Solar System were
such that Saturn began much closer to Jupiter (as in the Nice model, {\it Tsiganis
et al}., 2005), disk truncation would be accomplished jointly.
 
How the circumjovian gas disk evolves over time
is dependent on the level of both solar nebula and subnebula turbulence. It
is expected that as long as there is inflow through the giant planet gap, it will
drive subnebula evolution.  However, whether or not there is a sustained, intrinsic 
source of turbulence in the circumplanetary disk has a profound effect on the 
environment in which the Galilean satellites formed. We have offered examples of
different pathways to satellite growth and survival dependent on if one posits
sustained turbulence, or turbulence decays once the gas inflow subsides. 
The assumption leads to two qualitatively different formation environments
that can account for the angular momentum of the regular satellites.  
Either model can in principle allow for
a differentiated or partially differentiated Callisto.
 
In addition, both models
rely on planetesimal delivery mechanisms to provide the mass necessary
to form the satellites, and do not rely on dust entrained in the gas inflow
to deliver solids. The compositional diversity and
potential similarities in the primordial compositions of the satellites of
Jupiter and Saturn ({\it Hibbitts}, 2006) hint that Jupiter and its satellite 
system may be derived from planetesimals formed locally as well as in more distant
regions of the solar nebula. The overall implication then is that planetesimal delivery
mechanisms likely provide the bulk of material for satellite accretion, regardless
of the gas mass contained in the circumplanetary disk.

Finally, as the inventory of discovered extrasolar planets increases,
the subject herein gains in relevance.
In most cases, there are dynamical differences between these 
newly discovered planetary systems and our own. Yet, having an understanding of 
Jupiter's formation from its beginning to its late stages serves as a 
benchmark to our general understanding of giant planet formation,
and, by analogy, of the formation of the satellite systems which
likely await as secondary discoveries around these extrasolar giants.
 
\vspace{0.1in}
{\bf \it Acknowledgements.} We would like to thank D. Stevenson and S.
Weidenschilling for their reviews of this work. We also thank
B. McKinnon for a thorough reading of the manuscript and suggestions
on how to improve its exposition. P.R.E. acknowledges the support
of a grant from NASA's Origins of Solar Systems (SSO) program.

\vspace{0.1in}
{\bf REFERENCES}
\vspace{0.1in}

\noindent
Agnor, C. B., Canup, R. M., and Levison, H. F. (1999) On the
character and consequences of large impacts in the late stage of
terrestrial planet formation. {\em Icarus}, 142, 219-237.

\noindent
Agnor, C. B., and Asphaug, E. (2004) Accretion efficiency during planetary 
collisions. {\it Astrophys. J.}, 613, L157-L160.

\noindent
Agnor, C. B., and Hamilton, D. P. (2006) Neptune's capture of its moon 
Triton in a binary-planet gravitational encounter. {\it Nature}, 441, 192-194.

\noindent
Alibert, Y., Mousis, O., Benz, W. (2005a) Modeling the Jovian subnebula. I.
Thermodynamical conditions and migration of proto-satellites.
{\it Astron. Astrophys.}, 439, 1205-1213.

\noindent
Alibert, Y., Mousis, O., Mordasini, C. and Benz, W. (2005b) New
Jupiter and Saturn formation models meet observations. {\em Astrophys.
J.}, 626, L57-L60.

\noindent
Anderson, J. D., Schubert, G., Jacobson, R. A., Lau, E. L., Moore, W. B.,
and Sjogren, W. L. (1998) Distribution of rock, metals, and ices in Callisto.
{\it Science}, 280, 1573-1576.

\noindent
Anderson, J. D., Jacobson, R. A., McElrath, T. P., Moore, W. B., Schubert, G.,
and Thomas, P. C. (2001) Shape, mean radius, gravity field, and interior
structure of Callisto. {\it Icarus}, 153, 157-161.

\noindent
Anderson, J. D., and 11 colleagues (2005) Amalthea's density is less than that
of water. {\it Science}, 308, 1291-1293.
 
\noindent	
Andrews, S. M., and Williams, J. P. (2007) High-resolution submillimeter 
constraints on circumstellar disk structure. {\it Astrophys. J.}, 659,
705-728. 

\noindent
Artymowicz, P., and Lubow, S. H. (1996) Mass flow through gaps in
circumbinary disks. {\it Astrophys. J.}, 467, L77-L80.

\noindent
Afshordi, N., Mukhopadhyay, B., and Narayan, R. (2005) Bypass to turbulence in 
hydrodynamic accretion: Lagrangian analysis of energy growth. 
{\it Astrophys. J.}, 629, 373-382.

\noindent
Atreya, S. K., Wong, M. H., Owen, T. C., Mahaffy, P. R., Niemann, H. B.,
de Pater, I., Drossart, P., and Encrenaz, T. (1999) A comparison of the
atmospheres of Jupiter and Saturn: Deep atmospheric composition, cloud
structure, vertical mixing, and origin. {\it Planet. Space. Sci.}, 47,
1243-1262.

\noindent
Balbus, S. A., and Hawley, J. F. (1991) A powerful local shear instability
in weakly magnetized disks. I. Linear analysis. {\it Astrophys. J.},
376, 214-222. 

\noindent
Balbus, S. A., Hawley, J. F., and Stone, J. M. (1996) Nonlinear stability,
hydrodynamical turbulence, and transport in disks. {\it Astrophys. J.}, 467, 76-86.

\noindent
Bar-Nun, A., Herman, G., Laufer, D., and Rappaport, M. L. (1985) Trapping 
and release of gases by water ice and implications for icy bodies.
{\it Icarus}, 63, 317-332.

\noindent
Bar-Nun, A., Dror, J., Kochavi, E., and Laufer, D. (1987) Amorphous water 
ice and its ability to trap gases. {\it Phys. Rev. B}, 35, 2427-2435. 

\noindent
Bar-Nun, A., Kleinfeld, I., and Kochavi, E. (1988) Trapping of gas mixtures 
by amorphous water ice. {\it Phys. Rev. B}, 38, 7749-7754.

\noindent	
Bar-Nun, A., Notesco, G., and Owen, T. (2007) Trapping of N2, CO and Ar in 
amorphous ice. Application to comets. {\it Icarus}, 190, 655-659.

\noindent
Barranco, J. A., and Marcus, P. S. (2005) Three-dimensional Vortices in 
Stratified Protoplanetary Disks. {\it Astrophys. J.}, 623, 1157-1170. 

\noindent
Bate, M. R., Ogilvie, G. I., Lubow, S. H., and Pringle, J. E. (2002) The
excitation, propogation, and dissipation of waves in accretion disks: the
non-linear axisymmetric case. {\it Mon. Not. Roy. Astron. Soc.}, 332,
575-600.

\noindent
Bate, M. R., Lubow, S. H., Ogilvie, G. I., and Miller, K. A. (2003) 
Three-dimensional calculations of high- and low-mass planets embedded in
protoplanetary disks. {\it Rev. Modern Phys.}, 70, 1-53.

\noindent
Bell, K. R., and Lin, D. N. C. (1994) Using FU Orionis outbursts to 
constrain self-regulated protostellar disk models. {\it Astrophys. J.}, 
427, 987-1004

\noindent
Benz, W., Slattery, W. L., and Cameron, A. G. W. (1988) Collisional stripping
of Mercury's mantle. {\it Icarus}, 74, 516-528.

\noindent
Bodenheimer, P., and Pollack, J. B. (1986) Calculations of the accretion
and evolution of the giant planets: the effects of solid cores. {\it Icarus},
67, 391-408.

\noindent
Bodenheimer, P., Burket, A., Klein, R. and Boss, A.P. (2000a)
Multiple fragmentation of protostars. In {\em Protostars and Planets IV}
(V. Mannings, A. P. Boss, and S. S. Russell, eds.), pp. 675-701. 
University of Arizona Press, Tucson.

\noindent
Bodenheimer, P., Hubickyj, O., and Lissauer, J. J. (2000b) Models of the {\it in situ}
formation of detected extrasolar giant planets. {\it Icarus}, 143, 2-14.

\noindent
Boss, A. P. (2000) Gas Giant Protoplanet Formation: Disk
instability models with thermodynamics and radiative transfer. {\em
Astrophys. J.}, 536, L101-L104.

\noindent
Brownlee, D., and 182 colleagues (2006) Comet 81P/Wild 2 under a microscope.
{\it Science}, 314, 1711-1716.

\noindent
Bryden, G., Chen, X., Lin, D. N. C., Nelson, R. P., and Papaloizou, C. B.
(1999) Tidally induced gap formation in protostellar disks: Gap clearing and
suppression of protoplanetary growth. {\it Astrophys. J.}, 514, 344-367.

\noindent
Bryden, G., Rozyczka, M., Lin, D. N. C., and Bodenheimer, P. (2000) On the
interaction between protoplanets and protostellar disks. {\it Astrophys. J.},
540, 1091-1101.

\noindent	
Buriez, J. C., and de Bergh, C. (1981) A study of the atmosphere of Saturn 
based on methane line profiles near 1.1 microns. {\it Astron. Astrophys.}, 94, 
382-390.

\noindent
Cai, K., Durisen, R. H., Michael, S., Boley, A. C., Mejía, A. C., Pickett, M. K.,
and D'Alessio, P. (2006) The effects of metallicity and grain size on gravitational instabilities in protoplanetary disks. {\it Astrophys. J.}, 636, L149-L152.

\noindent
Calvin, W. M., Clark, R. N., Brown, R. H., and Spencer, J. R. (1995)  Spectra of the icy Galilean satellites from 0.2 to 5 $\mu$m:  A compilation, new observations, and a recent summary.  {\it J. Geophys. Res.-Planets}, 100, 19041-19048.

\noindent
Cameron, A. G. W., Decampli, W. M., and Bodenheimer, P. (1982) Evolution of giant 
gaseous protoplanets embedded in the primitive solar nebula. {\it Icarus}, 49,
298-312.

\noindent
Canup, R., and Ward, W. R. (2002) Formation of the Galilean satellites:
Conditions for accretion. {\it Astron. J.}, 124, 3404-3423.

\noindent
Canup, R., and Ward, W. R. (2009) Origins of the Galilean satellites. In {\it Europa},
(McKinnon, W. B., Pappalardo, R., and Khurana, K., eds.), University of Arizona Press,
Tucson.

\noindent
Cassen, P., and Pettibone, D. (1976) Steady accretion of a rotating fluid.
{\it Astrophys. J.}, 208, 500-511.

\noindent
Castillo-Rogez, J. C., Matson, D. L., Sotin, C., Johnson, T. V., Lunine, J.
I., and Thomas, P. C. (2007) Iapetus'
geophysics: Rotation rate, shape, and equatorial
ridge. {\it Icarus}, 190, 179-202.

\noindent
Chaban, G. M., Bernstein, M., and Cruikshank, D. P. (2007)  Carbon dioxide on planetary bodies:  Theoretical and experimental studies of molecular complexes. {\it Icarus}, 
187, 592-599.

\noindent
Chagelishvili, G. D., Zahn, J.-P., Tevzadze, A. G., and Lominadze, J. G. (2003)
On hydrodynamic shear turbulence in Keplerian disks: Via transient growth to 
bypass transition. {\it Astron. Astrophys.}, 402, 401-407.

\noindent
Chambers, J. E. (2001) Making more terrestrial planets. {\it Icarus}, 
152, 205-224.

\noindent
Charnoz, S. and Morbidelli, A. (2003) Coupling dynamical and collisional
evolution of small bodies: An application to the early ejection of 
planetesimals from the Jupiter-Saturn region. {\it Icarus}, 166, 141-156.

\noindent
Chiang, E. I., and Goldreich, P. (1997) Spectral energy distributions of T 
Tauri stars with passive circumstellar disks. {\it Astrophys. J.}, 490,
368-376.

\noindent
Chiang, E. I., Joung, M. K., Creech-Eakman, M. J., Qi, C., Kessler, J. E.,
Blake, G. A., and van Dishoeck, E. F. (2001) Spectral energy distributions
of passive T Tauri and Herbig Ae disks: Grain minerology, parameter dependences,
and comparison with Infrared Space Observatory LWS observations. {\it Astrophys. J.},
547, 1077-1089.

\noindent
Ciesla F. J., and Cuzzi, J. N. (2006) The evolution of the water distribution 
in a viscous protoplanetary disk. {\it Icarus}, 81, 178-204.

\noindent
Ciesla, F. J. (2007) Outward transport of high-temperature materials around 
the midplane of the solar nebula. {\it Science}, 318, 613-616.
 
\noindent
Coradini, A., Magni, G., and Federico, C. (1981) Gravitational instabilities 
in satellite disks and formation of regular satellites. 
{\it Astron. Astrophys.}, 99, 255-261.

\noindent
Coradini, A., Federico, C., and Luciano, P. (1982) Ganymede and Callisto:
Accumulation heat content. In {\it The Comparative Study of Planets},
(A. Coradini, and N. Fulchignoni, eds.), Dordrecht Reidel, Dordrecht.

\noindent
Coradini, A., Cerroni, P., Magni, G., and Federico, C. (1989) Formation of the
satellites of the outer solar system - Sources of their atmospheres. In
{\it Origin and Evolution of Planetary and Satellite Atmospheres},
(S. K. Atreya, J. B. Pollack, and M. S. Matthews, eds.), pp. 723-762,
University of Arizona Press, Tucson.

\noindent
Courtin, R., Gautier, D., Marten, A., Bezard, B., and Hanel, R. (1984)
The composition of Saturn's atmosphere at northern temperate latitudes from 
Voyager IRIS spectra - NH$_3$, PH$_3$, C$_2$H$_2$, C$_2$H$_6$, CH$_3$D, CH$_4$, 
and the Saturnian D/H isotopic ratio. {\it Astrophys. J.}, 287, 899-916. 

\noindent
Cruikshank, D. P., and 30 colleagues (2007a) Surface composition of Hyperion. 
{\it Nature}, 448, 54-56.

\noindent
Cruikshank, D. P., and 27 colleagues (2007b) Hydrocarbons on Saturn's 
satellites Iapetus and Phoebe.  {\it Icarus}, in press. 

\noindent
Cuzzi, J. N., Dobrovolskis, A. R., and Champney, J. M. (1993) Particle-gas
dynamics in the midplane of the protoplanetary nebula. {\it Icarus}, 106,
102-134.

\noindent
Cuzzi, J. N., Davis, S. S., and Dobrovolskis, A. R. (2003) Radial drift, 
evaporation, and diffusion: enhancement and redistribution of silicates, 
water, and other condensibles in the nebula (abstract). In {\it BAAS}, 35,
\# 27.02, pp. 964.

\noindent
Cuzzi, J. N., and Hogan, R. C. (2003) Blowing in the wind I. Velocities of 
chondrule-sized particles in a turbulent protoplanetary nebula. {\it Icarus}, 
164, 127-138.

\noindent
Cuzzi, J. N., and Zahnle, K. J. (2004) Material enhancement in protoplanetary 
nebulae by particle drift through evaporation fronts. {\it Astrophys. J.},
614, 490-496.

\noindent
Cuzzi, J. N., and Weidenschilling, S. J. (2004) Formation of planetesimals in 
the solar nebula. In {\it Protostars and Planets III}, ((E. H. Levy, and J. I. 
Lunine, eds.), pp. 1031-1060, University of Arizona Press, Tucson.

\noindent
Cuzzi, J. N., and Weidenschilling, S. J. (2006) Particle-gas dynamics and 
primary accretion. In {\it Meteorites and the Early Solar System II}, 
(D. S. Lauretta and H. Y. McSween Jr., eds.), pp. 353-381, University of
Arizona Press, Tucson.

\noindent
Cuzzi, J. N., Hogan, R. C., and Shariff, K. (2007) Towards a scenario for 
primary accretion of primitive bodies (abstract). In Lunar and Planetary 
Science XXXVIII, Abstract \# 1338, Lunar and Planetary Science Institute,
Houston (CD-ROM).

\noindent
D'Angelo, G. Henning, T., and Kley, W. (2002) Nested-grid calculations of
disk-planet interaction. {\it Astron. Astrophys.}, 385, 647-670.

\noindent
D'Angelo, G., Kley, W. and Henning, T. (2003a) Orbital migration
and mass accretion of protoplanets in three-dimensional global
computations with nested grids. {\em Astrophys. J.}, 586, 540-561.

\noindent	
D'Angelo, G., Henning, T., and  Kley, W. (2003b) Thermohydrodynamics of 
circumstellar disks with high-mass planets. {\it Astrophys. J.}, 599,
548-576.

\noindent
D'Angelo, G., Bate, M. R., and Lubow, S. H. (2005) The dependence of 
protoplanet migration rates on co-orbital torques. {\it Mon. Not. Roy.
Astron. Soc.}, 358, 316-332.

\noindent	
de Pater, I., and Massie, S. T. (1985) Models of the millimeter-centimeter 
spectra of the giant planets. {\it Icarus}, 62, 143-171.

\noindent
Dominik, C., Blum, J., Cuzzi, J. N., and Wurm, G. (2007) Growth of dust as the 
initial step toward planet formation. In {\it Protostars and Planets V}, 
(B. Reipurth, D. Jewitt,
and K. Keil, eds.), pp. 783-800. University of Arizona Press, Tucson.

\noindent
Dones, L., and Tremaine, S. (1993) On the origin of planetary spins. 
{\it Icarus}, 103, 67-92.
 
\noindent
Dullemond, C. P., and Dominik, C. (2005) Dust coagulation in protoplanetary 
disks: A rapid depletion of small grains. {\it Astron. Astrophys.}, 434, 
971-986.

\noindent
Dullemond, C. P., Hollenbach, D., Kamp, I., and D'Alessio, P. (2007) Models
of the structure and evolution of protoplanetary disks. In
{\it Protostars and Planets V}, (B. Reipurth, D. Jewitt,
and K. Keil, eds.), pp. 559-572. University of Arizona Press, Tucson.

\noindent
Durisen, R. H., Boss, A. P., Mayer, L., Nelson, A. F., Quinn, T., and
Rice, W. K. M. (2007) Gravitational instabilities in gaseous protoplanetary
disks and implications for giant planet formation. In
{\it Protostars and Planets V}, (B. Reipurth, D. Jewitt,
and K. Keil, eds.), pp. 607-622. University of Arizona Press, Tucson.
 
\noindent
Estrada, P. R. (2002) Formation of satellites around gas giant planets. Ph.D.
thesis, Cornell University, Ithaca. 326 pp.

\noindent
Estrada, P. R., and Mosqueira, I. (2006) A gas-poor planetesimal capture model
for the formation of giant planet satellite systems. {\it Icarus}, 181,
486-509.

\noindent
Fegley, Jr., B. (1993)  Chemistry of the solar nebula.  In {\it The Chemistry of 
Life’s Origins},  (J. M. Greenberg, C. X. Mendoza-G\'{o}mez, and V. Pirronello, eds.), Kluwer, pp. 75-147.

\noindent
Friedson, A. J., and Stevenson, D. J. (1983) Viscosity of rock-ice mixtures 
and applications to the evolution of icy satellites. {\it Icarus}, 56, 1-14.

\noindent
Gammie, C. F. (1996) Linear theory of magnetized, viscous, self-gravitating
gas disks. {\it Astrophys. J.}, 463, 725.

\noindent
Gammie, C. F., and Johnson, B. M. (2005) Theoretical studies of gaseous disk 
evolution around solar mass stars. In {\it Chondrites and the Protoplanetary Disk}, 
(A. N. Krot, E. R. D. Scott, and B. Reipurth., eds.), ASP Conference Series, 341,  
pp. 145.

\noindent
Garaud, P., and Lin, D. N. C. (2007) The effect of internal dissipation and 
surface irradiation on the structure of disks and the location of the snow line 
around Sun-like stars. {\it Astrophys. J.}, 654, 606-624.

\noindent
Gautier, D. Hersant, F., Mousis, O., and Lunine, J. I. (2001a) Enrichments
in volatiles in Jupiter: A new interpretation of the Galileo measurements.
{\it Astrophys. J.}, 550, L227-L230.

\noindent
Gautier, D. Hersant, F., Mousis, O., and Lunine, J. I. (2001b) Erratum:
Enrichments in volatiles in Jupiter: A new interpretation of the Galileo 
measurements. {\it Astrophys. J.}, 559, L183.

\noindent
Gladman, B. and Duncan, M. (1990) On the fates of minor bodies in the outer
solar system. {\it Astron. J.}, 100, 1680-1693.

\noindent
Goldreich, P., and Ward, W. R. (1973) The formation of planetesimals.
{\it Astrophys. J.}, 183, 1051-1062.

\noindent
Goldreich, P. and Tremaine, S. (1979) The excitation of density waves at the
Lindblad and co-rotation resonances by an external potential. {\it Astrophys.
J.}, 233, 857-871.

\noindent
Goldreich. P., Lithwick, Y., and Sari, R. (2002) Formation of Kuiper-belt 
binaries by dynamical friction and three-body encounters. {\it Nature}, 420, 
643-646.

\noindent
Goldreich, P., Lithwick, Y., and Sari, R. (2004) Final stages of planet 
formation. {\it Astrophys. J.}, 614, 497-507.

\noindent
Goodman, J., and Rafikov, R. R. (2001) Planetary torques as the viscosity of
protoplanetary disks. {\it Astrophys. J.}, 552, 793-802.

\noindent
Greenberg, R., Hartmann, W. K., Chapman, C. R., and Wacker, J. F. (1978)
Planetesimals to planets - Numerical siumulation of collisional evolution.
{\it Icarus}, 35, 1-26.

\noindent
Guillot, T. (2005) The interiors of giant planets: Models and outstanding 
questions. {\it Annual Rev. Astron. Astrophys.}, 33, 493-530.

\noindent
Hawley, J. F., Balbus, S. A., and Winters, W. F. (1999) Local hydrodynamic 
stability of accretion disks. {\it Astrophys. J.}, 518, 394-404.

\noindent
Hayashi, C. (1981) Structure of the solar nebula, growth and decay of magnetic
fields, and effects of magnetic and turbulent viscosities on the nebula.
{\it Prog. Theor. Phys. Suppl.}, 70, 35-53.

\noindent
Hersant, F., Gautier, D., and Lunine, J. I. (2004) Enrichment in volatiles
in the giant planets of the Solar System. {\it Planet. Space Sci.}, 52,
623-641. 

\noindent
Hibbitts, C. A., McCord, T. B., and Hansen, G. B. (2000) Distributions of 
CO$_2$ and SO$_2$ on the surface of Callisto. {\it J. Geophys. Res.}, 
105, 22541-22558.

\noindent
Hibbitts, C. A. (2006) Intriguing differences and similarities in the surface
compositions of the icy Saturnian and Galilean satellites (abstract). American
Geophysical Union, Fall Meeting 2006, abstract \#P32A-08.

\noindent
Hibbitts, C. A., Pappalardo, R., Klemaszewski, J., McCord, T. B., and Hansen, G. B. 
(2001). Comparing carbon dioxide distributions on Ganymede and Callisto.  
LPSC XXXII, Abstract 1263.pdf

\noindent
Hibbitts, C. A., Pappalardo, R. T., Hansen, G. B., and McCord, T. B. (2003)
Carbon dioxide on Ganymede. {\it J. Geophys. Res.}, 108, (2)1-22.
 
\noindent
Hubbard, W. B., Burrows, A., and Lunine, J. I. (2002) Theory of giant planets.
{\it Annual Rev. Astron. Astrophys.}, 40, 103-136.

\noindent
Hubickyj, O., Bodenheimer, P., and Lissauer, J. J. (2005) Accretion of the 
gaseous envelope of Jupiter around a 5-10 Earth-mass core. {\it Icarus},
179, 415-431.
 
\noindent
Ji, H., Burin, M., Schartman, E., and Goodman, J. (2006) Hydrodynamic 
turbulence cannot transport angular momentum effectively in astrophysical disks.
{\it Nature}, 444, 343-346.

\noindent	
Johansen, A., Oishi, J. S., Low, M.-M. M., Klahr, H., Henning, T., and Youdin, 
A. N. (2007) Rapid planetesimal formation in turbulent circumstellar disks.
{\it Nature}, 448, 1022-1025.

\noindent
Kary, D. M., Lissauer, J. J., and Greenzweig, Y. (1993) Nebular gas drag
and planetary accretion. {\it Icarus}, 106, 288.

\noindent
Kenyon, S. J., and Hartmann, L. (1987) Spectral energy distributions of T 
Tauri stars - Disk flaring and limits on accretion. {\it Astrophys. J.},
323, 714-733.

\noindent
Kenyon, S. J., and Luu, J. X. (1999)   Accretion in the early outer Solar 
System. {\it Astrophys. J.}, 526, 465-470.
 
\noindent
Kitamura, Y., Momose, M., Yokogawa, S., Kawabe, R., Tamura, M., and Ida, S.
(2002) Investigation of the physical properties of protoplanetary disks 
around T Tauri stars by a 1 arcsecond imaging survey: Evolution and diversity 
of the disks in their accretion stage. {\it Astrophys. J.}, 581, 357-380.

\noindent
Klahr, H. H., and Bodenheimer, P. (2003) Turbulence in accretion disks:
Vorticity generation and angular momentum transport via the global
baroclinic instability. {\it Astrophys. J.}, 582, 869-892.

\noindent
Klahr, H., and Kley, W. (2006) 3D-radiation hydro simulations of disk-planet 
interactions. I. Numerical algorithm and test cases. {\it Astron. Astrophys.},
445, 747-758.

\noindent
Kley, W. (1999) Mass flow and accretion through gaps in accretion discs.
{\it Mon. Not. Roy. Astron. Soc.}, 303, 696-710.

\noindent
Kley, W. (2000) On the migration of a system of protoplanets. 
{\it Mon. Not. Roy. Astron. Soc.}, 313, L47-L51.

\noindent
Kokubo, E., and Ida, S. (1998) Oligarchic growth of protoplanets.
{\em Icarus}, 131, 171-178.

\noindent
Kornet, K., R\'{o}zyczka, M., and Stepinski, T. F. (2004) An alternative 
look at the snowline in protoplanetary disks. {\it Astron. Astrophys.},
417, 151-158.

\noindent
Korycansky, D. G., Pollack, J. B., and Bodenheimer, P. (1991) Numerical 
models of giant planet formation with rotation. {\it Icarus}, 92, 234-251.

\noindent	
Kouchi, A., Yamamoto, T., Kozasa, T., Kuroda, T., and Greenberg, J. M.
(1994) Conditions for condensation and preservation of amorphous ice and 
crystallinity of astrophysical ices. {\it Astron. Astrophys.}, 290, 1009-1018.

\noindent
Kuramoto, K., and Matsui, T. (1994) Formation of a hot proto-atmosphere on the
accreting giant icy satellite: Implications for the origin and evolution of
Titan, Ganymede, and Callisto. {\it J. Geophys. Res.}, 99, 21183-21200.

\noindent
Laskar, J. (2000) On the spacing of planetary systems. {\em Phys.
Rev. Lett.}, 84, 3240-3243.
 
\noindent
Lesur, G., and Longaretti, P.-Y. (2005) On the relevance of subcritical 
hydrodynamic turbulence to accretion disk transport. 
{\it Astron. Astrophys.}, 444, 25-44. 

\noindent
Levin, B. J. (1978) Relative velocities of planetesimals and the early accumulation
of planets. {\it Moon Plan.}, 19, 289-296.

\noindent
Levison, H. F., and Morbidelli, A. (2003) The formation of the Kuiper belt 
by the outward transport of bodies during Neptune's migration.
{\it Nature}, 426, 419-421.

\noindent
Li, H., Finn, J. M., Lovelace, R. V. E., and Colgate, S. A. (2000) Rossby wave
instability of thin accretion disks. II. Detailed linear theory. 
{\it Astrophys. J.}, 533, 1023-1034.

\noindent
Lin, D. N. C., and Papaloizou, J. (1979) Tidal torques on accretion
discs in binary systems with extreme mass ratios. {\em Mon. Not. Roy.
Astron. Soc.}, 186, 799-812.

\noindent
Lin, D. N. C., and Papaloizou, J. (1980) On the structure and evolution 
of the primordial solar nebula. {\it Mon. Not. Roy. Astron. Soc.}, 191, 
37-48.
 
\noindent
Lissauer, J. J. (1987) Timescales for planetary accretion and the
structure of the protoplanetary disk. {\em Icarus}, 69, 249-265.

\noindent
Lissauer, J. J. (1993) Planet formation. {\em Ann. Rev. Astron.
Astrophys.}, 31, 129-174.

\noindent
Lissauer, J. J. (1995) Urey prize lecture: On the diversity of plausible
planetary systems. {\it Icarus}, 114, 217-236.

\noindent
Lissauer, J. J. (2007) Planets formed in habitable zones of M dwarf stars 
probably are deficient in volatiles. {\it Astrophys. J.}, 660, L149-L152.

\noindent
Lissauer, J.J., and Kary, D. M. (1991) The origin of the systematic component
of planetary rotation. I. Planet on a circular orbit. {\it Icarus}, 94,
126-159.
 
\noindent
Lissauer, J.J., and Stevenson, D. J. (2007) Formation of giant planets. In
{\it Protostars and Planets V}, (B. Reipurth, D. Jewitt,
and K. Keil, eds.), pp. 591-606. University of
Arizona Press, Tucson.

\noindent
Lissauer, J. J., Hubickyj, O., D'Angelo, G., and Bodenheimer, P. (2009) Models
of Jupiter's growth incorporating thermal and hydrodynamical constraints. 
Icarus, in press.
 
\noindent
Lubow, S. H., Seibert, M., and Artymowicz, P. (1999) Disk accretion onto
high mass planets. {\it Astrophys. J.}, 526, 1001.

\noindent
Lunine, J. I., and Stevenson, D. J. (1982) Formation of the Galilean
satellites in a gaseous nebula. {\it Icarus}, 52, 14-39. 

\noindent
Lunine, J. I., and Stevenson, D. J. (1985) Thermodynamics of clathrate 
hydrate at low and high pressures with application to the outer solar system.
{\it Astrophys. J. Supp. Ser.}, 58, 493-531.

\noindent
Machida, M. N., Kokubo, E., Inutsuka, S.-I., and Matsumoto, T. (2008) Angular
momentum accretion onto a gas giant planet. {\it Astrophys. J.}, 685, 1220-1236.

\noindent
Mahaffy, P. R., Niemann, H. B., Alpert, A., Atreya, S. K., Demick, J., Donahue, 
T. M., Harpold, D. N., and Owen, T. C. (2000) Noble gas abundance and isotope 
ratios in the atmosphere of Jupiter from the Galileo Probe Mass Spectrometer.
{\it J. Geophys. Res.}, 105, 15061-15072.

\noindent
Makalkin, A. B., Dorofeeva, V. A., and Ruskol, E. L. (1999) Modeling the 
protosatellite circum-jovian accretion disk: An estimate of the basic 
parameters. {\it Solar Sys. Res.}, 33, 456. 

\noindent
Malhotra, R. (1993) Orbital resonances in the solar nebula - Strengths and 
weaknesses. {\it Icarus}, 106, 264-273.

\noindent	
Marley, M. S., Fortney, J. J., Hubickyj, O., Bodenheimer, P., and Lissauer, J. J.
(2007) On the luminosity of young Jupiters. {\it Astrophys. J.}, 541-549.

\noindent
Marten, A., Courtin, R., Gautier, D., and Lacombe, A. (1980) Ammonia vertical 
density profiles in Jupiter and Saturn from their radioelectric and infrared 
emissivities. {\it Icarus}, 41, 410-422. 

\noindent
Mayer, L., Quinn, T., Wadsley, J., and Standel, J. (2002)
Formation of giant planets by fragmentation of protoplanetary disks.
{\em Science}, 298, 1756-1759.
 
\noindent
McCord, T. B., and 11 colleagues (1997) Organics and other molecules in the surfaces 
of Ganymede and Callisto. {\it Science}, 278, 271-275.

\noindent
McKinnon, W. B. (1997) NOTE: Mystery of Callisto: Is it undifferentiated?
{\it Icarus}, 130, 540-543.

\noindent
McKinnon, W. B. (2006) Differentiation of the Galilean satellites: It's different 
out there (abstract). Workshop on Early Planetary Differentiation, LPI Contribution 
No. 1335, pp. 66-67.

\noindent	
Mekler, Y., and Podolak, M. (1994) Formation of amorphous ice in the 
protoplanetary nebula. {\it Plan. Spa. Sci.}, 42, 865-870. 

\noindent 
Meyer, M. R., Backman, D. E., Weinberger, A. J., and Wyatt, M. C. (2007) 
Evolution of circumstellar disks around normal stars: Placing our solar
system in context. In {\it Protostars and Planets V}, (B. Reipurth, D. Jewitt,
and K. Keil, eds.), pp. 573-590. University of Arizona Press, Tucson.

\noindent
Miyoshi, K., Takeuchi, T., Tanaka, H., and Ida, S. (1999) Gravitational interaction
between a protoplanet and a protoplanetary disk. I. Local three-dimensional
simulations. {\it Astrophys. J.}, 516, 451-464.

\noindent
Mizuno, H., Nakazawa, K., and Hayashi, C. (1978) Instability of a gaseous
envelope surrounding a planetary core and formation of giant planets.
{\it Pro. Theor. Phys.}, 60, 699-710.

\noindent
Mizuno, H. (1980) Formation of the giant planets. {\em Prog.
Theor. Phys.}, 64, 544-557.

\noindent
Morbidelli, A., and Crida, A. (2007) The dynamics of Jupiter and Saturn in the 
gaseous protoplanetary disk. {\it Icarus}, 191, 158-171.

\noindent
Morfill, G. E., and V\"{o}lk, H. J. (1984) Transport of dust and vapor 
and chemical fractionation in the early protosolar cloud.
{\it Astrophys. J.}, 287, 371-395.

\noindent
Mosqueira, I., and Estrada, P. R. (2000) Formation of large regular satellites 
of giant planets in an extended gaseous nebula: Subnebula model and accretion 
of satellites. Technical Report, NASA Ames Research Center Moffett Field, CA.

\noindent
Mosqueira, I., Estrada, P. R., Cuzzi, J. N., and Squyres, S. W. (2001)
Circumjovian disk clearing after gap-opening and the formation of a partially 
differentiated Callisto (abstract). In Lunar and Planetary 
Science XXXII, Abstract \# 1989, Lunar and Planetary Science Institute,
Houston.
 
\noindent
Mosqueira, I., and Estrada, P. R. (2003a) Formation of the regular satellites
of giant planets in an extended gaseous nebula I: Subnebula model and
accretion of satellites. {\it Icarus}, 163, 198-231.

\noindent
Mosqueira, I., and Estrada, P. R. (2003b) Formation of the regular satellites
of giant planets in an extended gaseous nebula II: Satellite migration and
survival. {\it Icarus}, 163, 232-255.

\noindent
Mosqueira, I., and Estrada P. R. (2005) On the Origin of the Saturnian satellite 
system: Did Iapetus form in-situ? (abstract). In Lunar and Planetary Science 
XXXVI, Abstract \# 1951, Lunar and Planetary Institute, Houston (CD-ROM).

\noindent
Mosqueira, I., and Estrada, P. R. (2006) Jupiter's obliquity and a long-lived 
circumplanetary disk. {\it Icarus}, 180, 93-97.

\noindent
Mousis, O., and Gautier, D. (2004)  Constraints on the presence of volatiles 
in Ganymede and Callisto from an evolutionary turbulent model of the Jovian 
subnebula. {\it Plan. Spa. Sci.}, 52, 361-370.

\noindent
Movshovitz, N., and Podolak, M. (2007) The opacity of grains in protoplanetary
atmospheres. {\it Icarus}, 194, 368-378.

\noindent
Nakagawa, Y., Sekiya, M., and Hayashi, C. (1986) Settling and growth of dust 
particles in a laminar phase of a low-mass solar nebula. {\it Icarus}, 67, 
375-390.

\noindent
Natta, A., Testi, L., Calvet, N., Henning, T., Waters, R., and Wilner, D.
(2007) Dust in protoplanetary disks: Properties and evolution. In
{\it Protostars and Planets V}, (B. Reipurth, D. Jewitt,
and K. Keil, eds.), pp. 767-782. University of Arizona Press, Tucson.

\noindent
Nesvorn\'{y}, D., Alvarellos, J. L. A., Dones, L., and Levison, H. F. (2003)
Orbital and collisional evolution of the irregular satellites. {\it Astron. J.},
126, 398-429.
 
\noindent
Ohtsuki, K., Stewart, G. R., and Ida, S. (2002) Evolution of
planetesimal velocities based on three-body orbital integrations and
growth of protoplanets. {\em Icarus}, 155, 436-453.

\noindent
Ormel, C. W., and Cuzzi, J. N. (2007) Closed-form expressions for particle 
relative velocities induced by turbulence.  {\it Astron. Astrophys.}, 
466, 413-420.

\noindent
Ossenkopf, V. (1993) Dust coagulation in dense molecular clouds: The 
formation of fluffy aggregates. {\it Astron. Astrophys.}, 280, 617-646.

\noindent
Owen, T., and Encrenaz, T. (2003) Element abundances and isotope ratios in 
the giant planets and Titan. {\it Space. Sci. Rev.}, 106, 121-138. 

\noindent
Paardekooper, S.-J., and Mellema, G. (2006) Dust flow in gas disks in the 
presence of embedded planets. {\it Astron. Astrophys.}, 453, 1129-1140.

\noindent
Podolak, M., Pollack, J. B., and Reynolds, R. T. (1988) Interactions of 
planetesimals with protoplanetary atmospheres. {\it Icarus}, 73. 163-179.

\noindent
Pollack, J. B., and Reynolds, R. T.(1974) Implications of Jupiter's early
contraction history for the composition of the Galilean satellites. 
{\it Icarus}, 21, 248-253.

\noindent
Pollack, J. B., Grossman, A. S., Moore, R., and Graboske Jr., H. C. (1976)
The formation of Saturn's satellites and rings as influenced by Saturn's
contraction history. {\it Icarus}, 29, 35-48.

\noindent
Pollack, J. B., Grossman, A. S., Moore, R., and Graboske Jr., H. C. (1977)
A calculation of Saturn's gravitational contraction history. {\it Icarus},
30, 111-128.

\noindent
Pollack, J. B., Hollenbach, D., Beckwith, S., Simonelli, D. P., Roush, T., 
and Fong, W. (1994) Composition and radiative properties of grains in
molecular clouds and accretion disks. {\it Astrophys. J.}, 421, 615-639.

\noindent
Pollack, J. B., Hubickyj, O., Bodenheimer, P., Lissauer, J. J., Podolak, M.,
and Greenszeig, Y. (1996) Formation of the giant planets by concurrent
accretion of solids and gas. {\it Icarus}, 124, 62-85.

\noindent
Porco, C. C., and 34 colleagues (2005) Cassini imaging science: Initial 
results on Saturn's rings and small satellites. {\it Science}, 307, 1226-1236.

\noindent	
Prialnik, D., and Podolak, M. (1995) Radioactive heating of porous comet 
nuclei. {\it Icarus}, 117, 420-430.

\noindent
Quillen, A. C., and Trilling, D. E. (1998) Do proto-jovian planets drive
outflows? {\it Astrophys. J.}, 508, 707-713.

\noindent
Rafikov, R. R. (2002a) Nonlinear propagation of planet-generated tidal waves.
{\it Astrophys. J.}, 569, 997-1008.

\noindent
Rafikov, R. R. (2002b) Planet migration and gap formation by tidally induced
shocks. {\it Astrophys. J.}, 572, 566-579.

\noindent 
Rafikov, R. R. (2005) Can giant planets form By direct
gravitational instability? {\em Astrophys. J.}, 621, L69-L69.

\noindent
Rice, W. K. M., Armitage, P. J., Wood, K., and Lodato, G. (2006) Dust filtration 
at gap edges: implications for the spectral energy distributions of discs with 
embedded planets. {\it Mon. Not. Roy. Astron. Soc.}, 379, 1619-1626.

\noindent
Ruskol, E. L. (1981) Formation of planets. In {\it The Solar System and Its
Exploration}, ESA, pp. 107-113 (SEE N82-26087 16-88).

\noindent
Ruskol, E. L. (1982) Origin of planetary satellites. {\it Izv. Earth Phys.},
18, 425-433.

\noindent
Ryu, D., and Goodman, J. (1992) Convective instability in differentially rotating
disks. {\it Astrophys. J.}, 308, 438-450.

\noindent
Safronov, V. S. (1960) On the gravitational instability in
flattened systems with axial symmetry and non-uniform rotation. {\em
Annales d'Astrophysique}, 23, 979-982.

\noindent
Safronov, V. S. (1969) Evolution of the protoplanetary cloud and formation of
the Earth and planets. Nauka, Moscow. [Transl.: Israel Program for Scientific
Translations, 1972. NASA TTF-667].

\noindent
Safronov, V. S., and Ruskol, E. L. (1977) The accumulation of satellites. In
{\it Planetary Satellites}, (J. A. Burns, ed.), pp. 501-512. University of
Arizona Press, Tucson.

\noindent
Safronov, V. S., Pechernikova, G. V., Ruskol, E. L., and Vitiazev, A. V. (1986) 
Protosatellite swarms. In {\it Satellites}, (J. A. Burns, and M. S. Matthews, 
eds.), pp. 89-116, University of Arizona Press, Tucson.

\noindent
Sandford, S. A., and 54 colleagues (2006) Organics captured from comet 
81P/Wild 2 by the Stardust spacecraft. {\it Science}, 314, 1720-1725.

\noindent
Sari, R., and Goldreich, P. (2004) Planet-disk symbiosis. {\it Astrophys. J.},
606, L77-L80.

\noindent
Sari, R., and Goldreich, P. (2006) Spherical accretion. {\it Astrophys. J.}, 
642, L65-L67.

\noindent	
Sasselov, D. D., and Lecar, M. (2000) On the snow line in dusty protoplanetary 
disks. {\it Astrophys. J.}, 528, 995-998.

\noindent	
Schlichting, H. E., and Sari, R. (2007) Formation of Kupier belt binaries.
{\it Astrophys. J.}, in press.

\noindent
Schmidt, O. Yu. (1957) Four lectures on the theory of the origin of the Earth.
Third ed., NA SSSR Press, Moscow.

\noindent	
Schmitt, B., Espinasse, S., Grim, R. J. A., Greenberg, J. M., and Klinger, J.
(1989) Laboratory studies of cometary ice analogues. In ESA, Physics and Mechanics 
of Cometary Materials, pp. 65-69.

\noindent       
Schubert, G., Anderson, J. D., Spohn, T., and McKinnon, W. B. (2004) Interior 
composition, structure and dynamics of the Galilean satellites. In {\it Jupiter.
The planet, satellites and magnetosphere}, (F. Bagenal, T. E. Dowling, and W. B. 
McKinnon, eds.), pp. 281-306. Cambridge planetary science, Vol. 1, Cambridge, 
UK: Cambridge University Press.

\noindent
Scott, E. R. D., and Krot, A. N. (2005) Thermal processing of silicate dust 
in the solar nebula: Clues from primitive chondrite matrices. 
{\it Astrophys. J.}, 623, 571-578.

\noindent
Sekiya, M. (1998) Quasi-equilibrium density distributions of small dust 
aggregations in the Solar Nebula. {\it Icarus}, 133, 298-309.

\noindent
Shakura, N. I., and Sunyaev, R. A. (1973) Black holes in binary systems. 
Observational appearance. {\it Astron. Astrophys.}, 24, 337-355.

\noindent
Shen, Y., Stone, J. M., and Gardiner, T. A. (2006) Three-dimensional 
compressible hydrodynamic simulations of vortices in disks.
{\it Astrophys. J.}, 653, 513-524.

\noindent
Shoemaker, E. M. (1984) Kuiper Prize Lecture, 16th DPS meeting, Kona, HI.

\noindent
Showalter, M. R. (1991) Visual detection of 1981S13, Saturn's
eighteenth satellite, and its role in the Encke gap. {\em Nature}, 351,
709-713.

\noindent
Showman, A. P., and Malhotra, R. (1997) Tidal vvolution into the Laplace 
resonance and the resurfacing of Ganymede. {\it Icarus}, 127, 93-111.

\noindent
Shu, F. H., Johnstone, D., and Hollenbach, D. (1993) Photoevaporation of the
solar nebula and the formation of the giant planets. {\it Icarus}, 106, 92-101.
 
\noindent
Shu, F. H., Shang, H., Gounelle, M., Glassgold, A. E., and  Lee, T. (2001)
The origin of chondrules and refractory inclusions in chondritic meteorites.
{\it Astrophys. J.}, 548, 1029-1050.

\noindent
Sohl, F., Spohn, T., Breuer, D., and Nagel, K. (2002) Implications from Galileo 
observations on the interior structure and chemistry of the Galilean satellites.
{\it Icarus}, 157, 104-119.

\noindent
Spohn, T., and Schubert, G. (2003) Oceans in the icy Galilean satellites of 
Jupiter? {\it Icarus}, 161, 456-467.

\noindent
Stepinski, T. F., and Valageas, P. (1997) Global evolution of solid matter in 
turbulent protoplanetary disks. II. Development of icy planetesimals.
{\it Astron. Astrophys.}, 319, 1007-1019.

\noindent
Stern, S. A., and Weissman, P. R. (2001) Rapid collisional evolution of comets 
during the formation of the Oort cloud. {\it Nature}, 409, 589-591.

\noindent
Stern, S. A. (2003) The evolution of comets in the Oort cloud and Kuiper belt.
{\it Nature}, 424, 639-642.

\noindent
Stevenson, D. J. (1982) Formation of the giant planets. {\em
Plan. Space Sci.}, 30, 755-764.

\noindent
Stevenson, D. J., Harris, A. W., and Lunine, J. I. (1986) Origins of
satellites. In {\it Satellites}, (J. A. Burns, and M. S. Matthews, eds.),
University of Arizona Press, Tucson.

\noindent
Stevenson, D. J., and Lunine, J. I. (1988) Rapid formation of Jupiter by 
diffuse redistribution of water vapor in the solar nebula. {\it Icarus},
75, 146-155.

\noindent
Stevenson, D. J., McKinnon, W. B., Canup, R., Schubert, G., and Zuber, M.
(2003) The power of JIMO for determining Galilean satellite internal 
structure and origin (abstract). American Geophysical Union, Fall Meeting 
2003, abstract \#P11C-05.

\noindent
Stewart, G. R. and Wetherill, G. W. (1988) Evolution of
planetesimal velocities. {\em Icarus}, 74, 542-553.

\noindent
Takata, T., and Stevenson, D. J. (1996) Despin mechanism for protogiant planets and
ionization state of protogiant planetary disks. {\it Icarus}, 123, 404-421.

\noindent
Takeuchi, T., and Lin, D. N. C. (2002) Radial flow of dust particles in accretion 
disks. {\it Astrophys. J.}, 581, 1344-1355.

\noindent
Tanigawa, T., and Watanabe, S. (2002) Gas accretion flows onto giant protoplanets: 
High-resolution two-dimensional simulations. {\it Astrophys. J.}, 580, 506-518.

\noindent
Tsiganis, K., Gomes, R., Morbidelli, A., and Levison, H. F. (2005) Origin of the 
orbital architecture of the giant planets of the Solar System. {\it Nature}, 435,
459-461.

\noindent
Turner, N. J., Sano, T., and Dziourkevitch, N. (2007) Turbulent mixing and the dead 
zone in protostellar disks. {\it Astrophys. J.}, 659, 729-737.

\noindent	
Umurhan, O. M., and Regev, O. (2004) Hydrodynamic stability of rotationally 
supported flows: Linear and nonlinear 2D shearing box results. 
{\it Astron. Astrophys.}, 427, 855-872.

\noindent
V\"{o}lk, H. J., Morfill, G. E., Roeser, S., and Jones, F. C. (1980) 
Collisions between grains in a turbulent gas.
{\it Astron. Astrophys.}, 85, 316-325.

\noindent	
Wadhwa, M., Amelin, Y., Davis, A. M., Lugmair, G. W., Meyer, B., Gounelle, M., 
and Desch, S. J. (2007) From dust to planetesimals: Implications for the Solar 
protoplanetary disk from short-lived radionuclides. In {\it Protostars and
Planets V}. (B. Reipurth, D. Jewitt, and K. Keil, eds.), pp. 835-848. 
University of Arizona Press, Tucson.

\noindent
Ward, W. R. (1997) Protoplanet migration by nebula tides. {\it Icarus}, 126,
261-281.

\noindent
Weidenschilling, S. J. (1977a) Aerodynamics of solid bodies in the solar nebula.
{\it Mon. Not. Roy. Astron. Soc.}, 180, 57-70.

\noindent
Weidenschilling, S. J. (1977b) The distribution of mass in the planetary system
and solar nebula. {\it Astrophys. Space Sci.}, 51, 153-158.
 
\noindent
Weidenschilling, S. J. (1984) Evolution of grains in a turbulent solar nebula.
{\it Icarus}, 60, 553-567.

\noindent
Weidenschilling, S. J. (1988) Formation processes and timescales for meteorite
parent bodies. In {\it Meteorites and the Early Solar System}, (J. F. Kerridge,
M. S. Matthews, eds.), University of Arizona Press, Tucson.

\noindent
Weidenschilling, S. J. (1997) The origin of comets in the solar nebula: a
unified model. {\it Icarus}, 127, 290-306.

\noindent
Weidenschilling, S. J. (2002) Structure of a particle layer in the midplane 
of the Solar Nebula: Constraints on mechanisms for chondrule formation.
{\it Meteor. Plan. Sci.}, 37, Supp., A148

\noindent
Weidenschilling, S. J. (2004) From icy grains to comets. In {\it Comets II}, 
(M. C. Festou, H. U. Keller, and H. A. Weaver, eds.), pp. 97-104, University of Arizona 
Press, Tucson.

\noindent	
Weidenschilling, S. J., and Davis, D. R. (1985) Orbital resonances in the solar 
nebula - Implications for planetary accretion. {\it Icarus}, 62, 16-29.

\noindent
Weidenschilling, S. J., and Cuzzi, J. N. (1993) Formation of planetesimals in 
the solar nebula. In {\it Protostars and Planets
III}, (E. H. Levy and J. I. Lunine, eds.), pp. 1031-1060. University of 
Arizona Press, Tucson.

\noindent
Wetherill, G. W. (1980) Formation of the terrestrial planets. {\em
Ann. Rev. Astron. Astrophys.}, 18, 77-113.

\noindent
Wetherill, G. W. (1986) Accumulation of the terrestrial planets and 
implications concerning lunar origin. In {\it Origin of the Moon}, Lunar and
Planetary Institute, Houston, pp. 519-550.

\noindent
Wetherill, G. W. (1990) Formation of the Earth. {\em Ann. Rev.
Earth Planet. Sci.}, 18, 205-256.

\noindent
Wetherill, G. W. and Stewart, G. R. (1989) Accumulation of a swarm
of small planetesimals. {\em Icarus}, 77, 330-357.

\noindent
Wetherill, G. W. and Stewart, G. R. (1993) Formation of planetary embryos - 
Effects of fragmentation, low relative velocity, and independent variation 
of eccentricity and inclination. {\it Icarus}, 106, 190.

\noindent
Wong, M. H., Lunine. J. I., Atreya, S. K., Johnson, T., Mahaffy, P. R., Owen,
T. C., and Encrenaz, T. (2008) Oxygen and other volatiles in the giant planets and 
their satellites. In {\it Oxygen in the Solar System}, (G. MacPherson et al., eds.)
{\it Rev. Mineral. Geochem.}, 68, pp. 219-246.  

\noindent
Wuchterl, G., Guillot, T., and Lissauer, J. J. (2000) Giant planet formation.
In {\it Protostars and Planets IV} (V. Mannings, A. Boss, and S. Russel,
eds.), pp. 1081-1109. University of Arizona Press, Tucson.

\noindent
Wurm, G., and Blum, J. (1998) Experiments on preplanetary dust aggregation.
{\it Icarus}, 132, 125-136.
 
\noindent
Youdin, A. N., and  Shu, F. H. (2002) Planetesimal formation by gravitational 
instability. {\it Astrophys. J.}, 580, 494-505.

\noindent
Youdin, A. N., and Chiang, E. I. (2004) Particle pileups and planetesimal 
formation. {\it Astrophys. J.}, 601, 1109-1119.

\noindent
Young, R. E. (2003) The Galileo probe: how it has changed our
understanding of Jupiter. {\em New Astron. Revs.}, 47, 1-51.

\noindent
Zimmer, C., Khurana, K. K., and Kivelson, M. G. (2000) Subsurface oceans on Europa 
and Callisto: Constraints from Galileo magnetometer observations. {\it Icarus},
147, 329-347.

\noindent
Zolensky, M. E., and 74 colleagues (2006) Minerology and petrology of comet 
81P/Wild 2 nucleus samples. {\it Science}, 314, 1735-1741.

\end{document}